\documentclass[12pt]{JHEP3}
\pdfoutput=1

\newcommand{\ex}{\mathrm{e}}

\usepackage{amsmath,epsfig}
\usepackage{amssymb,amsfonts}
\usepackage{latexsym}
\usepackage[latin1]{inputenc}
\usepackage{slashed}
\usepackage{empheq}
\numberwithin{equation}{section}
\usepackage{dsfont}
\usepackage{cancel}
\usepackage{overpic}
\usepackage{subcaption}
\usepackage{subeqnarray}
\usepackage{xcolor}

\usepackage{graphicx}
\usepackage{longtable}
\usepackage{color}
\usepackage{bbm}
\usepackage{lipsum,float}
\usepackage[font=footnotesize,labelfont=bf]{caption}

\newcommand{\exclude}[1]{}
\def\L{\mathcal{L}}

\def\a#1{\alpha_{#1}}

\def\beq{\begin{equation}}
\def\eeq{\end{equation}}
\def\be{\begin{equation}}
\def\ee{\end{equation}}
\def\bea{\begin{eqnarray}}
\def\eea{\end{eqnarray}}
\def\bal{\begin{align}}
\def\eal{\end{align}}

\def\2b2[#1,#2][#3,#4]{\left( \begin{array}{cc} #1 & #2 \\ #3 & #4 \end{array}
\right)}
\def\3b3[#1,#2,#3][#4,#5,#6][#7,#8,#9]{\left( \begin{array}{ccc} #1 & #2 #3 \\
#4 & #5 & #6\\#7&#8&#9\end{array} \right)}

\newcommand\fverb{\setbox\pippobox=\hbox\bgroup\verb}
\newcommand\fverbdo{\egroup\medskip\noindent%
                        \fbox{\unhbox\pippobox}\ }
\newcommand\fverbit{\egroup\item[\fbox{\unhbox\pippobox}]}

\newcommand{\bear}{\begin{eqnarray}}

\newcommand{\eear}{\end{eqnarray}}

\newcommand{\bsea}{\begin{subeqnarray}}
\newcommand{\esea}{\end{subeqnarray}}
\newbox\pippobox

\def\f{\varphi}

\def\6{\partial}
\def\a{\alpha}

\def\e{\epsilon}

\def\sp{\;\;\;,\;\;\;}

\def\sq
\def\a{\alpha}
\def\b{\beta}
\def\l{\lambda}

\def\hri#1#2{\href{http://arxiv.org/abs/#1}{[ArXiv:#1]#2}}
\def\hre#1#2{\href{http://arxiv.org/abs/#1/#2}{[ArXiv:#1/#2]}}

\def\hree#1#2{\href{https://doi.org/#1}{#2}}

\def\e{\epsilon}

\def\L{\Lambda}

\def\D{\Delta}

\def\dd{{\rm d}}
\def\rr{{\cal R}}

\title{Holographic QFTs on S$^2\times $S$^2$, spontaneous symmetry breaking and Efimov saddle points}

\author{Elias Kiritsis$^{1,2}$, Francesco Nitti$^{1}$ and Edwan Pr\'eau$^{3}$
\\
~\\
$^{1}$ \href{http://www.apc.univ-paris7.fr}{APC, Universit\'e de Paris},
 CNRS/IN2P3, CEA/IRFU, Obs. de Paris, 10 Rue Alice Domon et L\'eonie
 Duquet,  75205, Paris Cedex 13, France (UMR du CNRS 7164).
\\
 \\
$^{2}$\href{http://hep.physics.uoc.gr/}{Crete Center for Theoretical Physics}, Institute for theoretical and Computational Physics,  Department of Physics, University of Crete
 71003 Heraklion, Greece
\\
 \\
$^{3}$\href{http://www.phys.ens.fr/}{International Center for Fundamental Physics},
D\'epartement de Physique, \'Ecole Normale Sup\'erieure, 24 rue Lhomond, 75231 Paris Cedex, France}

\preprint{CCTP-2020-5\\ ITCP-IPP-2020/5}

\abstract{Holographic CFTs and holographic RG flows on
  space-time manifolds which are $d$-dimensional products of
  spheres are investigated. On the gravity side, this corresponds to Einstein-dilaton
  gravity on  an asymptotically $AdS_{d+1}$ geometry,
  foliated by a product of spheres. We focus on holographic theories on
  $S^2\times S^2$, we show that the only regular five-dimensional bulk
  geometries have an IR endpoint where one of the sphere shrinks to zero size, while the
  other remains finite. In the  $Z_2$-symmetric limit, where the two spheres
  have the same UV radii, we show the existence of a infinite discrete
  set of regular solutions, satisfying an Efimov-like discrete scaling.   The
  $Z_2$-symmetric solution in which both spheres shrink to zero at the
  endpoint is singular, whereas the solution with lowest free energy
  is regular and  breaks $Z_2$ symmetry spontaneously. We
  explain this phenomenon analytically  by identifying an unstable
  mode in the bulk around the would-be $Z_2$-symmetric solution. The space of theories have two branches that are connected by a conifold transition in the bulk, which is regular and correspond to a quantum first order transition. Our
  results also imply that $AdS_5$ does not admit a regular slicing by
  $S^2\times S^2$.}

\begin{document}

\section{Introduction, summary of results and outlook}

Quantum field theories are usually studied in flat background space-time.
We can consider them, however, in background space-times that have non-trivial curvature. Space-time curvature is irrelevant in the UV, as at short distances any regular manifold is essentially flat. However, curvature is relevant in the IR and affects importantly the low-energy structure of the QFT.

There are several motivations to consider QFT in curved backgrounds.
\begin{itemize}

\item Many computations in CFTs and other massless QFTs (like that of supersymmetric indices) are well-defined when a (controllable) mass gap is introduced, and this can be generated by putting the theory on a positive curvature manifold, like a sphere.
    This has been systematically used in calculating supersymmetric indices in CFTs, \cite{komar} as well as regulating IR divergences of perturbation theory in QFT, \cite{Adler,JR,CW} and string theory, \cite{KK}.

\item Partition functions of QFTs on compact manifolds, like spheres, are important ingredients in the study of the monotonicity of the RG Flow and the definition of generalized C-functions, especially in odd dimensions, \cite{myers,J,F}.

\item Cosmology has always given a strong motivation to study QFT in curved space-time, \cite{Fulling,BD}.
    Especially, QFT in de Sitter or almost de Sitter space is motivated by early universe inflation as well as the current acceleration of the universe.

\item The issue of quantum effects in near de Sitter backgrounds is a controversial issue even today, \cite{Mottola}-\cite{GKNW}.

\item Partition functions of holographic QFTs on curved manifolds are important ingredients in the no-boundary proposal of the wave-function of the universe, \cite{hh}, and serve to determine probabilities for various universe geometries.

\item Curvature in QFT, although UV-irrelevant is IR-relevant and can affect importantly the IR physics. Among other, things it can drive (quantum) phase transitions in the QFT, \cite{C}.

\item Putting holographic QFTs on curved manifolds potentially leads to constant (negative) curvature metrics sliced by curved slices.
    The Fefferman-Graham theorem guarantees that such regular metrics exist near the asymptotically AdS boundary, \cite{PG}. However, it is not clear whether such solutions can be extended to globally regular solutions in the Euclidean case (which may have horizons in Minkowski signature). The few facts  that are known can be found in \cite{WY,Ander}.

Using holography, it may be argued, that as we can put any holographic CFT on any manifold we choose, there should be a related regular solution that is dual to such a saddle point. This quick argument has however a catch: it may be that for a regular solution to exist, more of the bulk fields need to be turned-on (spontaneously), via asymptotically vev solutions.
We shall see in this paper, a milder version of this phenomenon associated with spontaneous symmetry breaking of a parity-like $Z_2$ symmetry.

\end{itemize}

In this paper we are going to pursue a research program started in \cite{exotic} and \cite{C}, that investigates the general structure of
holographic RG flows for QFTs defined on various spaces that involve beyond flat space, constant curvature manifolds. The cases  analysed so far systematically concern the flat space case (or equivalently $(S^1)^d$), and the $S^d$, $dS_{d}$ and $AdS_d$ cases, although the results in \cite{C,F,GKNW} are valid for any d-dimensional Einstein manifold.

The case of $S^1\times S^{d-1}$ has also been studied extensively as it contains $AdS_{d+1}$ in global coordinates, and RG flows were also analysed in this case.
The general problem we are now interested in, is the case where the boundary is a product of constant (positive) curvature manifolds, which we shall take  without a loss of generality to be spheres.

In the CFT case, unlike the $S^d$ and $S^1\times S^{d-1}$ cases, there is no known slicing of $AdS_{d+1}$  by other sphere product manifolds , and even in this cases the solutions if they are regular must be non-trivial.

Solutions for CFTs on sphere product manifolds have been recently investigated in \cite{aharony} and phases transitions were found, generalizing the Hawking-Page transition (that is relevant in the $S^1\times S^{d-1}$ case), \cite{HP}.

In this work,  we  study four-dimensional holographic QFTs
on $S^2\times S^2$. The QFTs we
shall consider  are either CFTs, or RG-flows driven by
 a single scalar operator $O$ of dimension $\Delta$.

 We shall describe the
dual theory in the
 bulk, in terms of five-dimensional Einstein-dilaton gravity.  The relevant
geometries, describing the ground-state of the theory,  are then asymptotically $AdS_5$ space-times which admit a
radial  $S^2\times S^2$ foliation. In the case of CFTs, these
space-times are
solution of pure gravity with a negative cosmological constant. In
the case of RG flows, these are solutions of Einstein-dilaton gravity
with an appropriate scalar  potential.

The geometry  $S^2\times S^2$  is the only four-dimensional sphere product
manifold which has not been already studied in detail.
 It is also the only one that, as we shall see,  cannot be used to
 slice the AdS$_5$ metric.

Below,  we briefly summarize our setup and our main results.

We consider geometries whose metric is of the form:
\be \label{intro1}
ds^2 = du^2 + e^{2A_1(u)} \alpha_1^2 d\Omega_1^2 + e^{2A_2(u)} \alpha_2^2 d\Omega_2^2
\ee
where $\alpha_1$ and $\alpha_2$ are constants with dimensions of length, and $d\Omega_i^2$ are the metrics of $S^2$s with radius 1.
The generic space-time symmetry of the QFT, is $SU(2)\times SU(2)$ associated with the two spheres. When the spheres have equal size, $\alpha_1=\alpha_2$, then we have an extra $Z_2$ space-time symmetry that interchanges the two spheres.

The holographic coordinate $u$ in (\ref{intro1}) runs from the conformal boundary at $u = -\infty$,
corresponding to a UV fixed-point of the dual QFT,  to
an IR endpoint $u_0$ where the manifold ends regularly.

\paragraph*{The UV geometry and parameters.}
 In the near-boundary region, the space-time asymptotes $AdS_5$ with
 length-scale $\ell$.  The metric on the  boundary is (with an appropriate definition  of the scale
factors) conformally equivalent to the four-dimensional metric
\be\label{intro2}
ds_{bdr}^2 =  \alpha_1^2 d\Omega_1^2 + \alpha_2^2 d\Omega_2^2
\ee
This is the metric on the  $S^2\times S^2$ manifold on which the dual
UV field theory is defined.  Near the boundary, the scalar field
behaves at leading order as
\be\label{intro3}
\f \simeq \left(\f_-\ell^{\Delta_-}\right)\,  e^{\Delta_- u/\ell} + \ldots \qquad u \to -\infty
\ee
where $\f_-$ is a constant, $\ell$ the $AdS_5$ length, and   $\Delta_-=  4-\Delta >0$.

The UV theory is defined in terms of three sources,  entering equations (\ref{intro2}-\ref{intro3}):
\begin{itemize}
\item The scalar source $\f_-$ dual to the  relevant  coupling
  deforming the UV CFT;
\item The radii $\alpha_1$ and $\alpha_2$ of the two spheres, or
  equivalently their scalar curvatures $R_i^{UV} \equiv 2/\alpha_i^2$.
\end{itemize}
They can be combined into two dimensionless parameters,
\be \label{intro3-ii}
{\mathcal R}_1 = {R_1^{UV} \over \f_-^{2/\Delta_-}}, \quad {\mathcal
  R}_2 = {R_2^{UV} \over \f_-^{2/\Delta_-}}.
\ee

At subleading order in the UV expansions one finds two more
dimensionless integration constants of the bulk field equations, that we denote
$C_1$ and $C_2$ and are related to vevs of the field theory
operators. Schematically, they
enter   the scalar field  and the scale factors in the following way:
\be \label{intro4}
A_1 \sim \ldots + C_1 e^{4u/\ell} +\ldots , \quad A_2 \sim \ldots + C_2
e^{4u/\ell} +\ldots
\ee
\be\label{intro5}
 \f \sim \ldots + (C_1 + C_2)
e^{(4-\Delta_-) u /\ell} + \ldots
\ee

\begin{itemize}
\item The combination $C_1+C_2$ sets the vacuum expectation value of
  the scalar field dual to $\f$,
\be \label{intro6}
\langle O \rangle \propto  \f_-^{\Delta/\Delta_-}(C_1+C_2)
\ee
 This combination  also enters in the expectation
  value of the trace of the stress tensor.
\item The combination $C_1-C_2$ enters at order $e^{4u/\ell}$  in the
  difference of the scale factors,
\be \label{intro7}
A_1 -A_2 = \ldots + (C_1 - C_2) e^{4u/\ell} + \ldots \qquad  u\to -\infty
\ee
and it enters the difference in the stress tensor vevs along the
two spheres.

\item Additional terms in the stress tensor vev come from the
  curvatures $\mathcal{R}_i$, and they reproduce in particular the
  Weyl anomaly on $S^2\times S^2$.
\item In the limit where the source $\f_- \to 0$, one can find ``pure
  vev'' solutions, in which the leading  asymptotics of the scalar
  field are
\be \label{intro8}
\f \simeq \left(\f_+  \ell^{\Delta} \right)e^{\Delta  u/\ell} +\ldots , \qquad   u\to -\infty
\ee
In this case, only the combination $C_1-C_2$ is allowed to be
non-zero, and the scalar  vev parameter $C_1+C_2$ is replaced by the
constant parameter
$\f_+$. These solutions are dual to vev-driven  flows: the
source of the operator dual to $\f$ is set to zero, but a non-zero
condensate triggers a non-trivial RG flow.  These solutions have one
free parameter less than the source-driven flows, therefore, they are
generically  singular in the IR unless the bulk potential is  appropriately
fine-tuned \cite{C,Gur}. 
\end{itemize}

\paragraph*{IR Geometry.}  Regular solutions of the form (\ref{intro1}) have
an IR endpoint at some finite $u = u_0$, where the geometry has the
following properties:
\begin{itemize}
\item At the endpoint, one of the two scale factors $e^{A_i(u_0)}$
    vanishes, and the corresponding   sphere shrinks to zero size,
    while the other sphere stays at finite size. Near $u_0$, the metric has the
    form
\be \label{intro9}
ds^2 \simeq du^2 + (u-u_0)^2 d\Omega_1^2 + \alpha_{IR}^2  d\Omega_2^2,
\ee
and the geometry is isometric to  $R^3 \times S^2$. Therefore the topology of the solution is that of $D_3\times S^2$ where $D_3$ is a ``cigar" with $S^2$ slices or equivalently a hemisphere of an $S^3$.  Instead, any solution
in which the two spheres shrink to zero at the same point, has necessarily a
curvature singularity.
\item Regularity imposes two constraints on the four dimensionless UV
  parameters $\mathcal{R}_1,  \mathcal{R}_2, C_1, C_2$. Choosing  the
  sources $\mathcal{R}_1,  \mathcal{R}_2$ as independent free
  parameters, regularity fixes the vevs  $C_1, C_2$ as a function of
  the sources, as it always happens in holography:

\be \label{intro10}
C_{1,2} = C_{1,2}(\mathcal{R}_1,  \mathcal{R}_2)
\ee

\item From an analysis of the quadratic curvature invariants, we show
  that no regular slicing   of Euclidean $AdS_5$  by $S^2\times S^2$
  exists (unlike the known slicing by  $S^1\times S^3$, which
  corresponds to
  $AdS_5$ in global coordinates, and by $S^4$, discussed in \cite{C}). Instead,  it is
  possible to  find a regular slicing of  $AdS_5$ by $AdS_2\times S^2$.
\end{itemize}

\paragraph*{Efimov spiral and spontaneous $\mathbb{Z}_2$ breaking.}
If the UV radii of the two spheres are very different, there is a
single regular solution, in which it is  the sphere with the smallest UV
radius that shrinks to zero size in the IR. As the UV curvatures
become comparable however, multiple solutions start appearing with one
or the other sphere shrinking to zero in the IR.

 The limit in which  $\mathcal{R}_1 = \mathcal{R}_2$ is particularly
interesting. In this case, the two spheres start with the same size
in the UV, and the theory has a space-time $\mathbb{Z}_2$ symmetry under which
the two spheres are interchanged.

However, this symmetry is  broken by the dynamics. This is seen in the gravitational solution where in the bulk,  the symmetric
solution, in which $A_1 (u)= A_2(u)$ all the way to the IR endpoint,
is singular, as we have discussed above. Any regular solution  must
therefore break the $\mathbb{Z}_2$ symmetry by the presence of a non-zero vev
parameter $C_1- C_2$, which causes $A_1 (u) - A_2(u)$ to depart from
zero as we move away from the boundary, as in equation
(\ref{intro7}). This is similar to what happens with the singular
conifold in the Klebanov-Strassler solution  \cite{Klebanov:2000hb}, where the
non-zero vev which avoids the singularity, is associated to gaugino condensation. In our case however, it
corresponds to the spontaneous breaking of a discrete space-time
symmetry.

\begin{figure}[h]
\centering
\begin{overpic}
[width=0.65\textwidth]{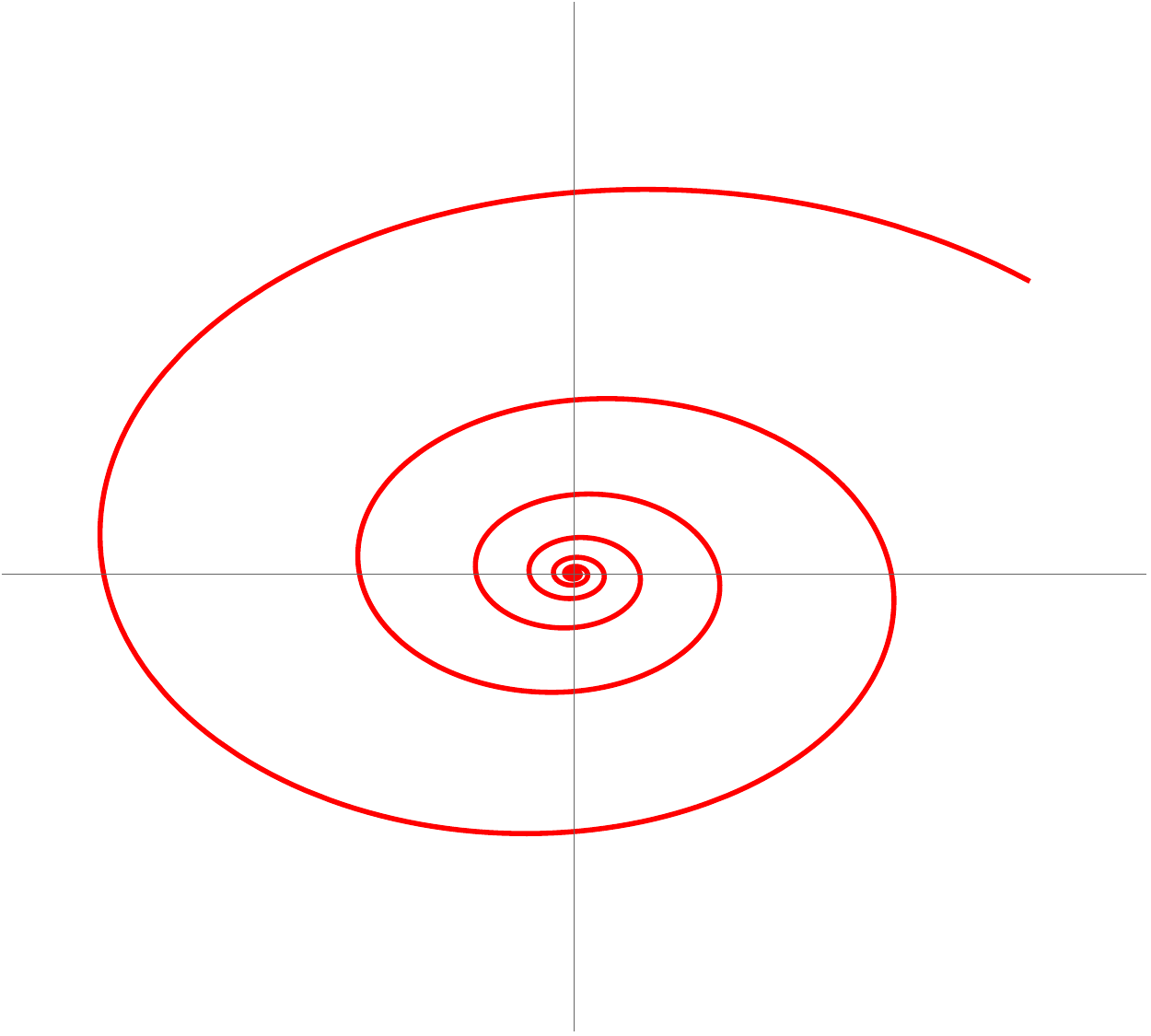}
\put(95,35){${\mathcal{R}_1\over \mathcal{R}_2}-1$}
\put(50,85){$C_1-C_2$}
\end{overpic}
\caption{Efimov spiral. Each point on the spiral represents a regular solution
  with  sphere 1 shrinking to zero size and sphere 2 remaining finite
  in the IR. As we proceed towards the center, the IR radius of sphere
  2 becomes smaller and smaller. The
  origin corresponds to  the singular solution, in which both spheres
  shrink to zero size.}
\label{fig:intro1}
\end{figure}

 This is indeed what happens: as $\mathcal{R}_1/\mathcal{R}_2 \to 1$, the
 theory develops an {\em infinite discrete set} of regular solutions,
 characterized by a smaller and smaller IR radius $\alpha_{IR}$ of the finite
 $S^2$, which approaches the singular solution  characterized by
 $\alpha_{IR}=0$ and $C_1=C_2$. In this regime, the solutions follow a discrete
 scaling law well described by an Efimov spiral in the plane
 $(\mathcal{R}_1/\mathcal{R}_2, C_1-C_2)$, given schematically by:
\be \label{Efimov}
{\mathcal{R}_2 \over \mathcal{R}_1} -1 = A \sin(s + \phi_1) e^{-b s},
\qquad C_1 - C_2 = B \sin(s + \phi_2) e^{-b s}
\ee
where $A,B$, $b$ and $\phi_1,\phi_2$ are constants, and   $s \sim \log(\ell/\alpha_{IR})$ runs to infinity in the singular limit\footnote{Here,  $\alpha_{IR}$ is the radius of
the sphere 2 (the one that does not shrink  to zero)  at the IR
endpoint, defined in equation (\ref{intro9}).}.   The schematic behavior is shown in figure
\ref{fig:intro1}, where each point of the spiral corresponds to a
regular  solution.  The solutions
lying on the vertical axis form an infinite countable set and correspond to a
symmetric UV boundary condition  $\mathcal{R}_1 =  \mathcal{R}_2$. The
center of the spiral corresponds to the singular solution with
$\mathcal{R}_1 =  \mathcal{R}_2$ and  $C_1=C_2$.

The symmetric solutions correspond to vanishing ${\mathcal{R}_2
  /\mathcal{R}_1} -1$ and one can find an  infinite number of them at
discrete values of $s$. The corresponding values of $C_1-C_2$ get
smaller and smaller as $s$ grows larger.

The  Efimov behavior  (\ref{Efimov}) is confirmed by numerical examples, both in the case of
a CFT (no scalar field, in which case it was already observed
in \cite{aharony})  and in the case of holographic RG-flows.

This type of  Efimov scaling has been observed in other contexts in
holography, \cite{JKST,Liu,JK}.
For example,  in holographic QCD-like
theories,  it is associated to a would-be IR fixed
point developing an instability to a violation of the BF bound
\cite{JK}. The associated symmetry that is broken is the chiral symmetry.
 Interestingly, a similar interpretation can be found in the
present context: we show that the scale factor difference $A_1-A_2$
behaves, away from the boundary, as an unstable perturbation. This
behavior  can be generalized to spheres of different dimensions, as discussed in Appendix \ref{sec:genprod}

\paragraph*{Conifold transition.} Having established that there may be
multiple solutions for a given  choice of the UV parameters
$\mathcal{R}_1$ and $\mathcal{R}_2$ (and even an infinite number for
the $\mathbb{Z}_2$-symmetric choice $\mathcal{R}_1= \mathcal{R}_2$) we analyse
which one is the dominant saddle-point solution. For this, we compute the free energy of the
regular solutions  as a function  of   $\mathcal{R}_1, \mathcal{R}_2$
by evaluating the Euclidean on-shell action. The solution with the
lowest free energy at fixed   $\mathcal{R}_1, \mathcal{R}_2$ is the
ground state of the system.

\begin{figure}[h!]
\centering
\begin{overpic}
[width=0.65\textwidth]{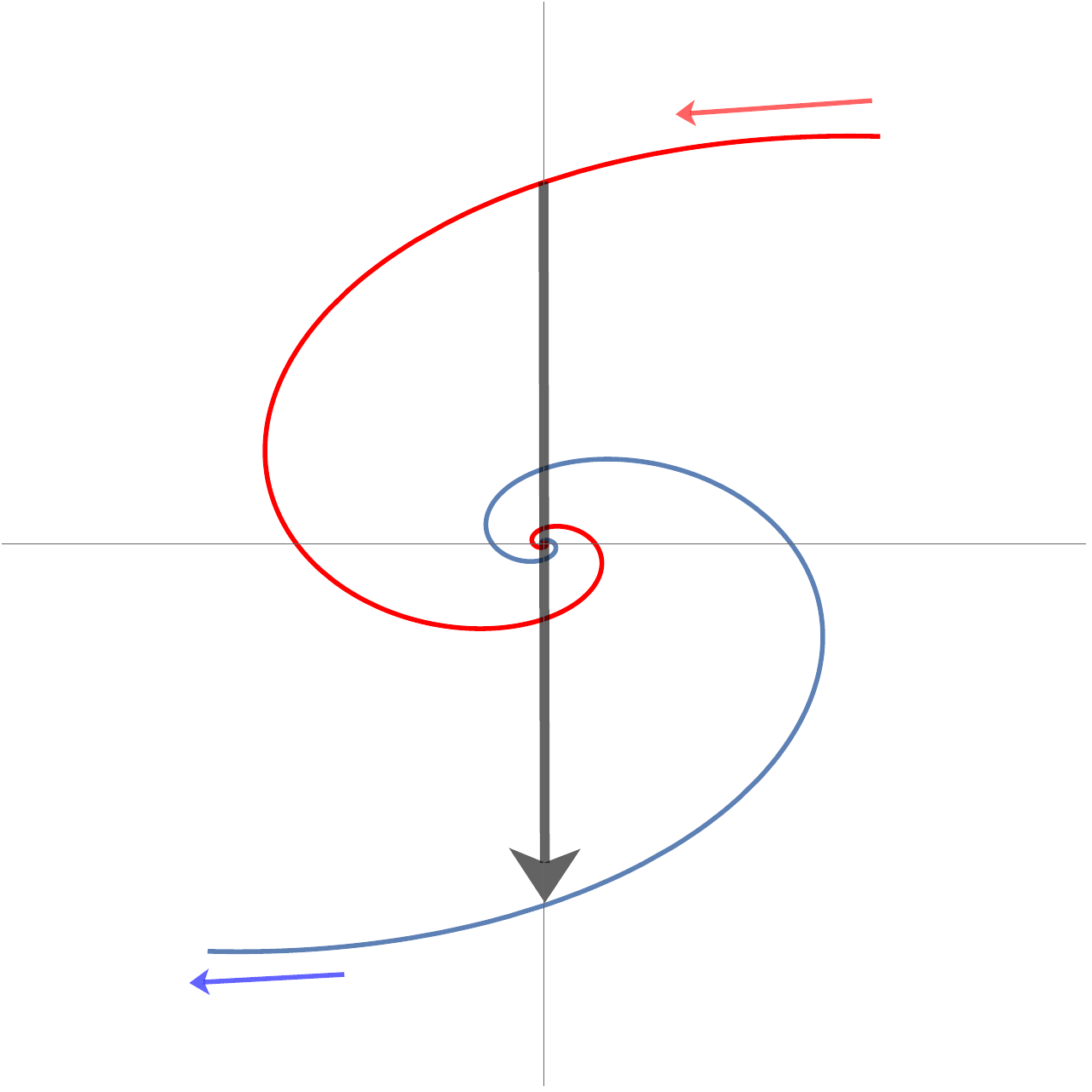}
\put(95,45){${\mathcal{R}_1\over \mathcal{R}_2}-1$}
\put(35,95){$C_1-C_2$}
\put(85,90){$\mathcal{R}_1 >  \mathcal{R}_2$}
\put(0,8){$\mathcal{R}_1 <  \mathcal{R}_2$}
\end{overpic}
\caption{The two spirals  correspond to the solutions in which sphere 1
shrinks to zero in the IR (red) and to those in which sphere 2 shrinks
(blue). As we cross the vertical axis, the dominant solution jumps
from the red to the blue spiral.}
\label{fig:intro2}
\end{figure}
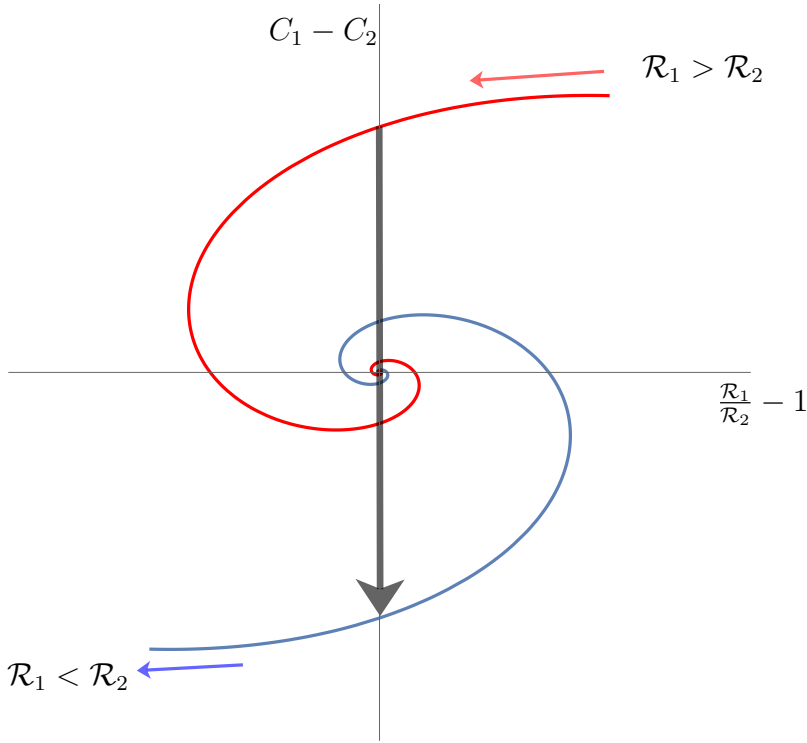
Both in the case of a CFT and of a non-trivial RG flow, we find that
the lowest free energy corresponds to the {\em first} occurrence along
the spiral of a given value of $\mathcal{R}_1/\mathcal{R}_2$, i.e. the
solution  which is farthest from the center.

The two dominant saddle points that exist for $\mathcal{R}_1 = \mathcal{R}_2$
correspond to the dominant Efimov solutions of the two branches $\mathcal{R}_1 > \mathcal{R}_2$ and $\mathcal{R}_1 < \mathcal{R}_2$.
They correspond to the two vacua of the theory with $\mathcal{R}_1 = \mathcal{R}_2$ and they are related by the spontaneously-broken $Z_2$ symmetry.

\begin{figure}[h!]
\centering
\begin{overpic}
[width=1.0\textwidth]{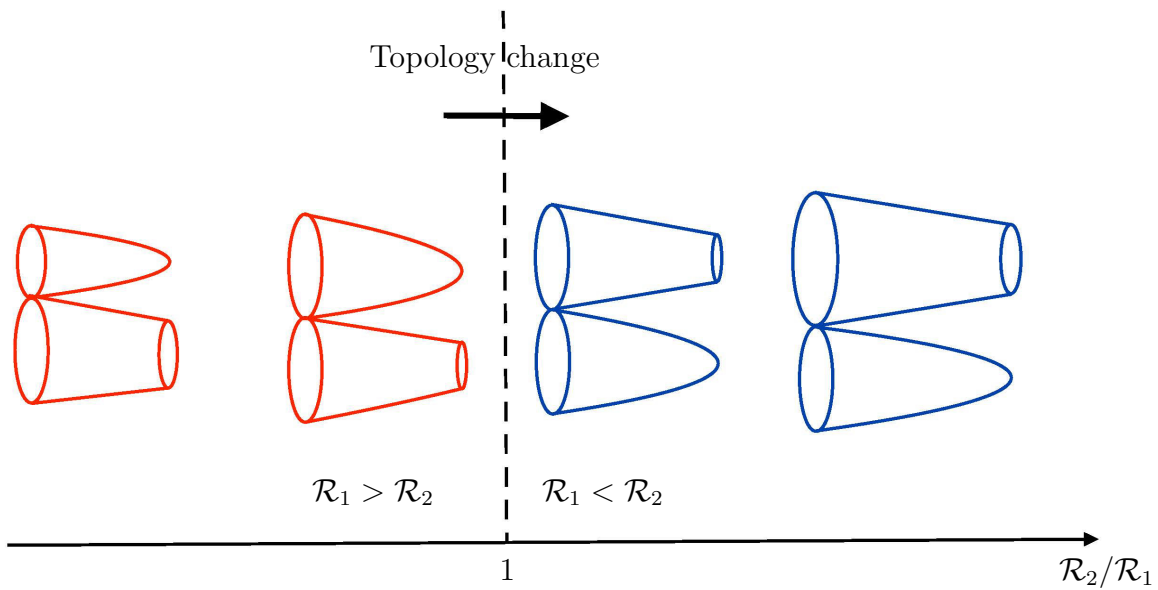}
\put(35,48){Topology change}
\put(46.3,3){1}
\put(95,3){${\mathcal{R}_2/\mathcal{R}_1}$}
\put(50,10){$\mathcal{R}_1 <  \mathcal{R}_2$}
\put(30,10){$\mathcal{R}_1 > \mathcal{R}_2$}
\end{overpic}
\caption{Conifold transition. The horizontal axis represents the
  ratio of the UV curvatures of the two spheres. In this figure the UV
  radius of sphere 2 is kept fixed, while that of sphere 1 is
  increased going from left to right.}
\label{fig:intro3}
\end{figure}

If we start in the regime  $\mathcal{R}_1 > \mathcal{R}_2$, on the
branch where sphere 1 shrinks to zero
and decrease the value of $\mathcal{R}_1$, at the
symmetric point
$\mathcal{R}_1/\mathcal{R}_2 =1$  the system undergoes   a first order
phase transition to the solution where the two spheres are
interchanged: decreasing  ${\mathcal R}_1$
further,
the dominant branch becomes the one in which sphere 2 shrinks to zero size
and sphere 1 remains finite. This  transition is shown
schematically in figure \ref{fig:intro2}. The classical saddle point
undergoes a topology-changing transition
similar to the conifold transition, see figure \ref{fig:intro3}.
We should stress that this conifold transition occurs via regular bulk solutions, and its signal in the boundary QFT is as a first order phase transition not unlike the one in large-N YM at $\theta=\pi$, \cite{witten}.
It seems to be quite distinct from the conifold transition of CY vacua in string theory, \cite{Stro}, where the singularity is resolved by non-perturbative effects in the string coupling.

\paragraph*{Outlook.} The study  of holographic solutions on products
of spheres can be extended to arbitrary dimension and arbitrary
number of spheres. The topology changing transition has the  potential
applications to unveil other (known or otherwise) transitions in
holography, when these can be embedded into a higher dimensional
theory by  the mechanism of generalized dimensional uplift
\cite{Gouteraux:2011ce, Gouteraux:2011qh}:
 it is known for example that reducing a higher
dimensional $AdS$ solution on a sphere yields a confining  holographic
theory in lower dimensions, \cite{Gouteraux:2011ce}. Starting from a product of spheres can be
useful to understand the phase structure of confining theories on
curved manifolds.

Another interesting question from the perspective of generalized
dimensional reduction/uplift is the origin of the discrete scaling
discussed in this paper: here,  we have identified an unstable mode
underlying this phenomenon, but it would be interesting to rephrase it
in terms of the more familiar language of scalar fields violating the
BF bound at some IR fixed point, as it was the case in
\cite{JKST,Liu,JK}. This formulation can be reached  using the method of
generalized dimensional reduction, starting from the sphere-product
ansatz in higher dimensions.

Another interesting direction  is to consider product manifolds which
have less symmetry, for example squashed spheres. These have found a
recent application in the context of cosmology, in
particular in the holographic approach to the no-boundary wave
function of the universe \cite{Conti:2017pqc}.

Finally, in this work we have not fully explored the case where one of
the factors in the product
manifold has negative curvature. Extending this work to that situation
can lead to  a better understanding of  AdS/AdS holography, which in
the case of RG flows presents some difficulty due to the appearance of
conical singularities on the boundary \cite{C}. It would be useful to
understand these situations by generalized uplift to a higher
dimensional AdS geometries foliated by a lower-dimensional AdS times
spheres or tori.

This work is organized as follows.  In Section \ref{sec:2}, we describe in detail the setup and derive  the
equations of motion for holographic RG flow solutions

In section \ref{sec:first} we develop a first-order formalism adapted
to the sphere-product ansatz, along the lines of the formalism
developed in \cite{C}. This is particularly useful for the calculation
of the on-shell action.

In section \ref{sec:nb} we discuss the UV asymptotics of the solution
and identify the parameters which correspond to dual field theory
quantities.

In section \ref{sec:IRbc} we discuss the IR side of the geometry and
we identify the conditions for the absence of singularities.

In section \ref{sec:FE} we find the expression for  the free energy,
a.k.a. the on-shell action,  in terms of the UV data (curvatures and
vev parameters), including the appropriate holographic
renormalization.

In section \ref{sec:CFT} we analyze in detail  solutions in
the case of a CFT:  we construct
solutions in numerical examples, and show that they  undergo a conifold phase transition. In this section we
also  discuss the Efimov scaling in the CFT case, of which we give an
analytical derivation in terms of an IR instability.

In section \ref{sec:rgflow} we present numerical examples of  the more
general case of holographic RG flows on $S^2\times S^2$, where we
again display the Efimov phenomenon and topology-changing transition.

Most technical details are presented in the Appendix.

\subsection*{Note added} During completion of this work we became
aware of the article \cite{aharony}, which addressed similar problems  and
reached some of our conclusions, in the absence of a bulk scalar
field.

\section{Holographic Theories on S$^2\times$ S$^2$} \label{sec:2}

In this section we  consider the bulk holographic description of RG flows of four-dimensional QFTs on $S^2\times S^2$. But before we specialize to this,
we shall first consider general holographic solutions that are cones over  a general product of spheres.

As a bulk holographic theory we consider an Einstein-scalar theory in (d + 1)-dimensions with Euclidean or Lorentzian metric (in the latter case we use the mostly plus convention). The two relevant bulk fields we keep are the metric, dual to the conserved QFT energy-momentum tensor, and a scalar field $\f$,  dual to a relevant scalar operator $O$ that is driving the RG flow of the boundary QFT\footnote{In general the bulk action contains many other scalars. We suppress their presence as we work with the single combination that is non-trivial for the flow.}.
The bulk theory in the infinite coupling limit, is described by the following two-derivative action (after field redefinitions):

\begin{align}
\label{eq:action} S[g, \f]= \int du \, d^ dx \, \sqrt{|g|} \left(R^{(g)} - \frac{1}{2} \partial_a \f \partial^a \f - V(\f) \right) + S_{GHY} \, ,
\end{align}
where we also included the Gibbons-Hawking-York term $S_{GHY}$. In the
Lorentzian signature we use the $(-,+\ldots +)$ convention for the
metric. The
Euclidean action is defined by setting $S_E = - S$ and changing the
metric to positive signature.

\subsection{The conifold ansatz}
Consider a  boundary (holographic) QFTs defined
on a space that is a product of Einstein manifolds. The natural bulk
metric ansatz in such a case, that preserves all the original
symmetries of the boundary metric, is given in terms of a domain wall holographic coordinate $u$
and a conifold  ansatz  (for both Euclidean and Lorentzian signatures):
\begin{align}
\label{eq:metric} \f = \f(u) \sp \dd s^2 = g_{\mu \nu} \dd{x^\mu} \dd{x^\nu} = \dd u^2 + \sum_{i=1}^n \mathrm{e}^{2A_i(u)} \zeta^i_{\alpha_i, \beta_i} \dd{x^{\alpha_i}} \dd{x^{\beta_i}} \, .
\end{align}
Here the constant $u$ slices are products of $n$ Einstein manifolds, each with metric $ \zeta^i_{\alpha_i, \beta_i}$, dimension $d_i$ and coordinates $x^{\alpha_i}$, $\a_i=1,2,\cdots,d_i$.
Each Einstein manifold is associated to a different scale factor $A_i(u)$, that depends on the coordinate $u$ only. Therefore,  every $d$-dimensional slice at constant $u$ is given by the product of $n$ Einstein manifolds of dimension $d_1, \ldots, d_n$.

The fact  that these are Einstein manifolds translates in the following relations\footnote{These can be either of Euclidean signature (like a $d_i$-dimensional sphere) or Lorentzian signature (for example  a $d_i$-dimensional de Sitter space in the maximal symmetry case)
In the rest of the paper we shall mainly refer to the Euclidean case
keeping in mind that the results also hold for Lorentzian signature.}
\begin{align}
\label{eq:Rzeta}
R^{(\zeta^i)}_{\mu \nu} = \kappa_i \zeta^i_{\mu \nu} \, , \quad R^{(\zeta^i)} = d_i \kappa_i \, , \quad
\end{align}
where $\kappa_i$ is the (constant) scalar curvature  scale of the $i$th manifold and no sum on $i$ is implied.
We have the identity
\be
\sum_{i=1}^n~d_i=d
\ee

In the case of maximal symmetry,
\be
\quad \kappa_i = \left\{
  \begin{array}{c l}\displaystyle{
   \hphantom{-} \frac{(d_i-1)}{\alpha_i^2} } & \quad \textrm{dS}_{d_i}~~{\rm or} ~~S^{d_i}\\
   0 & \quad \mathcal{M}_{d_i} \\
{\displaystyle- \frac{(d_i-1)}{\alpha_i^2}} & \quad \textrm{AdS}_{d_i}\\
  \end{array} \right. \, \ ,
  \ee
where $\alpha_i$ are associate radii and $\mathcal{M}_{d_i}$ denotes $d_i$-dimensional Minkowski space. \\[0.1cm]

In the following, we adhere to the following shorthand notation: derivatives with respect to $u$  are denoted by a dot while derivatives with respect to $\f$ are denoted by a prime, i.e.:
\begin{align}
\dot{f}(u) \equiv \frac{d f(u)}{du} \, , \qquad g'(\f) \equiv \frac{d g(\f)}{d \f} \, .
\end{align}

The non-trivial components of Einstein's equation are:
\be
\label{eq:EOM1bis} \left(\sum_{k=1}^n d_k \dot{A_k}\right)^2 -
\sum_{k=1}^n d_k \dot{A_k}^2 - \sum_{k=1}^n \mathrm{e}^{-2A_k}
R^{\zeta^k} - \frac{1}{2} \dot{\varphi}^2 + V  = 0, \qquad uu
\ee
\begin{equation} \label{eq:EOM4bis}
2(1 - \frac{1}{d}) \sum_{k=1}^n d_k \ddot{A_k} + \frac{1}{d} \sum_{i,
  j}d_id_j (\dot{A_i} - \dot{A_j})^2 + \frac{2}{d} \sum_{k=1}^n
\mathrm{e}^{-2A_k} R^{\zeta^k} + \dot{\varphi}^2 = 0, \qquad ii
\end{equation}
\begin{equation} \label{eq:EOM5bis}
\ddot{A_i} + \dot{A_i} \sum_{k=1}^n d_k \dot{A_k} - \frac{1}{d_i} \mathrm{e}^{-2A_i} R^{\zeta^i} = \ddot{A_j} + \dot{A_j} \sum_{k=1}^n d_k \dot{A_k} - \frac{1}{d_j} \mathrm{e}^{-2A_j} R^{\zeta^j}\sp i\not= j
\end{equation}
The Klein-Gordon equation reads:
\be
\label{eq:EOM3bis} \ddot{\varphi} + \left(\sum_{k=1}^n d_k \dot{A_k} \right) \dot{\varphi} - V'  = 0.
\ee

These equations are the same for both Lorentzian and Euclidean
signatures, so all our results  hold for both cases.

Holographic RG flows are in one-to-one correspondence with regular
solutions to the equations of motion
\eqref{eq:EOM1bis}--\eqref{eq:EOM3bis}. Hence, in the following we shall be
interested in the structure and properties of solutions to these equations for various choices of the bulk
potential $V(\f)$.

To be specific, we assume that $V(\f)$  has at least one
  maximum, where it takes a negative value. This ensures that there exists a
UV conformal fixed point, and a family of asymptotically AdS
solutions which correspond to deforming the theory away from the fixed
point by the relevant scalar operator dual to $\f$.

 In addition, $V(\f)$ may have other
maxima and/or minima (in the AdS regime, $V<0$) representing distinct UV or IR fixed points for
the dual QFT.

 Note that every equation can be associated with its equivalent in the case with a single sphere, \cite{C}, except for \eqref{eq:EOM5bis} that gives additional constraints on the solutions. One may also observe
 that these constraints are automatically  satisfied by $A_i = A(u)$
 for all $i$ and $R^{\zeta^i} = d_i \kappa$ for all $i$, where
 $\kappa$ is a constant and $A(u)$ is a function of $u$. In this case,
 the equations of motion \eqref{eq:EOM1bis}--\eqref{eq:EOM3bis} reduce
 to the equations one obtains for of a single sphere \cite{C}.
This could be foreseen since under these conditions there is only one scale factor and the product space is an single Einstein manifold.

In appendix \ref{A} we match these equations to some known special cases.

\subsection{The $S^2\times S^2$ case}
Now we restrict ourselves to the case $d=4$ and consider the product
of two 2-dimensional Einstein manifolds. In our ansatz, the metric reads:

\begin{align}
\label{eq:metricS2} \f = \f(u) \sp \dd s^2 = \dd u^2 + \mathrm{e}^{2A_1(u)} \zeta^1_{\alpha_1, \beta_1} \dd{x^{\alpha_1}} \dd{x^{\beta_1}} + \mathrm{e}^{2A_2(u)} \zeta^2_{\alpha_2, \beta_2} \dd{x^{\alpha_2}} \dd{x^{\beta_2}} \, .
\end{align}
where $\zeta^i_{\alpha_i \beta_i}$ is a fiducial, $u$-independent
$2$-dimensional metric of the each of the two  Einstein manifolds. In two dimensions, compact Einstein manifolds of positive curvature are spheres. On the other hand, if the  curvature is negative, there can be many Riemann surfaces with $g>1$.
We shall not consider the negative curvature case further as in that case the non-trivial holographic flows have extra singularities \cite{C}.

In the sequel we  assume that the slice manifold is $S^2\times S^2$.
The equations of motion specialize in this case to:

\begin{align}
\label{eq:EOM6bis} 2\dot{A_1}^2 + 2\dot{A_2}^2 + 8 \dot{A_1} \dot{A_2} -R^{\zeta_1}  \mathrm{e}^{-2A_1} - R^{\zeta_2}\mathrm{e}^{-2A_2} - \frac{1}{2} \dot{\varphi}^2 + V & = 0, \\
\label{eq:EOM7bis} 3\ddot{A_1} + 3\ddot{A_2} + 2\left(\dot{A_1} - \dot{A_2}\right)^2+ {R^{\zeta_1}\over 2}\mathrm{e}^{-2A_1} + {R^{\zeta_2}\over 2}\mathrm{e}^{-2A_2} + \dot{\varphi}^2 & = 0, \\
\label{eq:EOM8bis} \left(\ddot{A_1} + 2\dot{A_1}^2 - {R^{\zeta_1}\over 2}\mathrm{e}^{-2A_1} \right) - \left(\ddot{A_2} + 2\dot{A_2}^2 - {R^{\zeta_2}\over 2} \mathrm{e}^{-2A_2} \right) & = 0, \\
\label{eq:EOM9bis} \ddot{\varphi} + 2\left(\dot{A_1} + \dot{A_2} \right) \dot{\varphi} - V' & = 0.
\end{align}

As mentioned earlier, a trivial solution for the constraint \eqref{eq:EOM8bis} is given by $A_1 = A_2$.

If we set $A_1=A_2=A$ and $R^{\zeta_1}=R^{\zeta_2}=R$ we have the following set of equations:
\begin{align}\label{eq:A1}
6\ddot{A} + \dot{\varphi}^{2} + Re^{-2A}=0,\\ \label{eq:A2}
12\dot{A}^{2} -  \frac{1}{2}\dot{\varphi}^{2}  + V -   2Re^{-2A} = 0,\\
\ddot{\varphi} + 4\dot{\varphi}\dot{A} - V'  =0,
\end{align}
which is equivalent to the S$^4$ case, analyzed in \cite{C}, if we
make an appropriate constant shift in $A$. For $S^4$, as shown in \cite{C}, these
equations admit IR-regular solutions where $e^{A(u)}$ vanishes at an IR
endpoint $u_0$. As we shall see, the same solution is {\em singular} if
the slice manifold is $S^2\times S^2$ instead of $S^4$. We anticipate
that IR-regular solutions correspond to one of the two sphere
shrinking to zero size, while the other remaining finite.

\subsection{CFTs on $S^2\times S^2$}

Before analyzing RG-flow solutions, we conclude this section by
briefly discussing  equations (\ref{eq:EOM6bis}-\ref{eq:EOM9bis}) in the
special case of a conformal boundary theory. This amounts to setting $\f=$ constant
and $V'=0$ in  (\ref{eq:EOM6bis}-\ref{eq:EOM9bis}),  which leads to
\begin{align}
\label{eq:EOM6ter} 2\dot{A_1}^2 + 2\dot{A_2}^2 + 8 \dot{A_1} \dot{A_2} -R^{\zeta_1}  \mathrm{e}^{-2A_1} - R^{\zeta_2}\mathrm{e}^{-2A_2} + V_0 & = 0, \\
\label{eq:EOM7ter} 3\ddot{A_1} + 3\ddot{A_2} + 2\left(\dot{A_1} - \dot{A_2}\right)^2+ {R^{\zeta_1}\over 2}\mathrm{e}^{-2A_1} + {R^{\zeta_2}\over 2}\mathrm{e}^{-2A_2} & = 0, \\
\label{eq:EOM8ter} \left(\ddot{A_1} + 2\dot{A_1}^2 - {R^{\zeta_1}\over 2}\mathrm{e}^{-2A_1} \right) - \left(\ddot{A_2} + 2\dot{A_2}^2 - {R^{\zeta_2}\over 2} \mathrm{e}^{-2A_2} \right) & = 0
\end{align}
where $V_0$ now is a negative constant. These are two second-order
equations plus one first-order constraint for the functions $A_1(u),
A_2(u)$, depending on the two parameters $R^{\zeta_1},
R^{\zeta_2}$. The system has a total of three integration constants:
two of them are constant shifts of $A_1$ and $A_2$ which can be fixed
by requiring that $R^{\zeta_1}$ and $R^{\zeta_2}$ coincide with  the
actual curvatures of the manifold on which the  UV boundary theory is
defined  according to the holographic dictionary, i.e.
\be\label{FGexp}
ds^2 \to du^2 + e^{-2u/\ell}\left[ \zeta^1_{\alpha_1, \beta_1}
  \dd{x^{\alpha_1}} \dd{x^{\beta_1}} +  \zeta^2_{\alpha_2, \beta_2}
  \dd{x^{\alpha_2}} \dd{x^{\beta_2}}\right] + \text{subleading}, \qquad u\to
  -\infty,
\ee
where we set $V_0 = -{12\over \ell^2}$.

The remaining integration constant is the interesting one: it must
enter at subleading order in the UV expansion (since the leading order
is completely fixed by the condition  (\ref{FGexp}), and therefore it
corresponds to a vacuum expectation value. In particular, since the
only non-trivial bulk field is the metric, it must correspond to a combination of  the
vevs of the components of the  stress tensor. As we shall see, in the
symmetric case in which $R^{\zeta_1}=R^{\zeta_2}$, this combination is
the difference between the two (constant) expectation values of
$T_{\alpha\beta}$ along the two spheres, and it parametrizes the
difference between the two scale factors as we move towards the IR.

\section{The first order formalism and holographic RG flows} \label{sec:first}

To interpret the solutions to the equations of motion \eqref{eq:EOM6bis}-\eqref{eq:EOM9bis} in terms of RG flows, it will be convenient to rewrite the second-order Einstein equations as a set of first-order equations. This will allow an interpretation as gradient RG flows. Locally, this is always possible,
except at special points where $\dot{\f}=0$. Such points will be later be referred
to as "bounces", as previously observed in \cite{exotic,exotic2,C}. Given a solution,  as long as
$\dot{\f}(u)\neq 0$, we can invert the relation between $u$ and $\f(u)$ and define the following {\em scalar} functions of $\f$:

\be
\label{eq:defW1c} W_1(\f)  \equiv -2 \dot{A}_1 \sp
W_2(\f) \equiv -2 \dot{A}_2 \sp S(\f)  \equiv \dot{\f} \, ,
\ee
\be
\label{eq:defT1c}  T_1(\f) \equiv e^{-2A_1} R^{(\zeta^1)} \sp
 T_2(\f)  \equiv e^{-2A_2} R^{(\zeta^2)} \, .
\ee
where the expressions on the right hand side are evaluated at
$u=u(\f)$. In terms of the functions defined above, the equations of motion \eqref{eq:EOM6bis} --\eqref{eq:EOM9bis} become

\begin{align}
\label{eq:EOM11bis} W_1^2 + W_2^2 + 4 W_1 W_2 - S^2 - 2(T_1+T_2) + 2 V & = 0, \\
\label{eq:EOM12bis} S^2 - \frac{3}{2} S (W'_1 + W'_2) + \frac{1}{2} (W_1 - W_2)^2 + \frac{1}{2} (T_1 + T_2) & = 0, \\
\label{eq:EOM13bis} (-S W'_1 + W_1^2 - T_1) - (-S W'_2 + W_2^2 - T_2) & = 0, \\
\label{eq:EOM14bis} SS' - S(W_1 + W_2) - V' & = 0.
\end{align}
The last equation is not independent but it can be obtained by combining the derivative of
equation (\ref{eq:EOM11bis}) with equations  (\ref{eq:EOM12bis}) and
(\ref{eq:EOM13bis}). However it is convenient
to keep equation  (\ref{eq:EOM14bis}) , and  to eliminate instead $T_1$ and $T_2$, which only appear
algebraically and can be expressed in terms of the other
functions:
\begin{align}
T_1 & =-S^2 -W_2^2 +  W_1 W_2 +S(W'_1 +2 W'_2) \label{23} \\
T_2 & =-S^2- W_1^2+  W_1 W_2 + S(2W'_1 + W'_2)\label{24}
\end{align}
Inserting these relations in equations (\ref{eq:EOM11bis}) and
(\ref{eq:EOM14bis})  we are left with two
first order equations for $W_1,W_2$ and $S$,
\be
S^2 - 2 S (W'_1 + W'_2) + W_1^2 + W_2^2 + \frac{2}{3} V= 0, \label{29}
\ee
\be
SS' - S(W_1 + W_2) - V'  = 0.\label{30}
\ee
An additional equation is obtained  by using the relation
\be
{T_i'\over T_i}={W_i\over S} \sp i=1,2
\label{21}\ee
which follows from the definition (\ref{eq:defT1c}): differentiating
once equations (\ref{23}) and (\ref{24}), and using (\ref{21}) one
obtains:
\be
S^2(W_1''-W_2'')+SS'(W_1'-W_2')+(2W_1W_2-S^2)(W_1-W_2)+
\label{31}\ee
$$+S(W_2W_2'-W_1W_1'+2(W_1W_2'-W_2W_1'))=0
$$

Equations (\ref{29}), (\ref{30}) and (\ref{31}) will be the starting
point for our analysis of solutions. This system is  first order in $S$  and second order in
$W_{1,2}$ with a first order constraint (\ref{29}). Accordingly, there
are four integration constants.

These integration constants should match with the data of the dual QFT. On  the QFT side,
there are five dimensionful quantities which correspond to asymptotic
data of the solution:
\begin{itemize}
\item The UV coupling $\f_-$ which drives the flow away from the UV
  fixed point;
\item The two curvatures of the spheres,  $R^{\zeta_1}, R^{\zeta_2}$.
\item The vev of the deforming  operator, which by the trace identity
  is related to the trace of the stress tensor;
\item An additional vev parameter which controls the {\em difference}
  between the stress tensor components along the two spheres, and can
  vary independently of $T^\mu_\mu$.
\end{itemize}
Out of these five dimensionful quantities, we can construct four dimensionless ones
by measuring the curvatures and the vevs in units of the UV scale
$\f_-$.  These four  dimensionless parameters correspond to the four
integration constants\footnote{Notice that each equation is of
  homogeneous degree (namely 2 or 3) in $S, W_i$ so that each of them can be
  taken to be a dimensionless function times a fixed  scale determined
  by the potential $V$. Accordingly, all integration constants of the
  system can be taken to be dimensionless.} of the system of equations (\ref{29}),
(\ref{30}), (\ref{31}). The final integration constant $\f_-$
correspond to the initial condition we have to impose when we
integrate the equation for $\dot{\f} =S$ in order to write the
solution as a function of the $u$ coordinate\footnote{The
  corresponding integration constants arising in integrating
  $\dot{A_i} = -2 W_i$ are fixed by the requirement that the
  asymptotic expansion has the form (\ref{FGexp}) so that the curvatures
  $R^{\zeta_1}$ and $R^{\zeta_2}$ coincide with the curvatures of the
  space where the UV CFT lives. See the discussion in Section 2.3, or
  reference \cite{C}, for more details.}

This system displays an additional degree of freedom (an additional
integration  constant, beyond the extra curvature parameter, compared
with the $S^4$ case), which correspond to the last bullet point in the
list above. As we shall see in the next section, it controls how the relative sizes of the
spheres change as the theory flows to the UV.  Turning on this
parameter, allows us to obtain regular solutions in the IR, as we shall
see in section 5.

\section{The structure of solutions near the boundary} \label{sec:nb}
We now proceed to determining the near-boundary geometry in the
vicinity of an extremum of the potential.  This will allow us to
identify the integration constants in the bulk with the corresponding
parameters of the boundary field theory.

Without loss of generality, we take the extremum to be at $\f=0$. It will then be sufficient to consider the potential
\be
\label{potext} V = -\frac{d(d-1)}{\ell^2} - \frac{m^2}{2}\f^2 + \mathcal{O}(\f^3)
\ee
where $m^2 > 0$ for maxima and $m^2 < 0$ for minima.
A maximum of $V$ always corresponds to a UV fixed point. In contrast,  a minimum of the potential, in the flat slicing case, can be reached
either in the UV or in the IR, but as we shall see, the second
possibility does not arise when slices are curved\footnote{This was already the
case for maximally symmetric $S^4$ slicing,  as it was shown in
\cite{C}.}

In the following we solve equations \eqref{eq:EOM11bis}-\eqref{eq:EOM14bis} for $W(\f)$, $S(\f)$ and $T(\f)$ near $\f = 0$. The relevant calculations are presented in appendix \ref{bc}. Here we present and discuss the results.

Like in the case of a maximally symmetric boundary field theory
\cite{C}, there are two branches of solutions to equations
\eqref{eq:EOM11bis}-\eqref{eq:EOM14bis}, and we shall distinguish them
by the subscripts $(+)$ and $(-)$:
\bea
\label{eq:W1gensol}  W_1^{\pm}(\f) && = \frac{1}{\ell} \left[2 +
  \frac{\Delta_\pm}{6} \f^2 + \mathcal{O}(\f^3) \right]  + \frac{1}{6
  \ell}(2\mathcal{R}_1 - \mathcal{R}_2) \, |\f|^{\frac{2}{\Delta_\pm}}
\ [1+ \mathcal{O}(\f)] \\
&&   - \frac{1}{24\Delta_{\pm} \ell}(\mathcal{R}_1^2 - \mathcal{R}_2^2)|\f|^{\frac{4}{\Delta_\pm}} \log\left|\f\right|[1+ \mathcal{O}(\f)]
\nonumber\\
&& + \left[\frac{C_1}{\ell} \,
  +\frac{1}{144\ell}\left(\mathcal{R}_1^2+4\mathcal{R}_2^2-4\mathcal{R}_1\mathcal{R}_2\right)\right]
|\f|^{\frac{4}{\Delta_{\pm}}}[1+ \mathcal{O}(\f)] \, , \nonumber\\
&& + \mathcal{O}(C^2) + \mathcal{O}(\mathcal{R}^3)  \nonumber
\eea
\bea
\label{eq:W2gensol} W_2^{\pm}(\f) && = \frac{1}{\ell} \left[2 +
  \frac{\Delta_\pm}{6} \f^2 + \mathcal{O}(\f^3) \right]  + \frac{1}{6
  \ell}(-\mathcal{R}_1 + 2\mathcal{R}_2) \,
|\f|^{\frac{2}{\Delta_\pm}} \ [1+ \mathcal{O}(\f)]+ \\
&&
   + \frac{1}{24\Delta_{\pm} \ell}(\mathcal{R}_1^2 - \mathcal{R}_2^2) |\f|^{\frac{4}{\Delta_\pm}}\log\left|\f\right|[1+ \mathcal{O}(\f)]
\nonumber\\
&& + \left[\frac{C_2}{\ell}  +\frac{1}{144\ell}\left[4\mathcal{R}_1^2+\mathcal{R}_2^2-4\mathcal{R}_1\mathcal{R}_2\right)\right] |\f|^{\frac{4}{\Delta_{\pm}}}[1+ \mathcal{O}(\f)] \, ,
\nonumber \\
&& + \mathcal{O}(C^2) + \mathcal{O}(\mathcal{R}^3)  \nonumber
\eea
\bea
\label{eq:Sgensol} S_{\pm}(\f) && = \frac{\Delta_{\pm}}{\ell} \f \ [1+
\mathcal{O}(\f)] + \frac{6(C_1+C_2)}{\Delta_{\pm} \ell} \,
|\f|^{\frac{4}{\Delta_{\pm}}-1} \ [1+ \mathcal{O}(\f)]  + \\
&& + \mathcal{O}(C^2) + \mathcal{O}(\mathcal{R}^3)  \nonumber
\eea
\bea
\label{eq:T1gensol} T_1^{\pm}(\f) && = {\mathcal{R}_1 \over \ell^2} \, |\f|^{\frac{2}{\Delta_{\pm}}} [1+ \mathcal{O}(\f) ] +
 +
 \frac{1}{12\ell^2}(2\mathcal{R}_1^2-\mathcal{R}_1\mathcal{R}_2)\,|\f|^{\frac{4}{\Delta_{\pm}}}[1+
 \mathcal{O}(\f)] \,  \\
&& + \mathcal{O}(C^2) + \mathcal{O}(\mathcal{R}^3)  \nonumber
\eea
\bea
\label{eq:T2gensol} T_2^{\pm}(\f)  && =  {\mathcal{R}_2 \over \ell^2} \,
|\f|^{\frac{2}{\Delta_{\pm}}} [1+ \mathcal{O}(\f)  ] +
\frac{1}{12\ell^2}(2\mathcal{R}_2^2-\mathcal{R}_1\mathcal{R}_2)\,|\f|^{\frac{4}{\Delta_{\pm}}}[1+
\mathcal{O}(\f) ] \\
&& + \mathcal{O}(C^2) + \mathcal{O}(\mathcal{R}^3)  \nonumber
\eea

The  expressions (\ref{eq:W1gensol}-\ref{eq:T2gensol}) describe two
continuous families of solutions, whose structure is a
\textit{universal} analytic expansion in integer powers of $\f$, plus
a series of non-analytic, subleading terms which, in principle, depend
on four (dimensionless) integration constants $C_1,C_2$ and
$\mathcal{R}_1,\mathcal{R}_2$, consistently with the counting made at
the end of section 3. In these expressions, we write $\mathcal{O}(C)$
and $\mathcal{O}(\mathcal{R})$ to indicate terms which are subleading
because they are  accompanied by a higher  power in $\f$, and
completely determined by $C_{1,2}$ and $\mathcal{R}_{1,2}$. Each of
these terms multiplies its own analytic power series in $\f$.

Notice that, close to a minimum of the potential, $\Delta_-
<0$. Therefore, terms proportional to $|\f|^{2/\Delta_{-}} \to
  \infty$. The absence of these terms requires ${\mathcal  R}_1 =
  {\mathcal R}_2 = 0$: As  we shall see below,
the constants  ${\mathcal  R}_i$ give the curvature of each sphere in
units of the UV source $\f_-$. Therefore,
 no solution  with curved slicing  can reach the minimum of the
  potential. This is similar to  what was found in \cite{C} for the case of the
  $S^4$ slicing, i.e. only the flat-slicing
solution can reach a minimum of the potential. The same property is also true for $S^1\times S^3$ slices that includes the global AdS$_5$ case.

On the other hand, $\Delta_+ >0$ for both a maximum and a minimum, so
the $(+)$-branch can exist around a minimum of $V(\f)$ (in which case
it corresponds to a UV fixed point, as we shall see below).  All in all, an AdS UV boundary exists for both + and - solutions near a
maximum of the potential and for a + solution near a minimum of the
potential.

As discussed in appendix \ref{sec:sum}, in the $(-)$-branch $C_1$ and $C_2$ are
arbitrary, but in  the $(+)$-branch they are constrained to obey
$C_1+C_2 =0$. Therefore, the $(+)$-branch has only three dimensionless
integration constants, namely $\mathcal{R}_{1,2}$ and $C_1-C_2$.

 Given our results for $W_1,W_2$, $S$ and $T_1,T_2$
\eqref{eq:W1gensol}-\eqref{eq:T2gensol}, we can
solve for $\f(u)$ and $A_1(u),A_2(u)$ by integrating equations
\eqref{eq:defW1c}.  This  introduces three more
integration constants (i.e. initial conditions  for the first order
flows), which we call $\bar{A}_{1}, \bar{A}_{2}, \f_\pm$, where
$\pm$ refers to the $\pm$ branches.

In  the $(-)$-branch, the result is:
\begin{eqnarray}
\label{eq:phimsolS2} \f(u) & =& \f_- \ell^{\Delta_-}\ex^{\Delta_-u / \ell} \left[ 1+ \mathcal{O} \left(\ex^{2u/\ell} \right)  \right]+ \\
\nonumber & \hphantom{=}& \ + \frac{6(C_1+C_2) \, |\f_-|^{\Delta_+ /
    \Delta_-}}{\Delta_-(4-2 \Delta_-)} \, \ell^{\Delta_+}
\ex^{\Delta_+ u / \ell} \left[ 1+ \mathcal{O} \left(\ex^{2u/\ell}
  \right) \right]  \, , 
\end{eqnarray}
\begin{eqnarray}
\label{eq:A1msol} A_1(u) & =& \bar{A_1} -\frac{u}{\ell} - \frac{\f_-^2 \, \ell^{2 \Delta_-}}{24} \ex^{2\Delta_- u / \ell} \, [1+\mathcal{O}(\ex^{\Delta_-u/\ell})]+\\
\nonumber &\hphantom{=}&  -\frac{|\f_-|^{2/\Delta_-} \, \ell^2}{24} (2\mathcal{R}_1 - \mathcal{R}_2) \ex^{2u/\ell} \, [1+\mathcal{O}(\ex^{\Delta_-u/\ell})+\mathcal{O}(\ex^{(\Delta_+-\Delta_-)u/\ell})] +\\
 \nonumber & \hphantom{=}& + \frac{1}{192}(\mathcal{R}_1^2-\mathcal{R}_2^2)|\f_-|^{4/\Delta_-} \, \ell^4 \frac{u}{\ell} \, \ex^{4u/\ell} [1+\mathcal{O}(\ex^{\Delta_-u/\ell})]+\\
\nonumber & \hphantom{=}& \ - \frac{1}{8}|\f_-|^{4/\Delta_-} \, \ell^4\ex^{4u/\ell}\left( (C_1+C_2)\frac{\Delta_+}{4-2\Delta_-}+\frac{C_1-C_2}{2} \right. +\\
\nonumber & \hphantom{=}& +
\left. \frac{1}{48}\left[\frac{5}{6}(\mathcal{R}_1^2+\mathcal{R}_2^2)-\frac{4}{3}\mathcal{R}_1\mathcal{R}_2\right]
  -
  \frac{\mathcal{R}_1^2-\mathcal{R}_2^2}{24\D_-}\log\left(\f_-\ell^{\D_-}\right)\right)
+\ldots \, ,
\end{eqnarray}
\begin{eqnarray}
\label{eq:A2msol} A_2(u) & =& \bar{A_2} -\frac{u}{\ell} - \frac{\f_-^2 \, \ell^{2 \Delta_-}}{24} \ex^{2\Delta_- u / \ell} \, [1+\mathcal{O}(\ex^{\Delta_-u/\ell})]+\\
\nonumber &\hphantom{=}&  -\frac{|\f_-|^{2/\Delta_-} \, \ell^2}{24} (-\mathcal{R}_1 + 2\mathcal{R}_2) \ex^{2u/\ell} \, [1+\mathcal{O}(\ex^{\Delta_-u/\ell})+\mathcal{O}(\ex^{(\Delta_+-\Delta_-)u/\ell})] +\\
 \nonumber & \hphantom{=}& - \frac{1}{192}(\mathcal{R}_1^2-\mathcal{R}_2^2)|\f_-|^{4/\Delta_-} \, \ell^4 \frac{u}{\ell} \, \ex^{4u/\ell} [1+\mathcal{O}(\ex^{\Delta_-u/\ell})]+\\
\nonumber & \hphantom{=}& \ - \frac{1}{8}|\f_-|^{4/\Delta_-} \, \ell^4\ex^{4u/\ell}\left( (C_1+C_2)\frac{\Delta_+}{4-2\Delta_-}-\frac{C_1-C_2}{2} \right. +\\
\nonumber & \hphantom{=}& + \left. \frac{1}{48}\left[\frac{5}{6}(\mathcal{R}_1^2+\mathcal{R}_2^2)-\frac{4}{3}\mathcal{R}_1\mathcal{R}_2\right] + \frac{\mathcal{R}_1^2-\mathcal{R}_2^2}{24\D_-}\log\left(\f_-\ell^{\D_-}\right)\right)  +\ldots \, ,
\end{eqnarray}
and in the $(+)$-branch:
\begin{eqnarray}
\label{eq:phipsolS2} \f(u) & =& \f_+ \ell^{\Delta_+}\ex^{\Delta_+ u /
  \ell} \left[ 1+ \mathcal{O} \left( \ex^{2u/\ell} \right) \right]  \,
, \\
&& \nonumber \\
\label{eq:A1psol} A_1(u) & =& \bar{A_1} -\frac{u}{\ell} - \frac{\f_+^2 \, \ell^{2 \Delta_+}}{24} \ex^{2\Delta_+ u / \ell} \, [1+\mathcal{O}(\ex^{\Delta_+u/\ell})]+\\
\nonumber &\hphantom{=}&  -\frac{|\f_+|^{2/\Delta_+} \, \ell^2}{24}
(2\mathcal{R}_1 - \mathcal{R}_2) \ex^{2u/\ell} \,
[1+\mathcal{O}(\ex^{\Delta_+u/\ell})] +\\
 \nonumber & \hphantom{=}& + \frac{1}{192}(\mathcal{R}_1^2-\mathcal{R}_2^2)|\f_+|^{4/\Delta_+} \, \ell^4 \frac{u}{\ell} \, \ex^{4u/\ell} [1+\mathcal{O}(\ex^{\Delta_+u/\ell})]+\\
\nonumber & \hphantom{=}& \ - \frac{1}{8}|\f_+|^{4/\Delta_+} \, \ell^4\ex^{4u/\ell}\left( \frac{C_1-C_2}{2} \right. +\\
\nonumber & \hphantom{=}& +
\left. \frac{1}{48}\left[\frac{5}{6}(\mathcal{R}_1^2+\mathcal{R}_2^2)-\frac{4}{3}\mathcal{R}_1\mathcal{R}_2\right]
  -
  \frac{\mathcal{R}_1^2-\mathcal{R}_2^2}{24\D_+}\log\left(|\f_+|\ell^{\D_+}\right)\right)
+\ldots \, , \\
&& \nonumber \\
\label{eq:A2psol} A_2(u) & =& \bar{A_2} -\frac{u}{\ell} - \frac{\f_+^2 \, \ell^{2 \Delta_+}}{24} \ex^{2\Delta_+ u / \ell} \, [1+\mathcal{O}(\ex^{\Delta_+ u/\ell})]+\\
\nonumber &\hphantom{=}&  -\frac{|\f_+|^{2/\Delta_+} \, \ell^2}{24} (-\mathcal{R}_1 + 2\mathcal{R}_2) \ex^{2u/\ell} \, [1+\mathcal{O}(\ex^{\Delta_+u/\ell})] +\\
 \nonumber & \hphantom{=}& - \frac{1}{192}(\mathcal{R}_1^2-\mathcal{R}_2^2)|\f_+|^{4/\Delta_+} \, \ell^4 \frac{u}{\ell} \, \ex^{4u/\ell} [1+\mathcal{O}(\ex^{\Delta_+u/\ell})]+\\
\nonumber & \hphantom{=}& \ - \frac{1}{8}|\f_-|^{4/\Delta_-} \, \ell^4\ex^{4u/\ell}\left( -\frac{C_1-C_2}{2} \right. +\\
\nonumber & \hphantom{=}& + \left. \frac{1}{48}\left[\frac{5}{6}(\mathcal{R}_1^2+\mathcal{R}_2^2)-\frac{4}{3}\mathcal{R}_1\mathcal{R}_2\right] + \frac{\mathcal{R}_1^2-\mathcal{R}_2^2}{24\D_+}\log\left(|\f_+|\ell^{\D_+}\right)\right)  +\ldots \, ,
\end{eqnarray}

A few comments are in order.
\begin{itemize}
\item In each branch, the solutions depend on three more integration
  constants  $\bar{A}_{i}$ and $\f_{\pm}$.
\item According to the discussion above, both
$\Delta_\pm >0$. Since the results above  are supposed to be valid for small
$\f$,  these expansions holds in the limit
$u\to -\infty$,  which means that we are close to the AdS boundary (the scale
factors diverge).  The omitted terms in equations
(\ref{eq:phimsolS2}-\ref{eq:A2psol}) vanish as $u\to -\infty$.

\item For the $(-)$-branch of solutions, we identify $\f_-$ as the source for the scalar operator $\mathcal{O}$ in the boundary field theory associated with $\f$. The vacuum expectation value of $\mathcal{O}$ depends on $C_1+C_2$ and is given by
\begin{equation}
\left<\mathcal{O}\right>_- = \frac{6(C_1+C_2)}{\Delta_-}|\f_-|^{\Delta_+/\Delta_-}
\end{equation}
\item For the $(+)$-branch of solutions, the bulk field $\f$ is also associated with a scalar operator $\mathcal{O}$ in the boundary field theory. However, in this case the source is identically zero, yet there is a non-zero vev given by
\begin{equation}
\left<\mathcal{O}\right>_+ = (2\Delta_+ - 4)\f_+
\end{equation}
\item As explained at the end of appendix \ref{sec:sum}, the integration
constants ${\cal R}_i$ appearing in
(\ref{eq:W1gensol}-\ref{eq:T2gensol}) are identified as the
``dimensionless,'' UV curvature parameters,
\be \label{curlyRs}
{\mathcal R}_i = R_i^{UV} |\f_\pm|^{-2/\Delta_{\pm}}.
\ee
where $R_i^{UV}$ are the {\em physical} scalar curvature parameters in
the UV, i.e. the  Ricci curvature scalars  of the two 2-spheres on
which the dual QFT is defined.  If we make the choice $\bar{A}_1
=\bar{A}_2 = 0$, these coincide with the ``fiducial'' curvatures
$R^{(\zeta^i)}$ that we have introduced in the metric.

\item An interesting property of the solution for the S$^2
  \times$S$^2$ slicing, compared with the maximally symmetric case, is
  that, beyond the fact there are two UV integration constants
  corresponding to the curvatures of the spheres, there are also two
  independent vev parameters $C_1$ and $C_2$. As can be seen  from equation
  (\ref{eq:phimsolS2}),     $C_1+C_2$ is the only combination which enters
  in the near-boundary asymptotics of $\f(u)$.

\item The combination ($C_1-C_2$) instead only enters the difference
  of the scale factors,
\begin{align}
 \label{eq:A1mA2} A_1(u) - A_2(u) & = \bar{A_1} - \bar{A_2} -\frac{|\f_-|^{2/\Delta_-} \, \ell^2}{8} (\mathcal{R}_1 - \mathcal{R}_2) \ex^{2u/\ell} \, 
 \\
  \nonumber & \hphantom{=} + \frac{1}{96}(\mathcal{R}_1^2-\mathcal{R}_2^2)|\f_-|^{4/\Delta_-} \ell^4\,  \frac{u}{\ell} \, \ex^{4u/\ell}
  \\
 \nonumber & \hphantom{=} \ - \frac{1}{8} \, \left[(C_1-C_2) -
   {\mathcal{R}_1^2-\mathcal{R}_2^2 \over 24\Delta_-} \log(|\f_-| \ell^{\Delta_-})\right] |\f_-|^{4/\Delta_-}\ell^4 \ex^{4u/\ell} +\ldots \, ,
 \end{align}
 As  is  shown
  in appendix \ref{sec:Tij}, this corresponds to a vev of difference
  of the boundary stress tensors along  the two spheres. One finds that the full stress tensor vev has the form:
\be
\label{Tab}
 \left<T_{\a \b}\right> =  4(M_p \ell)^3 |\f_-|^{4/\Delta_-} \left[\frac{T}{4}\left( \begin{array}{cc}
 \zeta^1 & 0  \\
0 &  \zeta^2   \end{array} \right) + \hat{T} \left( \begin{array}{cc}
 \zeta^1 & 0  \\
0 & - \zeta^2   \end{array} \right) \right]
\ee
where $T$ is the trace part and  $\hat{T}$ is the traceless part, given by:
\be
\label{Taa}
 T  =  \left(\frac{1}{96}(\mathcal{R}_1^2 + \mathcal{R}_2^2 - 4 \mathcal{R}_1 \mathcal{R}_2) - \frac{\Delta_+}{4-2\Delta_-}(C_1+C_2)\right)
\ee
\be
\label{T-traceless}
 \hat{T}= \frac{1}{96}
\left(\frac{\mathcal{R}_1^2-\mathcal{R}_2^2}{2}-12(C_1-C_2) +
  {\mathcal{R}_1^2 - \mathcal{R}_2^2 \over \Delta_-} \log (|\f_-| \ell^{\Delta_-}) \right)
\ee
As expected from the trace identity
$$T \sim \text{Weyl anomaly} + \beta \langle O\rangle\;,$$
  the scalar vev $C_1+C_2$ enters in the expectation value
of the trace of the  stress tensor. The combination  $(C_1-C_2)$  contributes
instead to the difference between  the stress tensor components  along
spheres $1$ and $2$. In particular, when $\mathcal{R}_1 =
\mathcal{R}_2$, it is manifest that $C_1\neq C_2$ introduces a asymmetry
between the two spheres.  This leads to the {\em spontaneous}
breaking of the $\mathbb{Z}_2$ symmetry that exchanges the two
spheres.
\end{itemize}

To conclude this section:  as expected \textit{maxima of the potential are associated with UV fixed points}. The bulk space-time asymptotes to AdS$^5$ and reaching the maximum of the potential is equivalent to reaching the boundary. Moving away from the boundary corresponds to a flow leaving the UV. Flows corresponding to solutions on the $(-)$-branch are driven by the existence of a non-zero source $\f_-$ for the perturbing operator $\mathcal{O}$. Flows corresponding to solutions on the $(+)$-branch are driven purely by a non-zero vev for the stress tensor of the boundary theory.
As for \textit{minima} of the potential, they can only be associated
with UV fixed points, only when  the flow that leaves them is in the
$(+)$-branch of solutions. This is because of the result from the
previous section that minima of the potential cannot be IR end-points
of the RG flow.

\subsection{The flows associated to a CFT on $S^2\times S^2$} \label{CFT}

In this section we discuss the structure of the bulk solution for
holographic CFTs on S$^2\times$ S$^2$. This could arise either in the case where the  bulk
  potential is purely a  (negative) cosmological constant, or by
  taking a solution with $\f(u)=const.$ at an extremum of $V(\f)$. In
  either case,  for the ansatz \eqref{eq:metricS2}, Einstein's
  equations correspond to equations \eqref{eq:EOM6bis} -
  \eqref{eq:EOM8bis} with $\f$ set to a constant and $V = -{12\over
    \ell^2}$.

We still use the first order formalism with the superpotentials defined as functions of $u$,
\be
\label{eq:defW1cft} W_1(u)  \equiv -2 \dot{A}_1 \sp
W_2(u) \equiv -2 \dot{A}_2 \, ,
\ee
\be
\label{eq:defT1cft}  T_1(u) \equiv e^{-2A_1} R^{(\zeta^1)} \sp
 T_2(u)  \equiv e^{-2A_2} R^{(\zeta^2)} \, .
\ee
The equations of motion \eqref{eq:EOM6bis} - \eqref{eq:EOM8bis} become
\begin{align}
\label{eq:EOM1cft} W_1^2 + W_2^2 + 4 W_1 W_2 - 2(T_1+T_2) -\frac{24}{\ell^2} & = 0 \, , \\
\label{eq:EOM2cft} - \frac{3}{2} (\dot{W}_1 + \dot{W}_2) + \frac{1}{2} (W_1 - W_2)^2 + \frac{1}{2} (T_1 + T_2) & = 0 \, , \\
\label{eq:EOM3cft} (-\dot{W}_1 + W_1^2 - T_1) - (-\dot{W}_2 + W_2^2 - T_2) & = 0 \, .
\end{align}
We can solve algebraically for  $T_1$ and $T_2$ as
\begin{align}
T_1 & = -W_2^2 +  W_1 W_2 + \dot{W}_1 +2 \dot{W}_2 \label{t1cft} \, , \\
T_2 & = - W_1^2+  W_1 W_2 + 2\dot{W}_1 + \dot{W}_2\label{t2cft} \, .
\end{align}
The $T_i$ also satisfy from their definition
\be
{\dot{T}_i\over T_i}=W_i \sp i=1,2 \, . 
\label{reltwcft}\ee
The two independent differential equations for the two $W$
superpotentials are:
\be
\label{eqindcft1} -2(\dot{W}_1+\dot{W}_2) + W_1^2 + W_2^2 - \frac{8}{\ell^2} = 0 \, ,
\ee
\be
\label{eqindcft2} \ddot{W}_1 - \ddot{W}_2 - (W_1\dot{W}_1 - W_2\dot{W}_2) + 2(W_1\dot{W}_2 - \dot{W}_1W_2) + 2 W_1W_2(W_1 - W_2) = 0 \, ,
\ee

There are three integration constants in this system of one first order equation and one second order equation.
Two of them are the two (independent) curvatures of the two S$^2$s of the space-time manifold, $R_1^{UV}$ and $R_2^{UV}$.
These are sources in the holographic dictionary and give rise to one dimensionless number, that is the ratio of the curvatures. Only this ratio is a non-trivial parameter of the boundary CFT.
The other integration constant  of the system, which we denote by $C$,
represents  a vev in the QFT, and it
corresponds to the $C_1-C_2$ vev in the non-conformal case.
Close to the boundary, the superpotentials  $W_1$ and $W_2$ have the
following expansion:
\begin{eqnarray}
\label{eq:W1gencft} W_1(u) & =& \frac{2}{\ell} + \frac{1}{6 \ell}(2\mathcal{R}_1 - \mathcal{R}_2) \ex^{2u/\ell}  -\frac{1}{24\Delta_{\pm}\ell}(\mathcal{R}_1^2-\mathcal{R}_2^2)\frac{u}{\ell}\ex^{4u/\ell}\, \\
\nonumber & \hphantom{=}& \,  + \left[\frac{C}{2\ell}  +
  \frac{1}{144\ell}\left(\mathcal{R}_1^2+4\mathcal{R}_2^2-4\mathcal{R}_1\mathcal{R}_2\right)
\right] \ex^{4u/\ell}  \, +
\ldots , \\
\label{eq:W2gencft}W_2(u) & =& \frac{2}{\ell} + \frac{1}{6
  \ell}(2\mathcal{R}_2 - \mathcal{R}_1) \ex^{2u/\ell}  + \frac{1}{24\Delta_{\pm}\ell}(\mathcal{R}_1^2-\mathcal{R}_2^2)\frac{u}{\ell}\ex^{4u/\ell}\, \\
\nonumber & \hphantom{=}& \,  + \left[ - \frac{C}{2\ell}  +
  \frac{1}{144\ell}\left(4\mathcal{R}_1^2+ \mathcal{R}_2^2-4\mathcal{R}_1\mathcal{R}_2\right)
\right] \ex^{4u/\ell}  \, +
\ldots , \\
T_1 (u)& = & {\mathcal{R}_1 \over \ell^2} e^{2u/\ell} +
{2\mathcal{R}_1^2 - \mathcal{R}_1\mathcal{R}_2 \over 12 \ell^2}
e^{4u/\ell} + \ldots \label{eq:T1gencft} \\
T_2 (u)& = & {\mathcal{R}_2 \over \ell^2} e^{2u/\ell} +
{2\mathcal{R}_2^2 - \mathcal{R}_1\mathcal{R}_2 \over 12 \ell^2}
e^{4u/\ell} + \ldots \label{eq:T2gencft}
\end{eqnarray}

The asymptotic solution for $(A_1(u), A_2(u))$ can be obtained
integrating the first order equations (\ref{eq:defW1cft}),
\begin{align}
\label{eq:A1msolu-i} A_1(u) & = \bar{A_1} -\frac{u}{\ell} -\frac{1}{24} (2\mathcal{R}_1 - \mathcal{R}_2) \ex^{2u/\ell} \\
\nonumber & \hphantom{=} + \frac{1}{192}(\mathcal{R}_1^2-\mathcal{R}_2^2) \frac{u}{\ell} \, \ex^{4u/\ell} \\
\nonumber & \hphantom{=} \ - \frac{1}{8}\left( C + \frac{1}{48}\left[\frac{5}{6}(\mathcal{R}_1^2+\mathcal{R}_2^2)-\frac{4}{3}\mathcal{R}_1\mathcal{R}_2\right] \right) \ex^{4u/\ell} +\mathcal{O}(\ex^{6u/\ell}) \, , \\
\label{eq:A2msolu-i} A_2(u) & = \bar{A_2} -\frac{u}{\ell} -\frac{1}{24} (-\mathcal{R}_1 + 2\mathcal{R}_2) \ex^{2u/\ell} \\
\nonumber & \hphantom{=} - \frac{1}{192}(\mathcal{R}_1^2-\mathcal{R}_2^2) \frac{u}{\ell} \, \ex^{4u/\ell} \\
\nonumber & \hphantom{=} \ - \frac{1}{8}\left( -C+ \frac{1}{48}\left[\frac{5}{6}(\mathcal{R}_1^2+\mathcal{R}_2^2)-\frac{4}{3}\mathcal{R}_1\mathcal{R}_2\right] \right) \ex^{4u/\ell} + \mathcal{O}(\ex^{6u/\ell})
\end{align}
where $\bar{A}_{1,2}$ are  integration constants. As in the general
non-conformal case, the physical UV
curvatures $R^{UV}_{1,2}$  are related  to the fiducial curvatures
$R^{\zeta_{1,2}} = R_{1,2}^{UV}$  by:
\be\label{RUV}
R_{i}^{UV} = e^{-2\bar{A_i}}R^{\zeta_{i}}.
\ee
We can always choose integration constants $\bar{A_i} =0$ so that
the two coincide. We implement this choice in what follows.

The dimensionless curvature parameters $\mathcal{R}_i$ can be
related to $R^{UV}_i$ by comparing equations (\ref{eq:defT1cft})  and
(\ref{eq:T1gencft}-\ref{eq:T2gencft}), which leads to:
\be
{\mathcal R}_i = R^{UV}_i \ell^2
\ee
The solution depends on an additional integration constant $C$,  which
appears at order $e^{4u/\ell}$, and therefore
corresponds to  a combination of vevs of the stress tensor: this is
most clearly seen by going to the symmetric case $R^{UV}_1 =
R^{UV}_2$, in which we observe that $C$ parametrizes the difference
between scale factors:
\be \label{A1-A2}
A_1 - A_2 = -{C \over 4} e^{4u/\ell} + \ldots,
\ee
whereas $A_1+A_2 \sim -2u/\ell$ is independent of $C$. Accordingly,
$C$ here plays the same role as $C_1-C_2$ in the non-conformal case,
and parametrizes the difference in the vevs of the stress tensor components along
the two spheres,
 It can be related to a specific component of the stress tensor. In
 the conformal case, the
 latter has a similar form as (\ref{Tab}),
\be
\label{Tabcft}
 \left<T_{\a \b}\right> =  4(M_p \ell)^3  \left[\frac{T_{CFT}}{4}\left( \begin{array}{cc}
 \zeta^1 & 0  \\
0 &  \zeta^2   \end{array} \right) + \hat{T}_{CFT} \left( \begin{array}{cc}
 \zeta^1 & 0  \\
0 & - \zeta^2   \end{array} \right) \right]
\ee
where now the trace part $T_{CFT}$ and traceless part $\hat{T}_{CFT}$ are:
\be
\label{Taacft}
T_{CFT} =
 \frac{1}{96}\Big((R_1^{UV})^2 + (R_2^{UV})^2 - 4 R_1^{UV} R_2^{UV}\Big)
\ee
\be
\label{Tcft-traceless}
\hat{T}_{CFT} =
 \frac{1}{96}\left({(R_1^{UV})^2 - (R_2^{UV})^2 \over 2} - 24 { C
     \over \ell^4}\right)
\ee
where $C$  is the vev parameter which enters in the scale factor as
displayed  in \eqref{A1-A2}.

 In contrast, there is no analog integration constant
for the vev of the sum of the two components, $C_1+C_2$, i.e. the trace of the stress
tensor, which here is completely determined by the curvatures (as
expected from the trace anomaly, see appendix \ref{sec:Tij}).

\section{Regularity in the bulk}\label{sec:IRbc}
We shall study here the regularity of the bulk solutions, as well as
their structure near IR endpoints of the flow. These points are
identified  by a vanishing scalar field  derivative, which in the
superpotential language corresponds to $S=0$.  Both in the the
flat-sliced domain walls \cite{exotic} and in the maximally symmetric
curved-sliced domain walls \cite{C},  one finds $S=0$  either at
true IR endpoints, where the scale factor vanishes, or  at a
a {\em bounce}, where the scalar fields has a turning point but the flow
keeps going. The latter case is a singular point of the superpotential
description, as the scalar field ceases to be a good coordinate, but
the geometry is regular. On the contrary, the vanishing of the scale
factor signals the end of the flow in the Euclidean case (which
becomes a horizon in the Lorentzian case). The question we
address in this section is: what are the possible IR endpoints which
give rise to a regular geometry?

The starting point of the study of regularity of the solutions are the
bulk curvature invariants, that are calculated and analyzed in
appendix \ref{CurvInv}. As shown there, the curvature  invariants up to quadratic
order are given by the following expressions in terms of the superpotentials,
\be \label{Ricci}
R = \frac{S^2}{2} + \frac{5}{3} V, \qquad  R_{AB} R^{AB} = \left(\frac{S^2}{2}
  + \frac{V}{3} \right)^2 + \frac{4V^2}{9} ,
\ee
\bea
{\mathcal K} \equiv R_{ABCD} R^{ABCD} &&  = \left(\frac{S^2}{2} + \frac{V}{3} \right)^2 +\left( S(W_1'-W_2')-{W_1^2-W_2^2\over 2}\right)^2 + W_1^2 W_2^2
  \nonumber \\
&& + \left({W_1^2\over 2}+S^2+W_2^2-W_1W_2-S(W_1'+W_2')\right)^2
\nonumber \\
 && +\left({W_2^2\over
     2}+S^2+W_1^2-W_1W_2-S(W_1'+W_2')\right)^2 \label{Riemann}
\eea
Our goal is to find under which conditions the vanishing of  $S$
at a scalar field value  $\f=\f_0$ corresponds to a regular endpoint
 { As we shall see
shortly, regularity requires one of the two spheres to shrink to zero
size in a specific way, while the other keeps a finite size.}

 As one can observe from equation
(\ref{Ricci}),   for the Ricci scalar and the square of the Ricci
tensor  it is enough that $V(\f)$ be finite at the endpoint. We
assume that $V(\f)$ can only diverge as $\f\to \pm \infty$ as is standard in string theory effective actions.

The conditions for regularity  of the Kretschmann scalar, ${\mathcal K}$, written in equation
(\ref{Riemann}),  is not so straightforward, since it also involves
$W_i$ and their derivatives. Already in the $S^4$ case, a divergent
$W'$ does not necessarily signal a singularity (see e.g. \cite{C}).  The detailed analysis of the regularity conditions  is
presented in appendix \ref{geomint_a}. The result is that regularity
restricts the superpotential to have the following behavior near an
endpoint $\f_0$:
\begin{align}
W_1 &= \sqrt{\frac{2V^{'}(\varphi_0)}{3(\varphi - \varphi_0)}} - \left(\frac{1}{27}V(\varphi_0) + \frac{1}{9}T_{2,0}\right)\sqrt{\frac{6(\varphi - \varphi_0)}{V^{'}(\varphi_0)}} + \mathcal{O}((\varphi - \varphi_0)^{3/2}), \label{W1}\\
T_1 &= \frac{V^{'}(\varphi_0)}{3(\varphi - \varphi_0)} +
\frac{1}{27}V(\varphi_0) + \frac{1}{9}T_{2,0} + \mathcal{O}(\varphi -
\varphi_0), \label{T1} \\
W_2 &= -\left(\frac{2}{9}V(\varphi_0) -
  \frac{1}{3}T_{2,0}\right)\sqrt{\frac{6(\varphi -
    \varphi_0)}{V^{'}(\varphi_0)}} + \mathcal{O}((\varphi -
\varphi_0)^{3/2}), \label{W2} \\
T_2 &= T_{2,0} + \mathcal{O}(\varphi - \varphi_0), \label{T2}\\
S &= -V^{'}(\varphi_0)\sqrt{\frac{2(\varphi -
    \varphi_0)}{3V^{'}(\varphi_0)}} + \mathcal{O}((\varphi -
\varphi_0)^{3/2}) \label{S}
\end{align}
The expressions (\ref{W1}-\ref{S})  depend on two free parameters: the point $\f_0$
in field space where the flow ends and the constant $T_{2,0} =
T_2(\f_0)$ at that point. We observe that one of the superpotentials (in this case, $W_1$) diverges
at $\f_0$,  as does the corresponding function $T_1$, while $W_2$ and
$T_2$ stay finite.  This implies that, as  $\f\to \f_0$, the first sphere $S^2_1$ shrinks
to zero size, whereas the size of the second one   $S^2_2$ stays finite\footnote{Of
course, the opposite is also possible, in which case the roles of $W_1,
T_1$ and $W_2, T_2$ in equations (\ref{W1}-\ref{T2}) are
interchanged}, if we recall  the definitions
\be T_i = R^{(\zeta^i)}
e^{-2A_i},
\label{Tagain} \ee
where  $R^{(\zeta^i)} = 2/\alpha_i^2$ are the curvature scalars  of
the  fiducial metrics  $\zeta^i$ of the two spheres, and $\alpha_i$
their radii.

Following the results of appendix \ref{sec:IRbc_a}, we can write the
near-endpoint expression  for the scalar field and the scale factors in
terms of the domain wall coordinate $u$:
\be \label{phi-end}
\varphi(u) = \varphi_0 + \frac{1}{6}V^{'}(\varphi_0)(u-u_0)^2 + \mathcal{O}((u - u_0)^4),
\ee
\be
\label{ansAbis} A_1(u) =  \ln(\frac{u_0-u}{\ell})+A_{1,0} +
\mathcal{O}(u_0-u),  \qquad  A_2(u) =
A_{2,0} +\mathcal{O}(u_0-u)
\ee
where
\be \label{A-end}
 A_{1,0} =  {1\over 2}\log {R^{(\zeta^1)}
    \ell^2 \over 2},  \qquad  A_{2,0} = {1\over 2}\log
  {R^{(\zeta^2)} \over T_{2,0}}.
\ee
where $u_0$ is the coordinate at which the endpoint is reached.
From equation (\ref{ansAbis}) we can see explicitly that,  as $u\to
u_0$,  the sphere $S^1$ has vanishing scale factor, whereas  the free parameter $T_{2,0}$ controls the size of the sphere
$S^2_2$ which remains of finite size at the endpoint. More
specifically, the  radius of sphere 2 at the endpoint is simply:
\be\label{alphair}
\alpha_{IR}  = \sqrt{{2 \over T_{2,0}}}
\ee
as can be seen from equation (\ref{A-end}), the metric ansatz
(\ref{eq:metricS2})  between the curvature and the radius
$R^{(\zeta^2)} = 2/\alpha_2^2$.

 With similar reasoning,
taking $T_{2,0}$ to be negative one would find endpoints for a
domain-wall solution with $S^2\times AdS_2$ slicing, in which the
$S^2$ shrinks to zero size while the $AdS_2$ remains finite.

The value of the Kretschmann scalar in the interior (at $\varphi =
\varphi_0$) can  be explicitly computed inserting the expansions
(\ref{W1}-\ref{S}) in equation (\ref{Riemann}), which leads to
\begin{equation}
\label{Kint} \mathcal{K}(\varphi_0) = \frac{V(\varphi_0)^2}{3}\left(1 + \frac{1}{24}(T_{2,0}\ell^2)^2 + \frac{1}{6}T_{2,0}\ell^2 \right)
\end{equation}
Notice that finiteness of ${\mathcal K}$ requires a finite $T_{2,0}$,
i.e. a finite size for the sphere $S^2_2$ at the endpoint: if {\em
  both} spheres shrink at the same time, the space-time is
singular. This means that, even if we start with a symmetric solution
with $R_1^{UV}  = R_2^{UV}$ in the UV, for regular solutions to exist,
the sizes of the two spheres will necessarily  start to deviate as the
geometry flows towards the IR.

We briefly comment on the particular case of a CFT,  where $\varphi$ is a constant and
\be
V(\varphi) \equiv V_0 = -{d(d-1)\over \ell^2} = -{12\over \ell^2}\,.
\ee
In this case the expressions (\ref{W1}-\ref{S})  are ill defined, but
equation (\ref{Kint}) still holds\footnote{This  can  be seen by writing ${\cal
  K}$ as a function of the scale factors, see appendix
\ref{CurvInv}. }, and reduces to,
\begin{equation}
\label{KintSlicing} \mathcal{K}(u_0) = \mathcal{K}_{AdS^5} \left(1 + \frac{1}{20}(T_{2,0}\ell^2 + 2)^2 \right)\sp \mathcal{K}_{AdS^5} = {40\over \ell^4} = {5\over 18}V_0^2\,.
\end{equation}
In equation (\ref{KintSlicing}) , $\mathcal{K}_{AdS_5}$ is the
Kretschmann scalar for the AdS$^5$ space-time. Therefore, AdS$_5$ is
recovered in the IR only when $T_{2,0} = -2/\ell^2$, which corresponds
to the AdS$_2 \times$S$^2$ slicing of AdS$_5$, whose explicit form
can be found  in appendix \ref{regads}.

 For any other value of
$T_{2,0}$ (in particular for any positive value, corresponding to an
S$^2\times$S$^2$ slicing), the Kretschmann scalar differs from the
AdS$^5$ value. This implies that the space-time with the metric
\eqref{eq:metricS2} is an asymptotically AdS$_5$ manifold but it
deviates from AdS$^5$ in the interior, and that there is no $S^2\times
S^2$ slicing of $AdS_5$.

\section{The on-shell  action}
\label{sec:FE}
In this section we compute the on-shell action of the bulk theory
 for regular $S^2\times S^2$-sliced solutions.  This will be used in
 sections 7 and 8 to determine which is the dominant solution when
 several are present for the same boundary conditions, since the
 Euclidean on-shell action  equals the free energy.

Starting  with the action (\ref{eq:action}),
the  on-shell action   is computed by substituting a
solution to the bulk equations into the bulk action. The details of
the computations can be found  in appendix \ref{app:FE}, and the
result is:

\be
\label{eq:F} S_{on-shell}= 32\pi^2M_p^3 \left(3\left[\frac{W_1(\f)+W_2(\f)}{T_1(\f)T_2(\f)}\right]^{UV}+\int_{UV}^{IR}\mathrm{d\f\over S(\f)}\;\left[\frac{1}{T_1(\f)}+\frac{1}{T_2(\f)}\right]\right)
\ee

The above expression is obtained from the action (\ref{eq:action}),
which is   written for the Lorentzian signature.  For a static
solution, the Euclidean action (aka the free energy ${\cal F}$) is
given by the same expression but for an overall sign.
\be\label{FvsS}
{\mathcal F} = - S_{on-shell}.
\ee

It is convenient to write the second term in equation (\ref{eq:F}) also as
a UV boundary term. For this, paralleling the procedure used in \cite{C}
for the maximally symmetric slicing, we introduce two new
superpotentials $U_1$ and $U_2$ as solutions of the differential equations:
\be
\label{eqUi} SU_i' - W_iU_i = -1
\ee
This allows us to write the integrals  appearing in the on-shell
action (\ref{eq:F}) as boundary terms: writing
\be
\label{reld}\frac{\mathrm{d}\f}{S(\f)T_i(\f)} = -\mathrm{d}\left(\frac{U_i(\f)}{T_i(\f)}\right)
\ee
 makes it possible to integrate the second term in  (\ref{eq:F}) and express $\mathcal{F}$ as
\be
\label{eq:Fbis}\mathcal{F} = -32\pi^2M_p^3 \left(3\left[\frac{W_1(\f)+W_2(\f)}{T_1(\f)T_2(\f)}\right]^{UV}+\left[\frac{U_1(\f)}{T_1(\f)}+\frac{U_2(\f)}{T_2(\f)}\right]_{IR}^{UV}\right)
\ee
The functions $U_i$ are defined up to an integration constant each.
 However, different  choices of these integration constants does not
 change the effective
 action, as it is clear from the fact that, for {\em any}  choice of the
 solutions of equations (\ref{eqUi}), the integral in the second term
 of equation (\ref{eq:F}) coincides with the second term in equation
 (\ref{eq:Fbis})    (for more details, see appendix \ref{app:FE}).

Given this
 freedom, it is convenient to choose the integration constants of
 (\ref{eqUi}) in such a way that the IR contribution in equation
 (\ref{eq:Fbis}) vanishes, and one is left with a UV boundary
 term. One can see that this is possible by solving (\ref{eqUi}) close
 to an IR endpoint.   We insert the expansions \eqref{eq:W1gensol}-\eqref{eq:Sgensol}
 into \eqref{eqUi} and find upon integration:
\begin{align}
\label{U1IR}U_1(\f) & \underset{\f \to \f_0^-} = \frac{b_1}{\f-\f_0} + U_0\sqrt{|\f-\f_0|} + \mathcal{O}(|\f-\f_0|) \, , \\
\label{U2IR}U_2(\f) & \underset{\f \to \f_0^-} = b_2 + U_0\sqrt{|\f-\f_0|} + \mathcal{O}(|\f-\f_0|) \, .
\end{align}
with $b_1$ and $b_2$ two integration constants and
\be
\label{defU0}U_0 \equiv \sqrt{\frac{6}{|V'(\f_0)|}}
\ee
In particular, choosing $b_1=b_2=0$ fixes the solution completely and
in such a way that, with the behavior of $T_i$ given in equations
(\ref{T1}-\ref{T2}), we have  $U_i/T_i \to 0$ as $\f
\to \f_0$ and only the UV contribution remains in the second term of
equation  (\ref{eq:Fbis}).

In what follows we need  the expression for the  near-boundary
expansion of $U_i$.
It is obtained by  substituting \eqref{eq:W1gensol}-\eqref{eq:Sgensol}
into \eqref{eqUi}. As $\f \to 0$, we obtain:
\be
\label{U1UV}U_1(\f) \underset{\f \to 0^+} = \ell \left[\frac{1}{2} +
  \left(\mathcal{B}_1 +\frac{2\mathcal{R}_1-\mathcal{R}_2}{12\D_-}
    \log |\f| \right)|\f|^{2/\Delta_{\pm}}[1+ \mathcal{O}(\f)]\right]
\ee
\be
\label{U2UV}U_2(\f) \underset{\f \to 0^+} = \ell \left[\frac{1}{2} + \left(\mathcal{B}_2 +\frac{2\mathcal{R}_2-\mathcal{R}_1}{12\D_-}  \log |\f|\right)|\f|^{2/\Delta_{\pm}}[1+ \mathcal{O}(\f)]\right]
\ee
where $\mathcal{B}_1$ and $\mathcal{B}_2$ appear as new integration constants,
which however are completely fixed by the  choice we already made to
set $b_1=b_2=0$ in the IR expansion.  Therefore,  $\mathcal{B}_1$ and $\mathcal{B}_2$  are
completely determined by the other integration constants  appearing in
$W$ and $S$. Among these,  $C_1$ and  $C_2$  are fixed by regularity
in terms of the UV curvatures $\mathcal{R}_i$, therefore $\mathcal{B}_i =
\mathcal{B}_i(\mathcal{R}_1,\mathcal{R}_2)$.

Notice however that this
determination may not be unique: for a given choice
$\mathcal{R}_i$ however there still may be different (discretely many)
regular solutions characterized by different values of $C_i$ and
$\mathcal{B}_i$.

\subsection{The UV-regulated free energy }
The free energy is a divergent quantity, due to the infinite volume of
the solution  near the boundary. We make it finite  by evaluating the various quantities at the regulated boundary at $u/\ell = \log(\e)$ with $\e \ll 1$ and we define the dimensionless energy cutoff:
\be
\label{deflambda}\Lambda \equiv \left. \frac{\ex^{\frac{A_1(u)+A_2(u)}{2}}}{\ell (R_1^{UV}R_2^{UV})^{1/4}}\right|_{\frac{u}{\ell}=\log(\e)}
\ee

The UV-regulated  free energy is then given by:
\be
\label{eq:Fbisbis}\mathcal{F}
= -32\pi^2M_p^3 \left[3\frac{W_1(\f)+W_2(\f)}{T_1(\f)T_2(\f)}+\frac{U_1(\f)}{T_1(\f)}+\frac{U_2(\f)}{T_2(\f)}\right]^{\f(\log(\e) \ell)}
\ee
where we have made explicit the dependence on the dimensionless parameters which enter the superpotentials, as well as the cut-off.

As shown in Appendix \ref{app:FE}, the free energy can be organized in
an expansion in $\Lambda^{-1}$, which takes the following form:
\bea \label{FEexp}
{\cal F} && = \Lambda^4 {\cal F}_{4}(\mathcal{R}_1,\mathcal{R}_2) +
\Lambda^2 {\cal F}_{2}(\mathcal{R}_1,\mathcal{R}_2) + (\log \Lambda)
{\cal F}_{0}(\mathcal{R}_1,\mathcal{R}_2) + \nonumber \\
&& + \overline{\mathcal{F}}(\mathcal{R}_1,\mathcal{R}_2,
  C_1,C_2,\mathcal{B}_1,\mathcal{B}_2) + O(\Lambda^{-\Delta_-}) .
\eea
The explicit expression is given by equation (\ref{expUVFbisbis}). The important point is that the terms which are divergent as $\Lambda
\to +\infty$, i.e.  ${\cal F}_{4},  {\cal F}_{2},  {\cal F}_{0}$, are
universal\footnote{Their explicit expressions can be found in equation
(\ref{expUVFbisbis})}, i.e. they only  depend on $\mathcal{R}_1,\mathcal{R}_2$,
which are fixed by the boundary conditions. On the contrary,  the vev
parameters $ C_1,C_2,\mathcal{B}_1,\mathcal{B}_2$
only enter the finite term $\overline{{\cal F}}$.

As a consequence, the  free energy {\em difference}
between two solutions with the same boundary curvatures $\mathcal{R}_1$ and
$\mathcal{R}_2$, but different  sets of  vev parameters
$(\mathcal{B}_i, C_i)$ and
$(\tilde{{\mathcal B}}_i,\tilde{C}_i)$ ,  is finite\footnote{It is also
  scheme-independent, i.e. it is unaffected if we regulate the free energies using a
  different prescription, or use boundary counterterms.},
\be
\label{defDF} \Delta\mathcal{F} = \lim_{\Lambda\to +\infty} \Big[\mathcal{F} - \tilde{\mathcal{F}}\Big] \, ,
\ee
  and it reduces to the remarkably simple expression:
\begin{equation}
\label{DF} \Delta\mathcal{F} =
-\frac{32\pi^2M_p^3\ell^3}{\mathcal{R}_1\mathcal{R}_2}\,\left[(\mathcal{B}_1-\tilde{\mathcal{B}}_1)\mathcal{R}_2+(\mathcal{B}_2-\tilde{\mathcal{B}}_2)\mathcal{R}_1
  + 3\Big((C_1+C_2)-(\tilde{C}_1+\tilde{C}_2)\Big) \right] ,
\end{equation}

\subsection{The free energy for CFTs on S$^2\times$ S$^2$}
We now consider the special case of a  CFT on $S^2\times S^2$. The only two energy scales in the problem are the curvatures of the two $S^2$s and the only non-trivial dimensionless parameter is the ratio of the two curvatures.

In the conformal case,  the scalar field is constant and locked at an
extremum of the bulk potential. We can still  define dimensionless
curvatures in AdS units,
\be
\label{defRCFT} \mathcal{R}_1 = R_1^{UV}\ell^2 \sp \mathcal{R}_2 = R_2^{UV}\ell^2.
\ee
For the on-shell action, the same expression \eqref{eq:Fbisbis} can be used after replacing the superpotentials by their expression in terms of $A_1$ and $A_2$:
\be
\label{eq:Fnof}\mathcal{F}
= -32\pi^2M_p^3
\left[-6\frac{\dot{A}_1(u)+\dot{A}_2(u)}{R^{\zeta_1}R^{\zeta_2}\ex^{-2(A_1(u)+A_2(u))}}+
\frac{U_1(u)}{R^{\zeta_1}\ex^{-2A_1(u)}}+\frac{U_2(u)}{R^{\zeta_2}\ex^{-2A_2(u)}}\right]^{u=\log(\e)
  \ell}
\ee

In terms of these UV parameters, the near-boundary expansions of $A_1$
and $A_2$ is given in equations (\ref{eq:A1msolu-i}-\ref{eq:A2msolu-i}).  %
Again, we choose $\bar{A}_{1} = \bar{A}_{2} = 0$ to have $R^{\zeta_{1,2}} =
R_{1,2}^{UV}$. As we discussed in section \ref{CFT},  in this case  there  is only a
single vev parameter, denoted by  $C$. Its field theory interpretation
is given in equation (\ref{Tabcft}).

To rewrite the free energy in terms of boundary quantities, we can again define the functions $U_1(u)$ and $U_2(u)$, which now  are solutions of the following ODEs:
\be
\label{eqUiu} \dot{U}_i +2\dot{A}_i U_i = -1 \sp i=1,2
\ee
As before,  the choice integration constants of these equations is
not affecting the free energy, and we can choose them so that the IR
contribution   vanishes. This choice is implicit in equation
(\ref{eq:Fnof}).

 Near the boundary (in the limit $u \to -\infty$), the U superpotentials have the following expansion:
\be
\label{U1UVu}U_1(u) = \ell \left[\frac{1}{2} + \left(\mathcal{B}_1
    +\frac{2\mathcal{R}_1-\mathcal{R}_2}{12} \frac{u}{\ell}\right)
  \ex^{2u/\ell}[1 + \ldots]\right]
\ee

\be
\label{U2UVu}U_2(u) = \ell \left[\frac{1}{2} + \left(\mathcal{B}_2 +\frac{2\mathcal{R}_2-\mathcal{R}_1}{12} \frac{u}{\ell}\right) \ex^{2u/\ell}[1 + \ldots]\right]
\ee
where $\mathcal{B}_1$ and $\mathcal{B}_2$ are integration constants,
which are  chosen so that $U_i = 0$ a the IR endpoint. This fixes
$\mathcal{B}_1$ and $\mathcal{B}_2$ (up to possible discrete
degeneracies) in terms of the position of the endpoint, or
equivalently of the UV parameters $R_{1,2}^{UV}$.

We can expand the free energy in terms of the  dimensionless
cut-off  $\Lambda$, defined in equation (\ref{deflambda}). The
explicit expression is given in equation (\ref{expUVFu})  and has the
same structure as  (\ref{FEexp}), except that the finite part only contains
on the  $\mathcal{B}_i$ vevs.

The free energy difference between two different solutions with same parameters
$\mathcal{R}_2$ and $\mathcal{R}_1$ is
\be
\label{DFu} \Delta\mathcal{F} = -{32\pi^2M_p^3\ell^3 \over
  \mathcal{R}_1 \mathcal{R}_2}\,\left[\mathcal{R}_2(\mathcal{B}_1-\widetilde{\mathcal{B}_1})+
      \mathcal{R}_1(\mathcal{B}_2-\widetilde{\mathcal{B}_2})\right]
\ee

\section{Holographic CFTs on  S$^2\times$ S$^2$ and Efimov phenomena}

\label{sec:CFT}

We are now ready to analyse in detail the full 2-sphere-flow geometries,
the various branches of the solutions, and the phase transitions
between various branches.

We start with the case of a CFT on S$^2\times$ S$^2$. This case already
displays a very interesting phase diagram, and it will give an insight
on what occurs for non-trivial RG flows on S$^2\times$S$^2$.

That the constant scalar field is already non-trivial is expected: we
already know from the discussion on the Kretschmann scalar in section
\ref{sec:IRbc} that the solution is an
asymptotically-AdS$^5$-space-time, but that is not AdS$^5$
everywhere.
 In contrast, the geometry corresponding to a CFT on $S^4$
{\em is} $AdS_5$ in a different coordinate system.

Since we could not solve Einstein's equations analytically,  we
employ a numerical approach.
We shall proceed by solving  equations \eqref{eq:EOM6ter} - \eqref{eq:EOM6ter} numerically for $A_1(u)$ and $A_2(u)$.
As it  happens in similar cases, solving the equations starting near the boundary, we generically end up with singular solutions in the bulk.
It is easier to specify boundary conditions for $A_1$ and $A_2$ at the
point where space-time ends in the interior,  as this is a
potential singularity. Demanding the absence of a singularity gives us
special initial conditions.

As we have seen in section 5, regularity demands that one of the
spheres shrink to zero size,  while the other one stays finite. In the following we assume that it is the sphere 1 that shrinks in the interior, at some value of $u$ that we call $u_0$.

The relevant regular boundary conditions on $W_1$,$W_2$,$T_1$ and
$T_2$ are described in section \ref{sec:IRbc}. In the case of a CFT,
we have
\be \label{wiaiti}
W_i = -2\dot A_i(u), \quad T_i = R^{UV}_i\mathrm{e}^{-2A_i(u)}, \qquad i=1,2
\ee
and their  expansions near a regular endpoint can be obtained from
equations the behavior of $A_1$ and $A_2$,
\be\label{ansAter}
A_1 (u) \simeq \ln {u_0-u\over \ell}, \qquad A_2(u) \simeq A_{2,0},
\qquad u \to u_0,
\ee
see equation  (\ref{ansAbis}).

There are two free dimensionless parameters in the IR: these are $u_0/\ell$ and
$A_{2,0}$ defined in \eqref{ansAter}.This  means that enforcing
regularity in the interior imposes one constraint on the three
boundary integration constants $C, R_1^{UV}, R_2^{UV}$. This
constraint can be written as $C = C(R_1^{UV},R_2^{UV})$. Since the
theory is conformal, $C$ actually only depends on the dimensionless
ratio $R_1^{UV}/R_2^{UV} = {\mathcal R}_1/ {\mathcal R}_2$.

The dependence of the three boundary integration constants on
$u_0/\ell$ can actually be deduced from the behavior  of the equations
of motion \eqref{eq:EOM6ter} - \eqref{eq:EOM6ter} under a shift of
$u$: Near the boundary, $A_{i}(u) \simeq -u/\ell$, implies  that
\be\label{scaling-u0}
R_{1,2}^{UV}\propto \ex^{-2u_0/\ell}, \qquad C\propto \ex^{-4u_0/\ell}
\ee
This implies in particular that the dimensionless quantities
$R_2^{UV}/R_1^{UV}$ and $C/((R_1^{UV})^2\ell^4)$ do not depend on
$u_0/\ell$.
We are therefore left with one  dimensionless parameter in the IR, that
 is $A_{2,0}$, which completely fixes the solution up to a choice of
 overall scale. The two dimensionless  UV parameters
 $R_2^{UV}/R_1^{UV}$ and $C/((R_1^{UV})^2\ell^4)$  are fixed by the
 choice of $A_{2,0}$.  Rather than $A_{2,0}$, for numerical purposes, we find it more
   convenient to work with  $T_{2,0}$  as an
   IR parameter independent of $u_0/\ell$, defined by
\be \label{T20}
T_{2,0} = R^{(\zeta_2)}e^{-2A_{2,0}}.
\ee

\begin{figure}[h!]
\begin{center}
\includegraphics[scale=0.80]{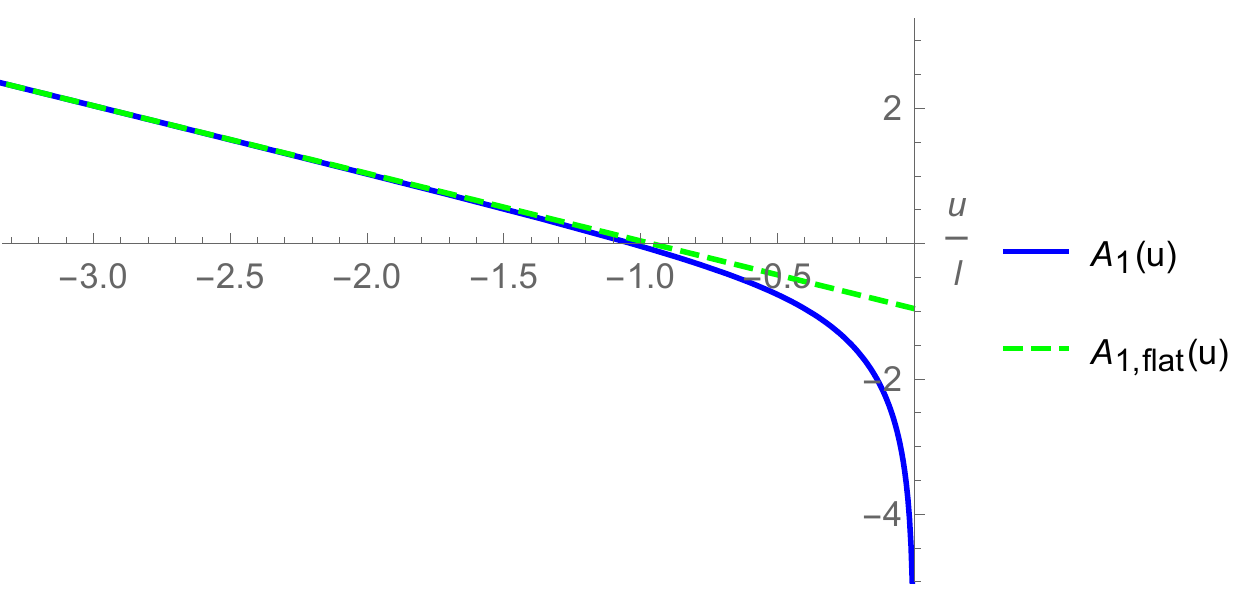} \hfill \includegraphics[scale=0.80]{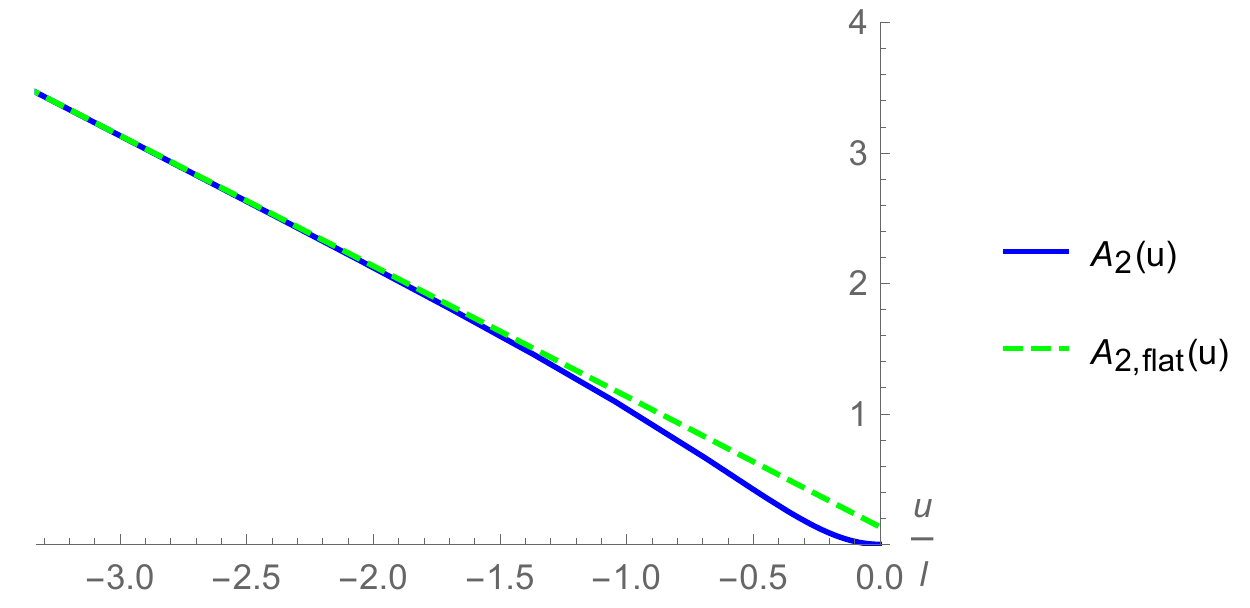}
\end{center}
\caption{The blue solid lines show the scale factors $A_1(u)-A_{1,0}$
  (top panel)  and $A_2(u)-A_{2,0}$ (bottom panel) in the case where the sphere 1
  shrinks for $u_0/\ell = 0$. The constant shifts $A_{1,0}$ is given
  in   equation (\protect\ref{A-end}), whereas $A_{2,0}$,  related to
  $T_{2,0}$ in (\protect\ref{A-end}), fixes   the initial
  condition in the IR (see equation (\protect\ref{ansAter})).  It is set  so that $T_{2,0} \ell^2 = 20$.  The green dashed line represents the same functions, in the case  of a flat boundary. }
\label{fig:Amain}
\end{figure}

A typical  numerical solution for $A_{1,2}(u)$ is presented in Figure
\ref{fig:Amain}. The initial conditions are given in the IR: we pick an
arbitrary $u_0$ then fix $A_1, \dot{A_1}$ and $A_2$ at a point $u$
slightly shifted from the IR endpoint ($u=u_0-\epsilon$),
so that the solution behaves as in (\ref{ansAter}). This fixes all initial
conditions of the system (\ref{eq:EOM6ter}-\ref{eq:EOM6ter}). While $A_1$ and $\dot{A_1}$ at the
endpoint are fixed by regularity, the value $A_{2,0}$ of  $A_2$ at
$u_0-\epsilon$ is free and we vary it to scan over different
solutions. For each solution, we then  read-off the boundary
parameters by analysing the asymptotic behavior as $u\to \infty$, as will be
discussed in detailed in subsection 7.1

\begin{figure}[h!]
\begin{center}
\includegraphics[scale=1]{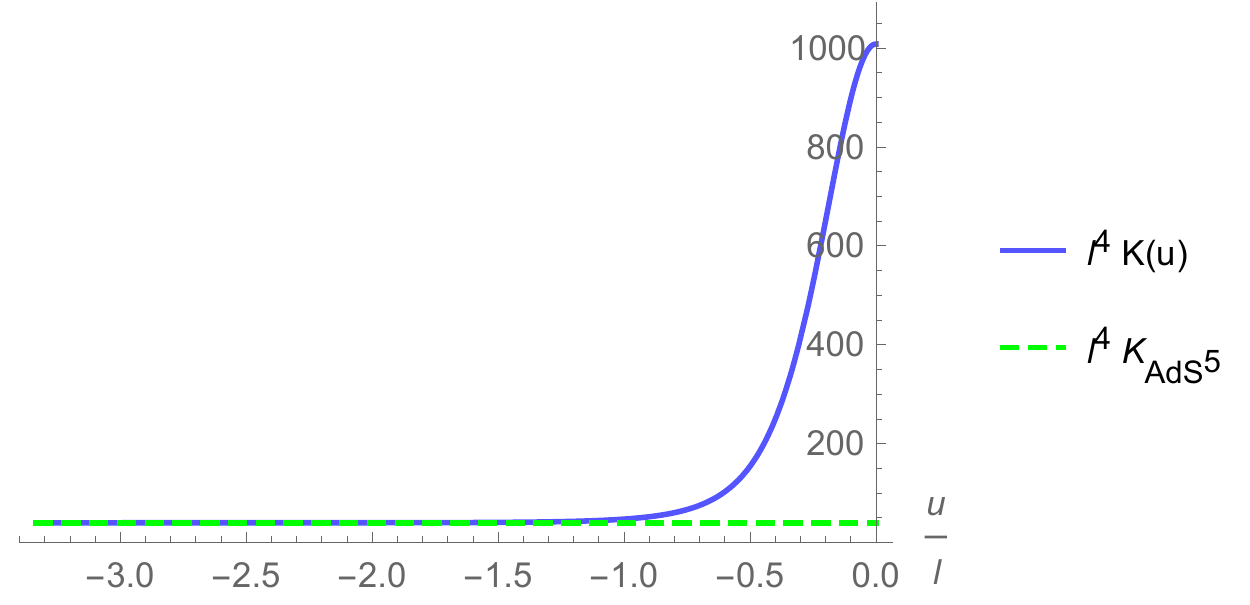}
\end{center}
\caption{The dimensionless Kretschmann scalar $\ell^4 \mathcal{K}(u) = \ell^4 R_{ABCD}R^{ABCD}$ (expressed in terms of the holographic variables in \protect\eqref{bb5}) as a function of $u$ for $u_0 = 0$. To be specific we set $T_{2,0}\ell^2 = 20$.}
\label{fig:K_main}
\end{figure}

When $u \to - \infty$, we approach the AdS boundary and the solutions
are described by the asymptotic form \eqref{eq:A1msolu-i} and
\eqref{eq:A2msolu-i}.
One interesting quantity is the bulk Kretschmann scalar
$\mathcal{K}(u) = \ell^4 R_{ABCD}R^{ABCD}$, whose expression is given
by equation  (\ref{Riemann}) (with $S=0$ for the CFT case).  For the solution corresponding to Figure \ref{fig:Amain},
$\mathcal{K}(u)$ is   shown  in Figure \ref{fig:K_main}: as expected,
as we move towards the interior it deviates from its constant $AdS_5$
value (which is attained  as $u\to -\infty$).  We have verified  that the value obtained  at $u_0$ is in agreement with \eqref{KintSlicing}.
\subsection{The UV parameters}
Given a numerical solution, we can extract the corresponding values of $\mathcal{R}_1$, $\mathcal{R}_2$ and $C$  explicitly by fitting the UV region with the asymptotics \eqref{eq:A1msolu-i} and \eqref{eq:A2msolu-i}.

\begin{figure}[h!]
\begin{center}
\includegraphics[scale=1]{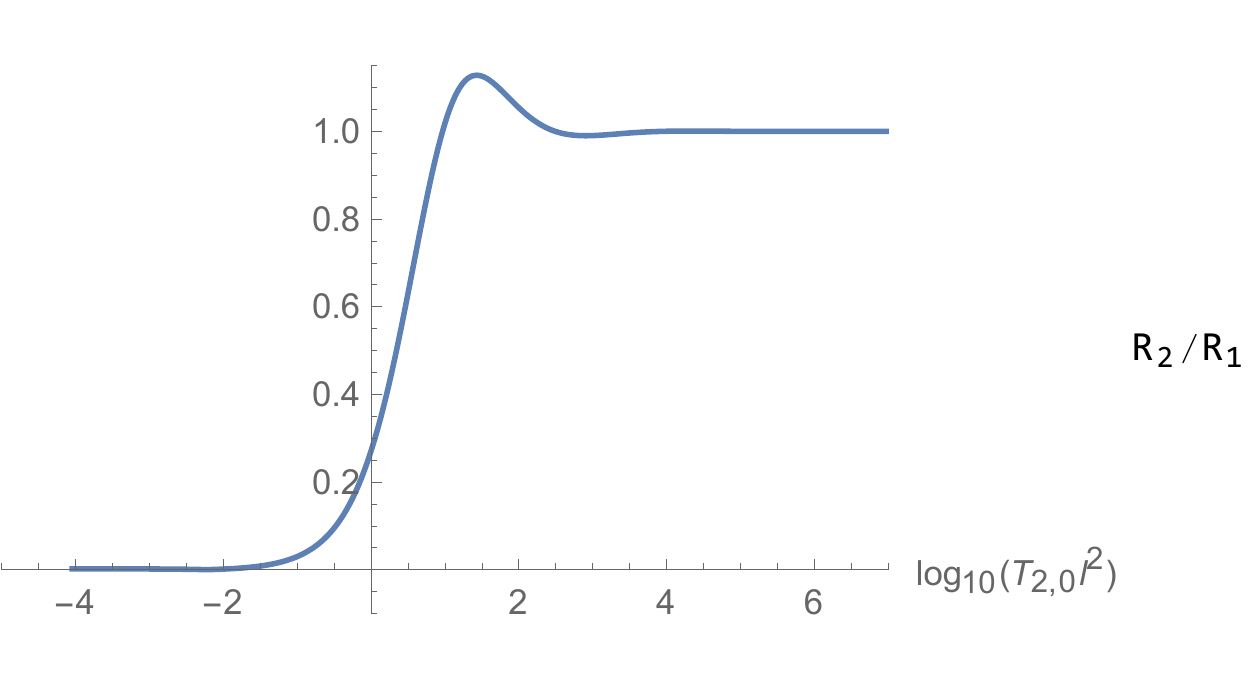}
\end{center}
\caption{$\frac{\mathcal{R}_2}{\mathcal{R}_1}(T_{2,0}\ell^2)$ in the case where the sphere 1 shrinks. The ratio is independent of $u_0$.}
\label{fig:R2sR1u_main}
\end{figure}

\begin{figure}[h!]
\begin{center}
\includegraphics[width=11cm]{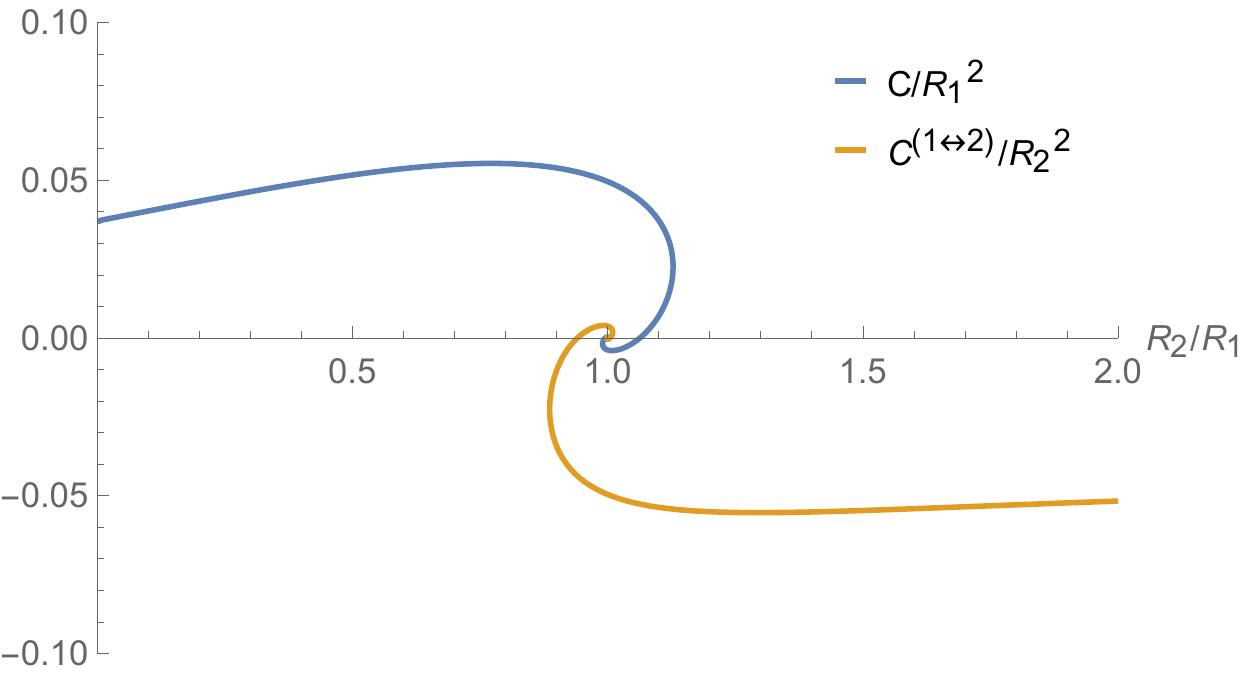}
\vskip 1.5cm
 \includegraphics[width=11cm]{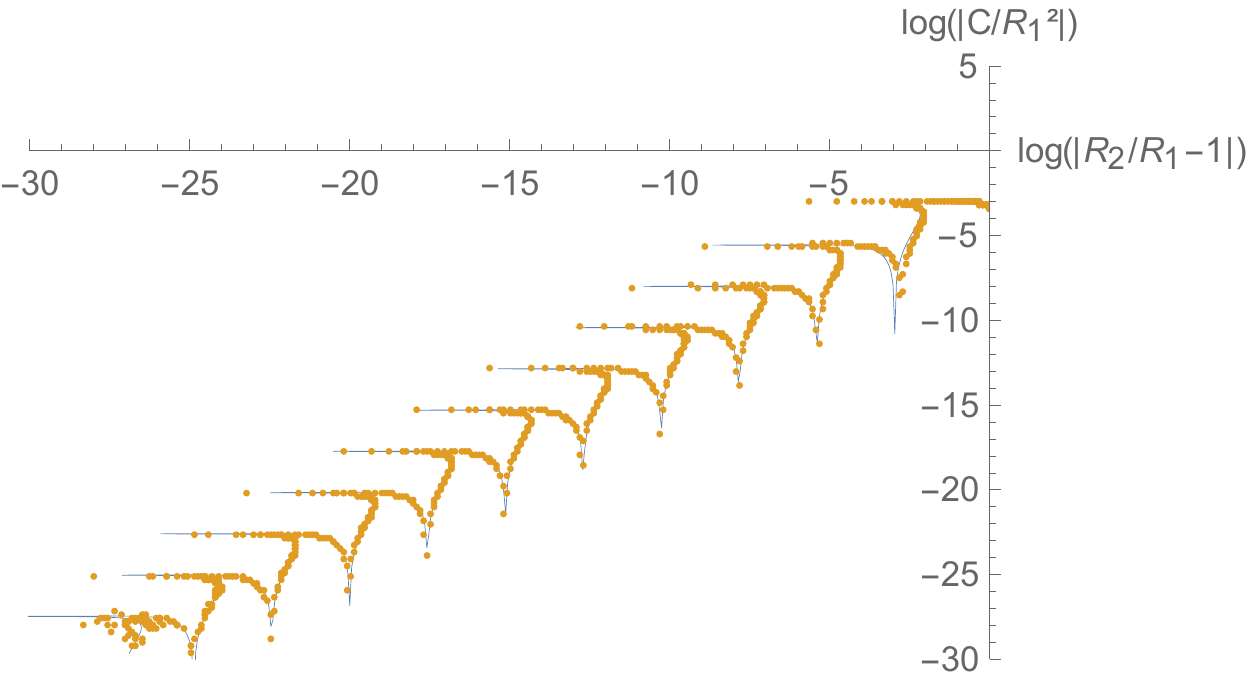}
\end{center}
\caption{{\em Top:} $C/\mathcal{R}_1^2$ in both the case where the sphere 1
  shrinks (blue) and the case where the sphere 2 shrinks
  (orange). {\em Bottom:} The same plot in the case where sphere 1 shrinks, where we represent the logarithm of the distance to the center of the spiral with coordinates $(1,0)$ for each quantity. The orange dots are given by numerical computation while the blue curve is the fit found in \protect\eqref{R2sR1vs}-\protect\eqref{Cvs}.}
\label{fig:RCvR_main}
\end{figure}

Let us first clarify the influence of $T_{2,0}\ell^2$ on the UV
parameters $\mathcal{R}_2/\mathcal{R}_1$ and
$C/\mathcal{R}_1^2$. Figure \ref{fig:R2sR1u_main} shows the evolution
of $\mathcal{R}_2 / \mathcal{R}_1$  (recall that this is independent
of $u_0$) when $T_{2,0}\ell^2$ varies from $0$ to $+\infty$. From the
figure, we observe the following facts:
\begin{itemize}
\item Each choice of $T_{2,0}\ell^2$ uniquely fixes the value
$\mathcal{R}_2 / \mathcal{R}_1$.
\item
When the ratio of curvatures  is far from unity,  increasing
$T_{2,0}\ell^2$  essentially amounts to increasing the ratio
$\mathcal{R}_2 / \mathcal{R}_1$.
\item As $T_{2,0}\ell^2 \to +\infty$,  $\mathcal{R}_2 / \mathcal{R}_1
  \to 1$ in a  non-monotonic way: the curvature ratio follows dampened
oscillations around the asymptotic value.  Thus, there is an infinite
number of values of  $T_{2,0}\ell^2$  for which
$\mathcal{R}_2/\mathcal{R}_1 = 1$. We shall see later that the dampened
oscillations are directly linked to a discrete scaling of the type of an {\em Efimov spiral.}
\end{itemize}
The other UV parameter $C/\mathcal{R}_1^2$  follows the same kind of dampened oscillation behavior. Figure \ref{fig:RCvR_main} gives a complete description of the solutions in the parameter space, that is the plane $\left(\mathcal{R}_2/\mathcal{R}_1, C/\mathcal{R}_1^2\right)$, both in the case where the sphere 1 shrinks to zero size in the IR and in the symmetric case where the sphere 2 does. The parameter that parametrizes  the curve is $T_{2,0}\ell^2$ (which increases as one follows the curve from the point $\mathcal{R}_2 = 0$).

Interestingly, we observe that there can be several possible values of
$C$ for a given value of the ratio $\mathcal{R}_2/\mathcal{R}_1$. The
resulting figure in the
$(\mathcal{R}_2/\mathcal{R}_1,C/\mathcal{R}_1^2)$ plane is a spiral
that shrinks exponentially as is apparent from the logarithmic plot in
Figure \ref{fig:RCvR_main}. This kind of behavior has been already observed in D-brane models, \cite{Liu} and holographic V-QCD, \cite{JK}, and is known as the  Efimov spiral.

Remarkably, if  both spheres have the same radius in the UV, an
infinite number of solutions exist, in which one of the two spheres
shrinks but not the other, corresponding to either positive or
negative $C$. This  translates into a   spontaneous breaking of the
$\mathbb{Z} ^2$-symmetry that exchanges the two spheres, which is a
symmetry in the UV for ${\mathcal R}_1 = {\mathcal R}_2$. The solution
in which the symmetry is unbroken corresponds to the end of the
spiral: $C=0$ and $T_{2,0} \to +\infty$. In this case, both spheres shrink, but the solutions is singular in the bulk.

\subsection{Efimov spiral}
\label{spiralcft}
We now investigate the origin of the  Efimov spiral, which arises
due to the appearance of multiple solutions as the UV radii of the two
spheres  ${\mathcal R_1}$ and ${\mathcal R_2}$ get close to each
other.

We follow a reasoning similar to what was done in \cite{JK}. The idea is to consider the behavior of a quantity in the bulk
theory when its source asymptotes  to zero, and look for signs of an
instability which will trigger a non-zero vev.   Here the relevant source is
$\mathcal{R}_2-\mathcal{R}_1$ and the corresponding quantity $A_1-
A_2$. The latter  has the following behavior in the  UV, which can be
deduced from equations  (\ref{eq:A1msolu-i}-\ref{eq:A2msolu-i}):
\be
\label{A1mA2sp}A_1(u) - A_2(u) \underset{u\to -\infty}{=} \bar{A}_1 -\bar{A}_2 +\frac{1}{8} (\mathcal{R}_2 - \mathcal{R}_1) e^{2u/\ell}
\ee
$$
+ \frac{1}{96}(\mathcal{R}_1^2-\mathcal{R}_2^2)\,  \frac{u}{\ell} \, \ex^{4u/\ell}[1+\mathcal{O}(\ex^{\Delta_-u/\ell})] - \frac{C}{4} e^{4u/\ell}  +\mathcal{O}(\ex^{6u/\ell}) \, .
$$
For simplicity we choose $R^{\zeta^{1}}=R^{\zeta^{2}}=2/\ell^2$. Note that $C$ is indeed the associated vev (this is how it is defined), which is consistent with the spiral appearing in the plane $(\mathcal{R}_2/\mathcal{R}_1,C/\mathcal{R}_1^2)$.

We now consider the case where:
\begin{enumerate}
\item $\mathcal{R}_1 = \mathcal{R}_2$ which is equivalent to $\bar{A}_1=\bar{A}_2$.
\item  $A_1 - A_2$ is infinitesimal, and we define
\be
\label{defe}\e = A_1 - A_2
\ee
\item  We are away from the UV regime, so that
\be \label{defe2}
\dot{A}_1 \simeq
  1/(u-u_0)
\ee
(as implied by the asymptotic behavior (\ref{ansAter}),
  but not too close to the IR end-point $u_0$ (where we know that $A_1
  - A_2 \to -\infty$), so that we can consider $\epsilon$ small.
\end{enumerate}
Condition 1 amounts to choosing $Z_2$-symmetric boundary conditions in
the UV; in this case, condition 2 certainly holds close to the
boundary (by  equation (\ref{A1mA2sp}) and condition 1) and down to the
point where the radii of the two spheres start deviating due to the
non-zero vev $C$. Condition  3 identifies an intermediate region
between the UV and the IR, as we explain below.

More precisely, the last condition holds in an intermediate region
\be
\label{aUV}
\alpha_{IR} \ll u_0-u \ll \alpha_{UV} , \qquad  \left\{\begin{array}{l}\a_{UV}
    = \ell \\ \a_{IR} = \sqrt{\frac{2}{T_{2,0}}} \end{array} \right. \, .
\ee
In other words,  $\alpha_{UV}$ is the UV $AdS$ radius and
$\alpha_{IR}$ the IR radius of sphere 2 (see equation (\ref{alphair})).

The  range (\ref{aUV}) for the validity of the last condition  can be
understood as follows. From the asymptotic behavior (\ref{ansAter}),
the assumption $A_1 - A_2 \ll 1 $ is violated in the interior starting
from the point $u_{IR}$ where  $\log((u_0-u_{IR})/\ell) \sim A_{2,0}$,
so that  (using the relation (\ref{A-end})):
\be
u_{IR} \approx u_0 - \ell \ex^{A_{2,0}} = u_0 - \ell \sqrt{R^{\zeta_2} / T_{2,0}}
\ee
 is the typical IR boundary of the region satisfying condition 3. Since we
are working with  $R^{\zeta_2} = 2/\ell^2$, this leads to the identification
 $\alpha_{IR}$ in equation (\ref{aUV}).

In the UV, the condition
$$\dot{A}_1 \sim {1\over (u-u_0)}$$  is violated
starting from the point $u_{UV}$ such that
$$-{1\over \ell} \sim {1\over (u_{UV}-u_0)}\;,$$
 where we used the fact that the leading UV
behavior of $A_1(u)$ is  $A_1(u) \sim -{u\over \ell}$. Therefore,  in the UV,
condition 3 is valid starting approximately at  $u_{UV} \approx u_0 -
\ell$. This leads to the upper bound in equation (\ref{aUV}).

The ratio of the IR length-scale to UV length-scale in (\ref{aUV})  is then given by:
\be
{\a_{IR}\over \a_{UV}}=\sqrt{\frac{2}{T_{2,0}\ell^2}}
\ee
Note that a condition for the validity of our analysis is that $\alpha_{IR} \ll
\alpha_{UV}$. This is automatic in the limit $T_{2,0} \to \infty$.

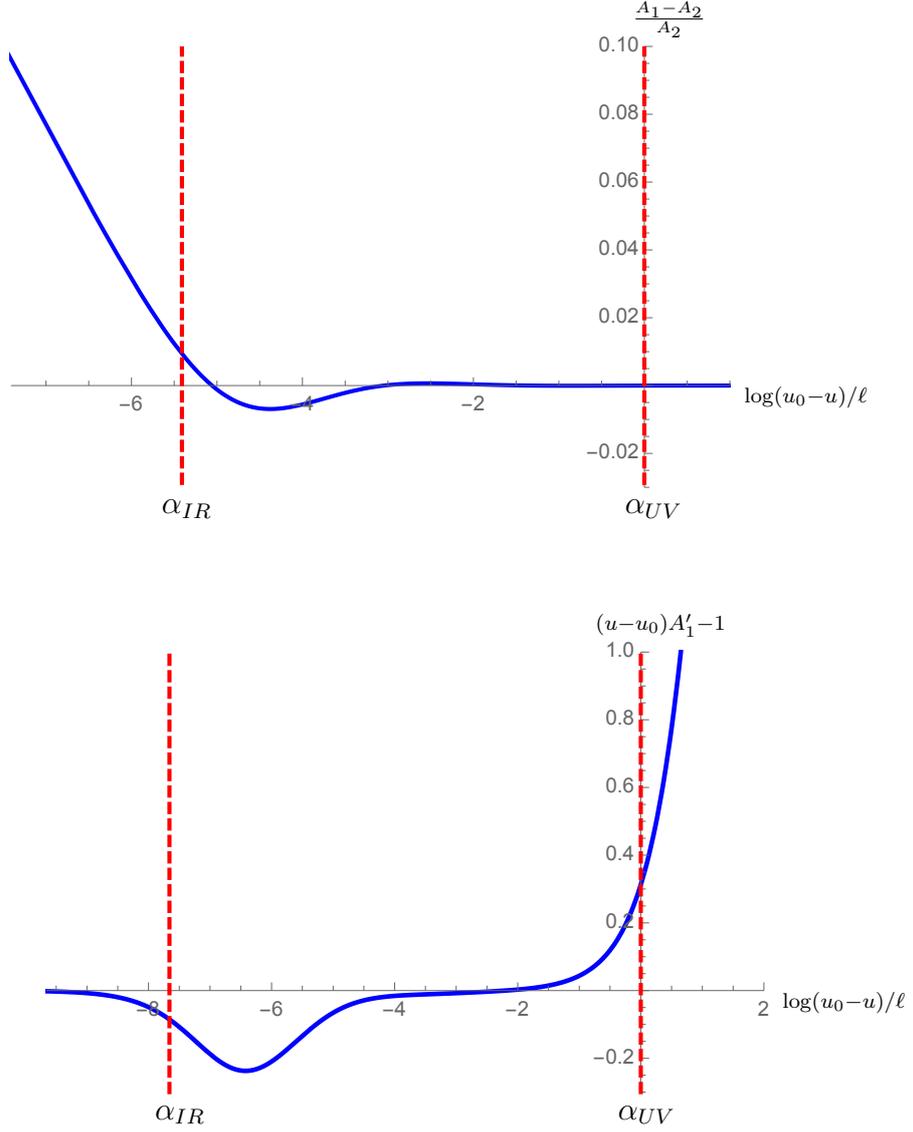
\begin{figure}[h!]
\centering
\begin{overpic}
[width=0.65\textwidth]{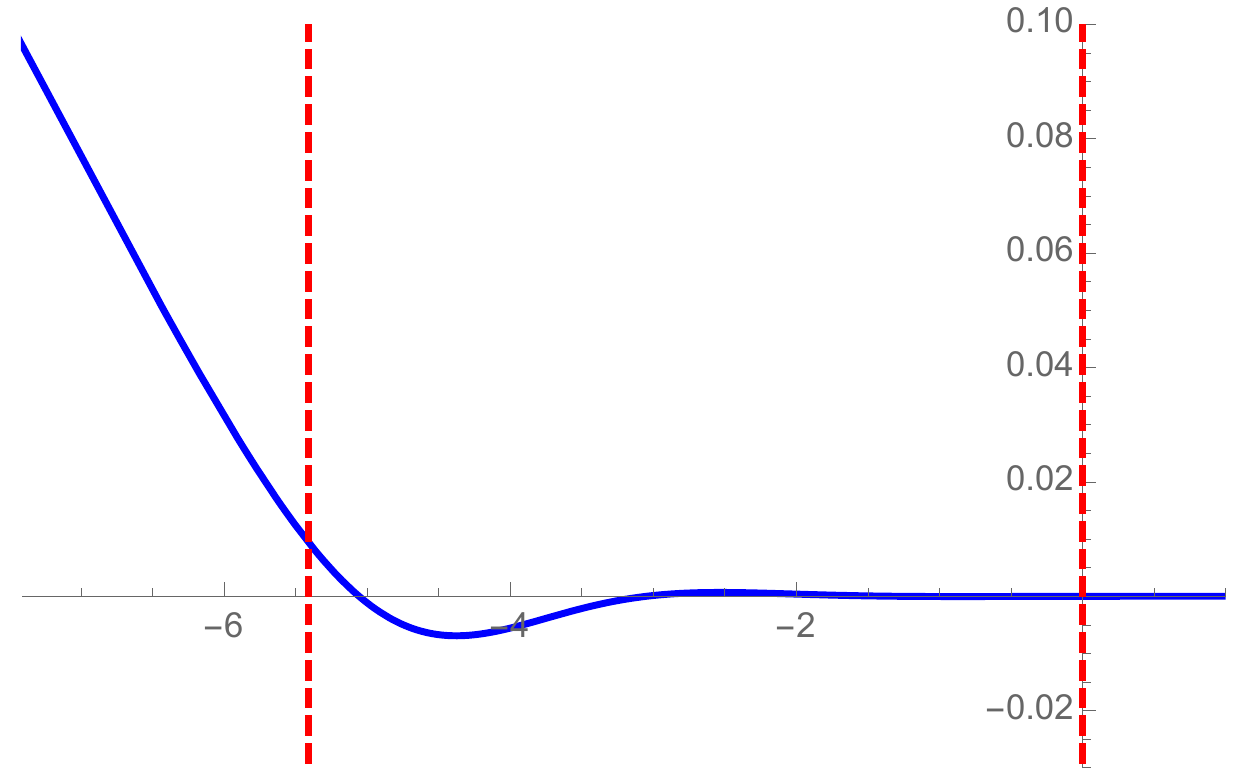}
\put(100,12){$\scriptstyle{\log(u_0-u)/\ell}$}
\put(85,63){$\scriptstyle{{A_1-A_2 \over A_2}}$}
\put(84,-3){$\alpha_{UV}$}
\put(22,-3){$\alpha_{IR}$}
\end{overpic}
\\
\vskip 2cm
\hspace{1cm}
\begin{overpic}
[width=0.65\textwidth]{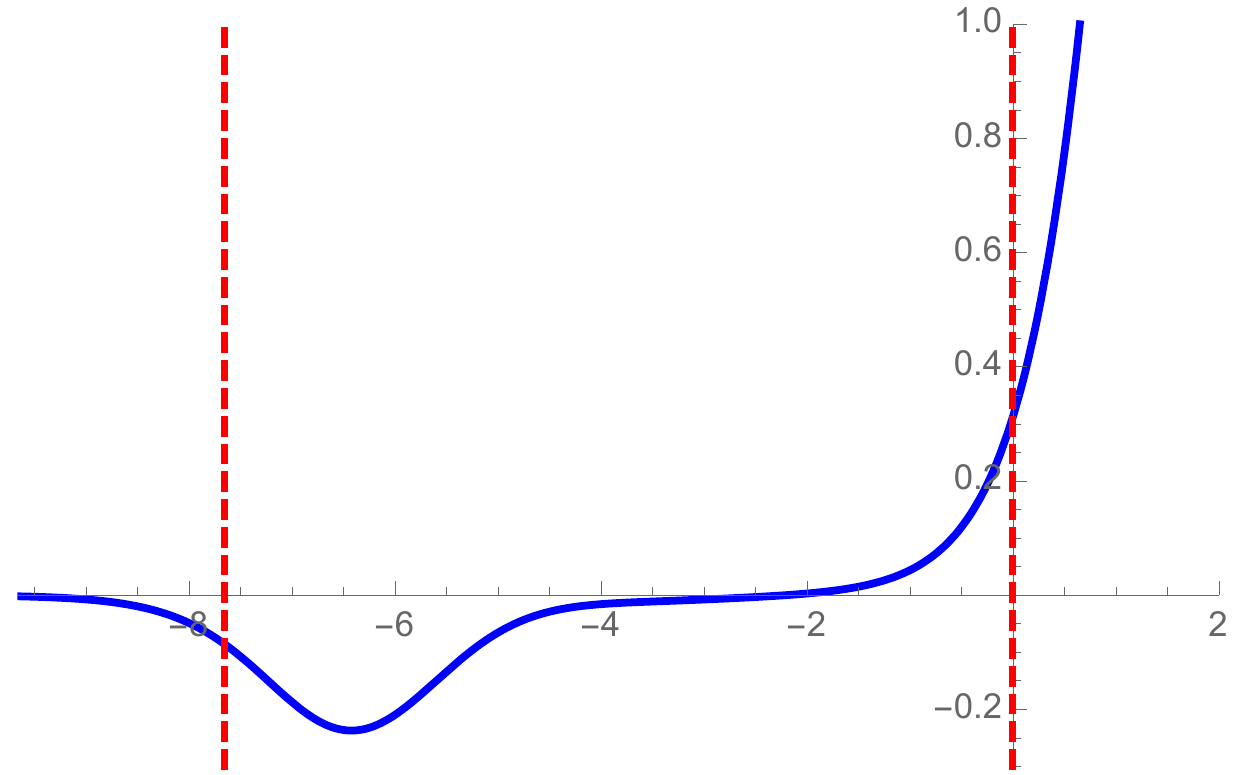}
\put(100,12){$\scriptstyle{\log(u_0-u)/\ell}$}
\put(75,63){$\scriptstyle{(u-u_0) A'_1 - 1}$}
\put(78,-3){$\alpha_{UV}$}
\put(16,-3){$\alpha_{IR}$}
\end{overpic}
\vspace{1cm}
\caption{These figures show the region of validity of equation
  (\protect\ref{eqe}), delimited by the red dashed lines
corresponding to the boundary values in equation
(\protect\ref{aUV}). The upper figure is a plot of  the relative difference
between  $A_1$ and $A_2$, while the lower figure shows the relative
difference between $\dot{A}$ and $(u-u_0)^{-1}$. Both are small in the
region delimited by the dashed lines, confirming the validity of
equation   (\protect\ref{eqe}) in this regime.}
\label{fig:defe}
\end{figure}

We have checked numerically  that in the range (\ref{aUV}) both
conditions 2 and 3 are satisfied: in this range, both  (\ref{defe}) and
(\ref{defe2}) hold,  as shown in figure \ref{fig:defe}.

Under the assumptions (\ref{defe}) and (\ref{defe2}),  we rewrite the EoM's
\eqref{eq:EOM6ter}-\eqref{eq:EOM8ter} as an expansion in $\epsilon$. In particular,
\eqref{eq:EOM8ter} reads, to linear order in $\epsilon$:
\be
\label{eqe1}
\ddot{\epsilon} + 4\frac{\dot{\e}}{u-u_0}(1+\cdots) + T_1(\e +
\mathcal{O}(\e^2)) = 0,
\ee
where the dots refer to subleading terms in the expansion in
$u-u_0$. The quantity $T_1$ is given by \eqref{t1cft} and under the
present assumption it reads:
\be
\label{T1s} T_1 = 4\dot{\e}\dot{A}_1 - 6\ddot{A}_1 + \mathcal{O}(\ddot{\e}) = \frac{6}{(u-u_0)^2} + \cdots + \mathcal{O}(\e) \, .
\ee
At leading order, the equation for $\e$ is therefore:
\be
\label{eqe}\ddot{\epsilon} + 4\frac{\dot{\e}}{u-u_0} + 6\frac{\e}{(u-u_0)^2} = 0 \, ,
\ee
whose solution is given by
\be
\label{esol}\epsilon(u) = \left(\frac{u_0-u}{\alpha}\right)^{-3/2} \sin\left(\frac{\sqrt{15}}{2}\log\left(\frac{u_0-u}{\alpha}\right) + \phi \right) \, ,
\ee
where $\alpha$ and $\phi$ are integration constants.
The solution displays two important properties:
\begin{itemize}
\item  It increases in amplitude as  $u_0-u \to 0$, which is an expected behavior in
  the IR. Eventually it diverges, as expected,   although this regime lies outside
  of our linear approximation, and it should only be taken qualitatively.
\item It oscillates an infinite number of times close to $u_0$. This
  is at the origin of the  Efimov spiral behavior  \cite{JK}.
\end{itemize}
From the above analysis, we conclude that the  singular background with
$A_1=A_2$ all the way to the IR (i.e. $\epsilon(u) \equiv 0$) has a
tachyonic instability, signaled by the unbounded  growth of the linear
perturbation $\epsilon$ in the IR. This instability points to the
existence of other solutions where $\epsilon \sim \mathcal{O}(1)$, which have a non-vanishing
vev for $A_1-A_2$. When this parameter is turned on,  the system  avoids the singular solution.

Now that we have identified the unstable mode at the origin of the
spiral, we can keep following the analysis of \cite{JK}. The relevant parameter that runs along the spiral is:
\be
\label{defs} s = \log\left(\frac{T_{2,0}\ell^2}{2}\right) = \log\left(\left(\frac{\alpha_{UV}}{\alpha_{IR}}\right)^2\right)
\ee
which is defined independently of $u_0$ as it should. The next step is
to connect the UV parameters
$(\mathcal{R}_2/\mathcal{R}_1,C/\mathcal{R}_1^2)$ with the IR
parameter $T_{2,0}\ell^2$. Note that we consider $C/\mathcal{R}_1^2$
instead of simply $C$ as the former is the $u_0$-independent quantity
that enters in the free energy \eqref{eq:Fnof}.

To see explicitly how the spiral behavior arises, we need to match
solutions \eqref{esol} on both edges of the domain of validity,
i.e. for $u_0-u \approx \alpha_{UV}$ and  $u_0-u \approx
\alpha_{IR}$.

We first consider  the UV regime, we know that $A_1-A_2$ has a
source and a vev term: the former is proportional to $\mathcal{R}_2 -
\mathcal{R}_1$, the latter is proportional to $C$ (see equation
(\ref{A1mA2sp})):
\be\label{A1A2UV}
A_1-A_2 \simeq \bar{A}_{1} - \bar{A}_2 + \left(\frac{\mathcal{R}_2}{\mathcal{R}_1}-1\right)e^{2u/\ell}\left(1
+ \ldots \right) + C e^{4u/\ell} \left(1 + \ldots\right), \qquad u \to -\infty \, ,
\ee
where $\bar{A}_{1} - \bar{A}_2 = \mathcal{O}\left(\frac{\mathcal{R}_2}{\mathcal{R}_1}-1\right)$. This regime  can be connected to the upper region ($u_0-u \approx
\alpha_{UV}$) of regime of validity of equation
(\ref{eqe}),  where the solution reads:
\begin{align}
\label{defeuv} \e_{UV} & = K_{R} \; \left(\frac{\mathcal{R}_2}{\mathcal{R}_1}-1\right)\left(\frac{u_0-u}{\alpha_{UV}}\right)^{-3/2} \sin\left(\frac{\sqrt{15}}{2}\ln\left(\frac{u_0-u}{\alpha_{UV}}\right) + \phi_{R} \right) \\
\nonumber & \hphantom{=} \ + K_{C} \; \frac{C}{\mathcal{R}_1^2}\left(\frac{u_0-u}{\alpha_{UV}}\right)^{-3/2} \sin\left(\frac{\sqrt{15}}{2}\ln\left(\frac{u_0-u}{\alpha_{UV}}\right) + \phi_{C} \right)
\end{align}
where $K_R$, $K_C$, $\phi_R$ and $\phi_C$ are some constants, which
are fixed by matching the solution in the UV to (\ref{A1A2UV}). Note that thus written, this expression for $\e_{UV}$ implies that $\alpha_{UV}$ is the length from which $\dot{\e}$ starts vanishing and should therefore be matched to its UV behavior \eqref{A1A2UV}. Because it is the same scale as the one for which $\dot{A}_1$  should be matched to its UV behavior (which is $\alpha_{UV}$ by definition), the presence of $\alpha_{UV}$ here is justified.

On the other hand, the solution when $u_0-u \approx \alpha_{IR}$ reads:
\be
\label{defeir} \e_{IR}  = K_{IR} \;
\left(\frac{u_0-u}{\alpha_{IR}}\right)^{-3/2}
\sin\left(\frac{\sqrt{15}}{2}\ln\left(\frac{u_0-u}{\alpha_{IR}}\right)
  + \phi_{IR} \right),
\ee
where $K_{IR}$ is another constant. The presence of $\alpha_{IR}$ is justified here by the fact that it is the scale in the IR for which $\e$ should reach $\mathcal{O}(1)$.
The two expressions (\ref{defeuv}) and
(\ref{defeir}) are valid in the same region, therefore we can match the
coefficients: this gives  an expression for $\mathcal{R}_2/\mathcal{R}_1$ and $C/\mathcal{R}_1^2$ as functions of $s$ (defined in \eqref{defs}):
\begin{eqnarray}
\label{R2sR1vs} \frac{\mathcal{R}_2}{\mathcal{R}_1}-1 & =& \frac{K_{IR}}{K_R} \; \frac{\sin\left(\phi_{IR}-\phi_{C} + \frac{\sqrt{15}}{4}s\right)}{\sin(\phi_R - \phi_C)}\; \ex^{-3/4\;s} \, , \\
\label{Cvs} \frac{C}{\mathcal{R}_1^2} & = & \frac{K_{IR}}{K_C} \; \frac{\sin\left(\phi_{IR}-\phi_{R} + \frac{\sqrt{15}}{4}s\right)}{\sin(\phi_C - \phi_R)}\;  \ex^{-3/4\;s} \, .
\end{eqnarray}
These expressions reproduce the spiral behavior: as $s$ increases, the
IR radius of the sphere 2 becomes smaller and smaller, and it reaches
zero in the singular limit ($s\to \infty)$. At the same time, the vev
parameter $C$ decreases and  the UV
ratio ${\mathcal R}_2/{\mathcal R}_1$ oscillates, crossing unity an infinite number
of times. Therefore,  if we consider the symmetric UV boundary
condition ${\mathcal R}_2/{\mathcal R}_1 =1 $, we find an infinite
number of solutions, for the values of $s$ which correspond to the
vanishing of the
$sin$  function in the numerator of equation (\ref{R2sR1vs}) . 

Note that because the spiral turns clockwise, $C/\mathcal{R}_1^2$ is ahead of $\mathcal{R}_2/\mathcal{R}_1-1$, which means that $\sin(\phi_R-\phi_C)<0$. \\
The fit corresponding to those solutions is plotted in Figure \ref{fig:RCvR_main}.

\subsection{The dominant vacuum}

The degeneracy which originates from the Efimov spiral close to the
singular point indicates that for a  boundary CFT characterized by a
given value of the ratio $\mathcal{R}_2/\mathcal{R}_1$, there are
several possible values for the vev $C$, that is, several possible
vacuum states (saddle points). The number of possible vacua increases when the ratio
tends to 1, asymptoting  to infinity for $\mathcal{R}_2 = \mathcal{R}_1$. For
each fixed value of $\mathcal{R}_2/\mathcal{R}_1$ the
dominant vacuum is the one with the lowest free energy (on-shell action).

\begin{figure}[h!]
\centering
\begin{overpic}
[width=0.65\textwidth]{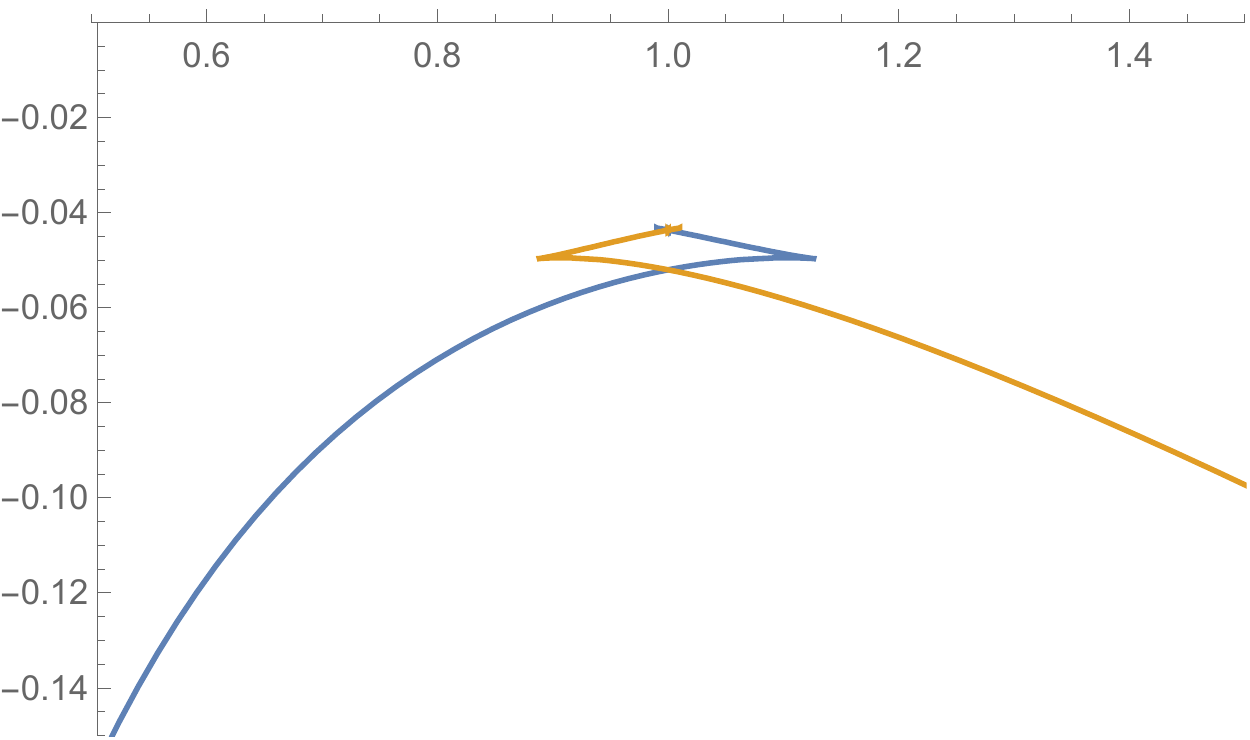}
\put(104,56){${\mathcal{R}_2\over \mathcal{R}_1}$}
\put(4,58){$\bar{\mathcal{F}}$}
\end{overpic}
\caption{Finite part $\bar{\mathcal{F}}$ of the the free energy \protect\eqref{expUVFu} as a function of the ratio $\mathcal{R}_2/\mathcal{R}_1$, in both the case where the sphere 1 shrinks (blue) and the case where the sphere 2 shrinks (orange). We normalize by the overall volume factor $32\pi^2M_p^3\ell^3$. The point where the two curves first cross at $\mathcal{R}_1 = \mathcal{R}_2$ corresponds to a bifurcation, where the sphere that shrinks is exchanged and the vev changes sign.}
\label{fig:FU}
\end{figure}

Numerical evaluation of the free energy  of the solution using equation
(\ref{DFu}) shows that the dominant vacuum is the first one  reached (for
a given  $\mathcal{R}_2/\mathcal{R}_1$) when moving  towards the
center of the spiral. This is displayed in Figure \ref{fig:FU}, which
shows the finite part of the free energy (i.e. the term
$\overline{\mathcal{F}}$ in equation (\ref{FEexp})) as a function of  $\mathcal{R}_2/\mathcal{R}_1$.  As a result, the Efimov degeneracy is broken and there is only one possible vacuum for every value of $\mathcal{R}_2/\mathcal{R}_1$, except at the critical point $\mathcal{R}_2 = \mathcal{R}_1$ where both $C$ and $-C$ are possible, corresponding to the spontaneous breaking of the $\mathbb{Z}^2$-symmetry that exchanges the two spheres. The system therefore exhibits a bifurcation at this point, where the vev $C$ changes sign and the sphere that shrinks in the IR is exchanged.

\begin{figure}[h!]
\begin{center}
\begin{overpic}[scale=0.80]{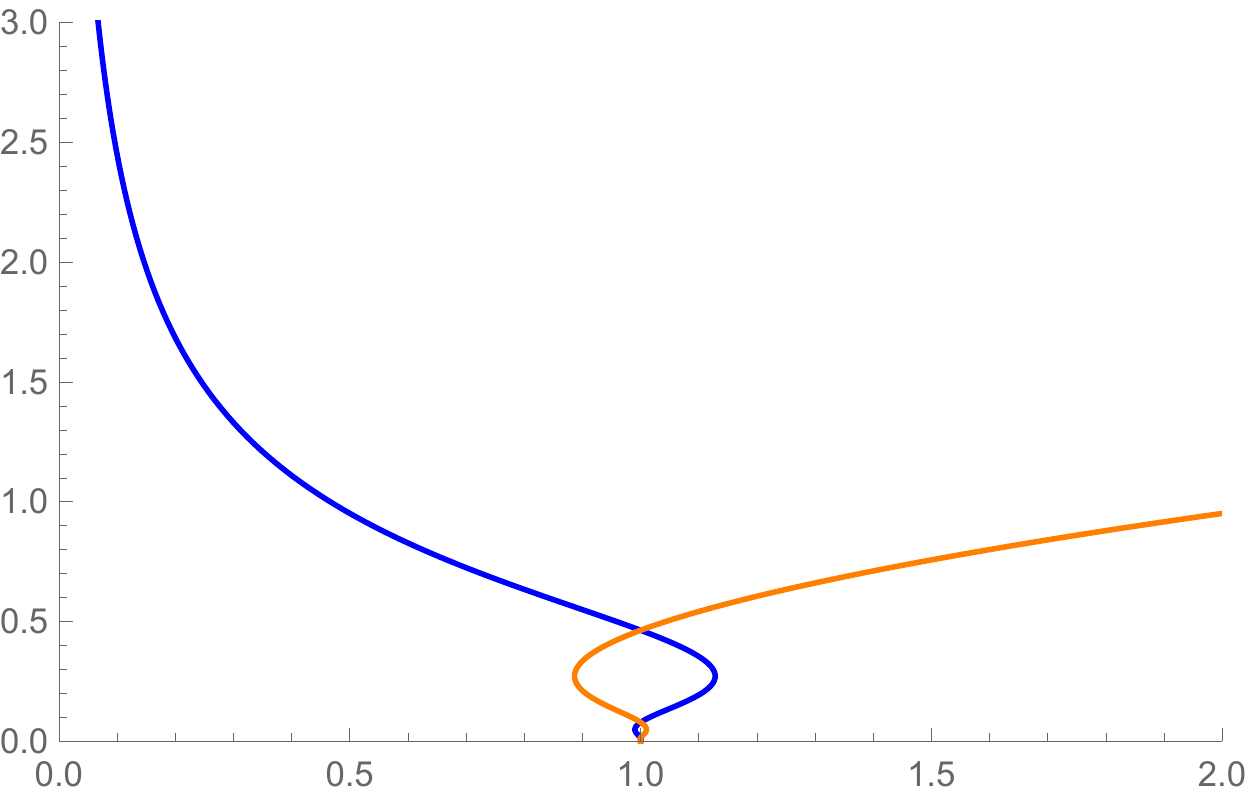}
\put(101,4){${\mathcal{R}_2\over \mathcal{R}_1}$}
\put(-2,66){$\alpha_{IR}/\ell$}
\end{overpic}
\end{center}
\caption{The IR endpoint radius $\alpha_{IR}$ of the  sphere which
  stays finite at the end-point of the flow, as a function of the curvature ratio
  $\mathcal{R}_2/\mathcal{R}_1$. The blue curve corresponds to
  $\alpha^{IR}_2$ in the
  case where  sphere 1 shrinks to zero size in the IR; the yellow
  curve to  $\alpha^{IR}_1$ in the  solution where the sphere  2
  shrinks to zero. The transition which exchanges the two spheres
  corresponds to the point where the curves cross at the
  largest value for the radius.
    }
\label{fig:qavR}
\end{figure}

Another point that deserves attention is the behavior of the radius of
the sphere that does not shrink to zero size in the IR (the sphere 2
here). To be more specific,   we computed numerically the  radius
of the non-vanishing sphere at the endpoint,  given in equation
(\ref{alphair}), as a function of $R_2/R_1$. The result is displayed  in figure \ref{fig:qavR}.

\section{Holographic RG-flows on S$^2\times$ S$^2$}
\label{sec:rgflow}

We now move to  consider RG-flow geometries, where the scalar field
is not constant. They originate in the UV from a maximum of the
potential (at $\f=0$)  and end regularly   when one of the spheres
shrinks to zero size. At this point,  the scalar reaches a value $\f=\f_0$
which lies in the region between this maximum and (typically) the nearest
minimum.

We consider solutions where  $\varphi$ changes monotonically along the flow
from UV to IR. Therefore one can use $\f$ as a coordinate along the
flow. We still assume that sphere 1 shrinks in the IR (i.e.
$A_1(\varphi) \xrightarrow[\varphi \to \varphi_0]{} -\infty$ which
implies that $A_2$ remains finite to have regularity in the interior
according to \ref{CurvInv}).

These solutions are generic,  as they arise for generic
potentials as long as they possess at least one maximum and one
minimum. The simplest such potential  is  the following quadratic-quartic function:

\begin{equation}
\label{V24} V(\varphi) = -\frac{12}{\ell^2} - \frac{m^2}{2}\varphi^2 + \lambda \varphi^4
\end{equation}

This potential has one maximum at $\varphi_{max}=0$. For purposes of illustration we choose $\lambda = m^2/4$ so that the minima occur at
$\varphi_{min} = \pm 1$. The qualitative features of the solutions do not depend on this choice.

 We then proceed by solving
\eqref{eq:EOM11bis}-\eqref{eq:EOM14bis} numerically for
$W_1(\varphi),W_2(\f)$ and $S(\varphi)$. Like in the case with no
scalar field, to impose regularity  we specify boundary conditions for
$W_1,W_2$ and $S$ at the IR end point $\varphi_0$, as prescribed by
equations (\ref{W1}-\ref{S}), with $T_{2,0}$ as a free parameter.  Given the symmetry of the setup, we restrict our attention to flows that end in the region $\varphi_0 \in [0,1]$.

\subsection{General properties of the RG flow}

We first discuss general properties of the flow solution of the
equations of motion \eqref{eq:EOM6bis}-\eqref{eq:EOM9bis} obtained
numerically.  The  IR parameters one can vary  are $\f_0$ and
$T_{2,0}\ell^2$. These determine all other terms in the IR expansion,
as well as all the dimensionless UV data. More specifically,  in the vicinity of the UV fixed point the solutions are described
by the family of solutions collectively denoted by $W_1^-$ and $W_2^-$
in section \ref{sec:nb}. These solutions depend on the four independent
dimensionless parameters $\mathcal{R}_1$, $\mathcal{R}_2$, $C_1$ and
$C_2$. There is one more parameter than  in  the CFT case (where $C_1
+ C_2 = 0$),  corresponding to the fact that the vev of the scalar operator is now a free parameter of the UV theory.

Below, we analyze separately the dependence on each of the two IR parameters $\f_0$ and
$T_{2,0}\ell^2$.

\subsubsection*{Fixed $T_{2,0}\ell^2$}
\begin{figure}[h!]
\begin{center}
\includegraphics[scale=1]{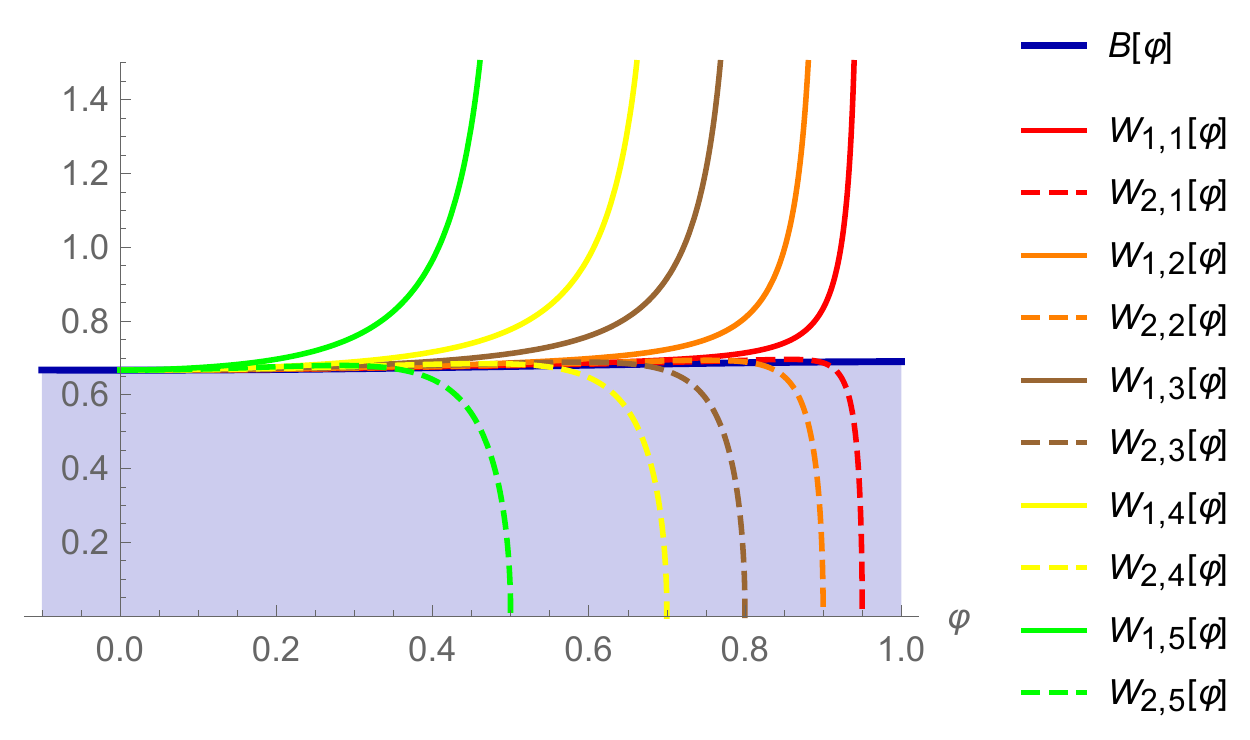}
\end{center}
\caption{Solutions $W_1(\varphi)$ and $W_2(\varphi)$ for the potential \protect\eqref{V24} with $\Delta_- = 1.2$ in the case where the sphere 1 shrinks. The five solutions $(W_{1,i},W_{2,i})$ with $i = 1,\dots ,5$ differ in the value of their IR endpoint $\varphi_0$ for $T_{2,0}\ell^2 = 4.5$. The critical curve is defined as $B(\varphi) \equiv \sqrt{-V(\varphi)/3}$. In the case of $S^4$ \protect\cite{C}, the superpotential $W(\varphi)$ cannot enter the area below the critical curve, which is depicted as the shaded region.}
\label{fig:WT05_m}
\end{figure}
In Figure \ref{fig:WT05_m} we exhibit solutions for the
superpotentials $W_1(\varphi)$ and $W_2(\varphi)$ corresponding to
generic RG flows for a bulk potential given by \eqref{V24} with
$T_{2,0}\ell^2$ fixed, and for different values of the endpoint
$\f_0$.  To be specific, we have set $\Delta_- = 1.2$ and $T_{2,0}\ell^2=4.5$ but our observations hold more generally.
\begin{itemize}
\item The main result is that for every value of $\varphi_0$ between $\varphi_{max}=0$ and $\varphi_{min} = 1$ there exists a unique solution to the superpotential equations \eqref{eq:EOM11bis}-\eqref{eq:EOM14bis} (remember that $T_{2,0}\ell^2$ has been fixed) corresponding to an RG flow originating from the UV fixed point at $\varphi_{max}=0$ and ending at $\varphi_0$.
\item Note that whereas $W_1(\varphi)$ diverges like $(\varphi_0 -
  \varphi)^{-1/2}$ when approaching the IR end point $\varphi_0$,
  $W_2(\varphi) \to  0$. This is in agreement  with the analytical
  results found in Section \ref{sec:IRbc}.
\item  The counting of parameters is as expected: picking a solution with the regular IR behavior for a RG flow
  fixes two combinations of the four UV parameters;  the remaining freedom
  is then equivalent to the choice of IR end point $\varphi_0$,
  together with the choice of the radius in the interior of the sphere
  that does not shrink at $\varphi_0$ (here the sphere 2), which is
  given by $T_{2,0}$. Therefore, regularity plus a choice of  $\varphi_0$,  $T_{2,0}$
  uniquely determines the solution.
\item The solution is then matched to the UV asymptotics
  (\ref{eq:W1gensol}-\ref{eq:T2gensol})  to extract the UV
  quantities ${\mathcal R}_{1,2}$ and $C_{1,2}$. The  two IR
  parameters $\varphi_0, T_{2,0}$ can then be traded for the two
  independent  UV parameters ${\mathcal R}_{1,2}$, and the vev
  parameters $C_{1,2}$ can then be expressed as functions of
  ${\mathcal R}_{1,2}$.

\end{itemize}

\subsubsection*{Fixed $\varphi_0$}
\begin{figure}[h!]
\begin{center}
\includegraphics[scale=1]{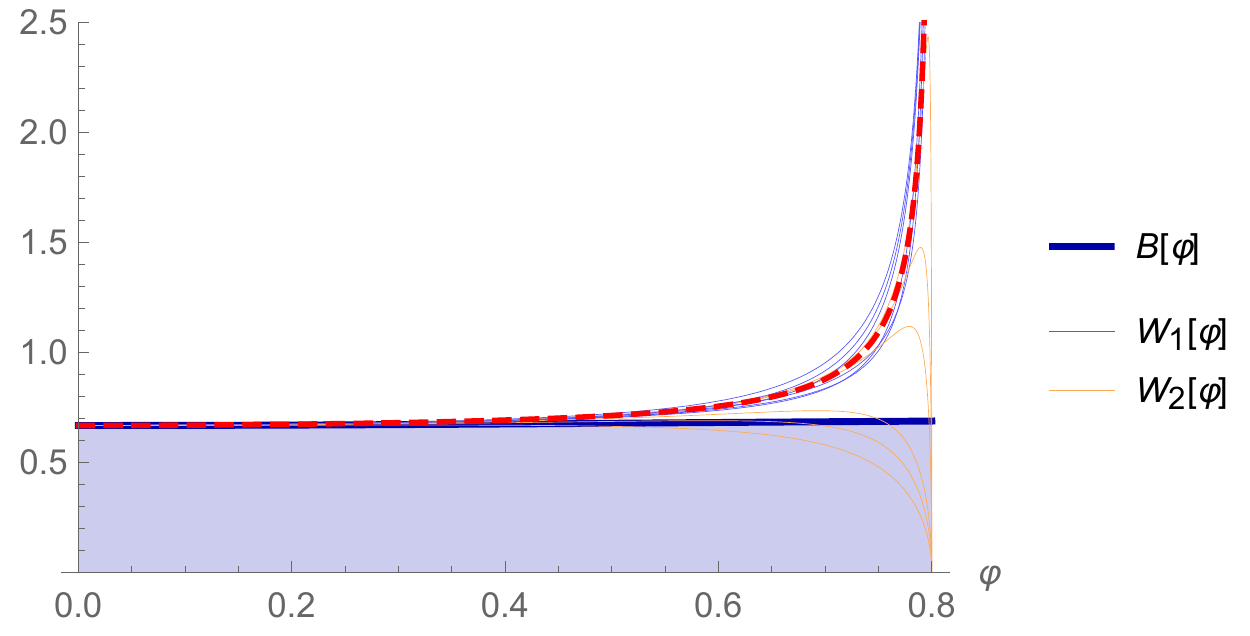}
\end{center}
\caption{Solutions $W_1(\varphi)$ and $W_2(\varphi)$ for the potential \protect\eqref{V24} with $\Delta_- = 1.2$ in the case where the sphere 1 shrinks. The six solutions differ in the value of $T_{2,0}$ whereas $\varphi_0=0.8$ is fixed. From the lowest curve for $W_2$ to the highest one, $T_{2,0}\ell^2$ increases, taking successively the values: $0.9,4.5,13,5,90,270$. The red dashed curve is the one towards which $W_2$ tends when $T_{2,0}\ell^2 \to +\infty$ on every interval in $[0,\varphi_0[$. The critical curve is defined as $B(\varphi) \equiv \sqrt{-V(\varphi)/3}$. In the case of $S^4$ \protect\cite{C}, the superpotential $W(\varphi)$ cannot enter the area below the critical curve, which is depicted as the shaded region.}
\label{fig:Wphi008_m}
\end{figure}

\begin{figure}[h!]
\begin{center}
\includegraphics[scale=0.55]{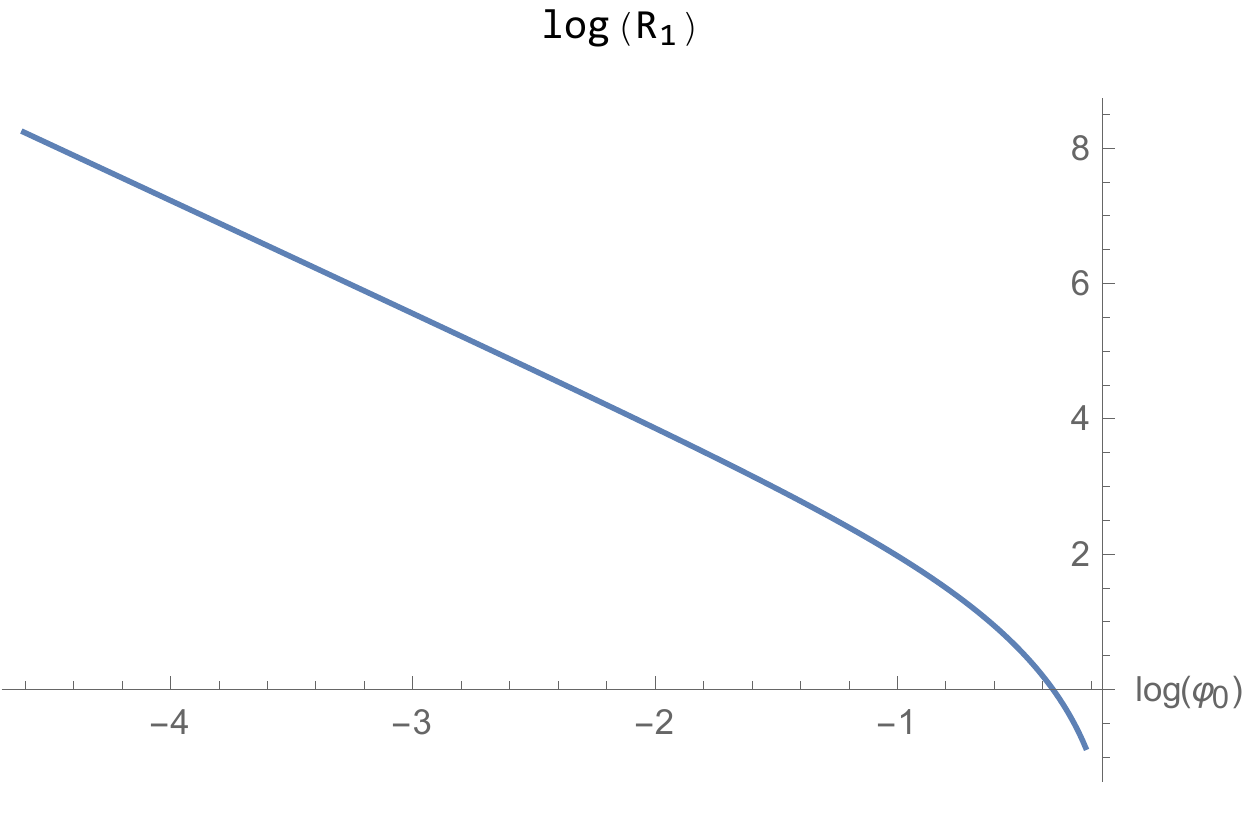} \hfill \includegraphics[scale=0.55]{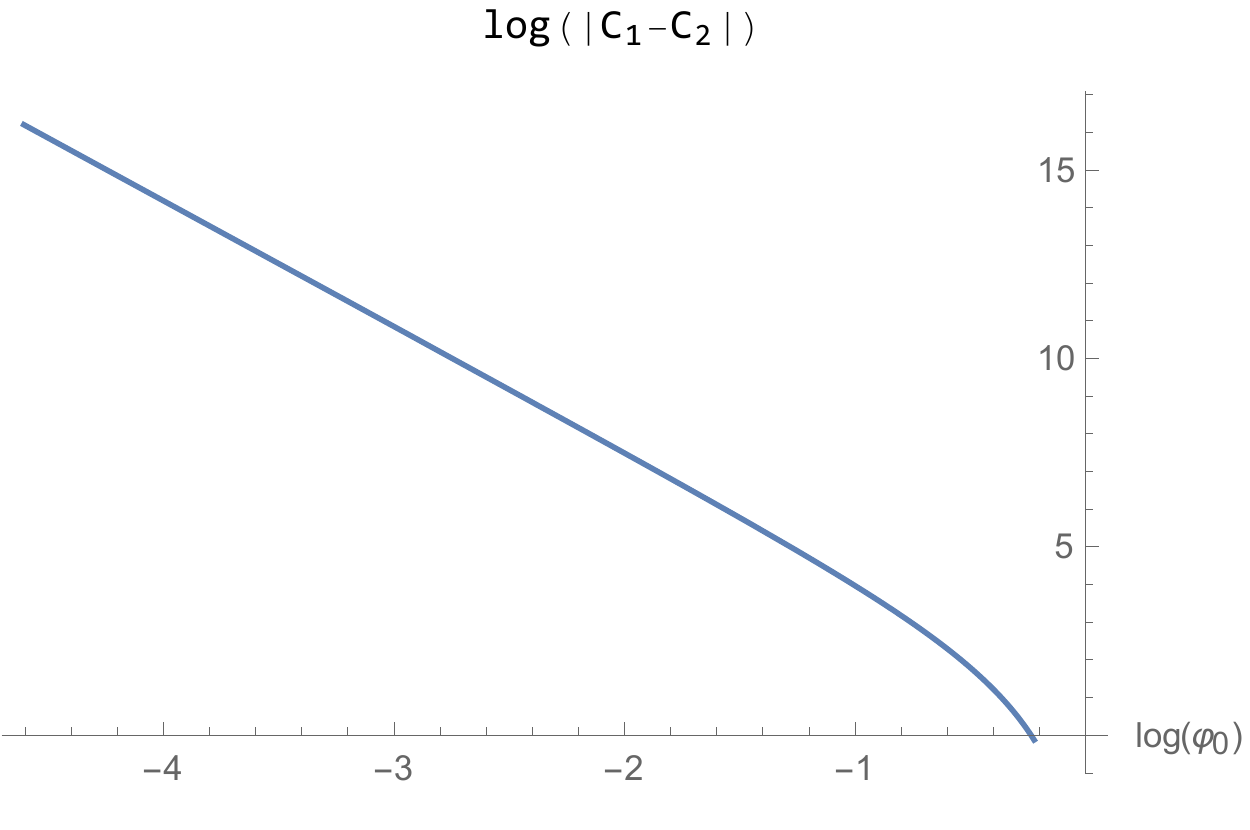}
\end{center}
\caption{$\mathcal{R}_1(\f_0)$ and $C_1(\f_0)-C_2(\f_0)$ in the case where the sphere 1 shrinks and for $T_{2,0} = 0$. In this situation the space 2 is flat ($\mathcal{R}_2=0$). }
\label{fig:RCmf}
\end{figure}

We now keep $\varphi_0$ fixed and let  $T_{2,0}$ vary. In Figure \ref{fig:Wphi008_m} we exhibit solutions for the superpotentials $W_1(\varphi)$ and $W_2(\varphi)$ corresponding to generic RG flows for a bulk potential given by \eqref{V24} when $\varphi_0$ is fixed. To be specific we have set $\Delta_- = 1.2$ and $\varphi_0 = 0.8$ but our observations  hold more generally.
\begin{itemize}
\item Because $\f_0$ is fixed, there is a unique solution for every $T_{2,0}\ell^2$.
\item As $T_{2,0}$ grows, $W_2$ goes to a value ever higher in the
  interior before diving to 0 to have regularity. More precisely we
  observe that both $W_1$ and $W_2$  approach a single curve (the
  red-dashed curve in Figure \ref{fig:Wphi008_m})  as
  $T_{2,0}\ell^2\to \infty$. This curve corresponds to the singular
  solution, where both $W_1$ and $W_2$ diverge in the IR and both
  spheres shrink to zero size.
\end{itemize}
\paragraph*{UV parameters.}Given a numerical solution, we can extract
the corresponding values of $\mathcal{R}_1$, $\mathcal{R}_2$, $C_1$
and $C_2$ explicitly by fitting the UV region with the asymptotics
\eqref{eq:W1gensol}-\eqref{eq:W2gensol}. Figure \ref{fig:RCmf}
represents $\mathcal{R}_1$ and $C_1-C_2$ as functions of $\f_0$ when
$T_{2,0} = 0$. In this section,  $C_1-C_2$ is defined as the full vev term for
$A_1 - A_2$, which corresponds to make the following redefinition in equation (\ref{eq:A1mA2})
\be \label{C1C2redef}
C_1 \to C_1 + \frac{\mathcal{R}_1^2-\mathcal{R}_2^2}{24\D_{-}}
\log(|\f_-| \ell^{\Delta_-}),
\quad C_2 \to C_2 -
\frac{\mathcal{R}_1^2-\mathcal{R}_2^2}{24\D_{-}}\log(|\f_-| \ell^{\Delta_-}).
\ee
This redefinition does not affect $C_1+C_2$.

We observe that for small $\f_0$, $\mathcal{R}_1\propto
\f_0^{-2/\Delta_-}$ and $C_1-C_2 \propto \f_0^{-4/\Delta_-}$. These
properties can once again be deduced from the scaling properties of
the equations of motion \eqref{eq:EOM11bis} - \eqref{eq:EOM14bis}
under a shift of $u\to u+ \delta u$: under such shift, in the UV we have:
Indeed, $\f(u) \sim \f_- \ell^{\Delta_-} \ex^{\Delta_-(\delta
  u/\ell)}$ in the UV. This implies that, if $u_0$ is the IR
coordinate of the endpoint, then
\be
\label{eqf0}\f_0 \propto \f_- \ell^{\Delta_-} \ex^{\Delta_{-
}u_0/\ell  }
\ee
The above  equation gives either the dependence of $\f_-$ on $u_0$ when $\f_0$ is
fixed, or the dependence of $\f_0$ when $\f_-$ is fixed. It is
apparent from $\f_0 \xrightarrow[u_0\to +\infty]{} 1$ that
\eqref{eqf0} is only valid for $\f_0 \ll 1$.

\begin{figure}[h!]
\begin{center}
\includegraphics[scale=0.8]{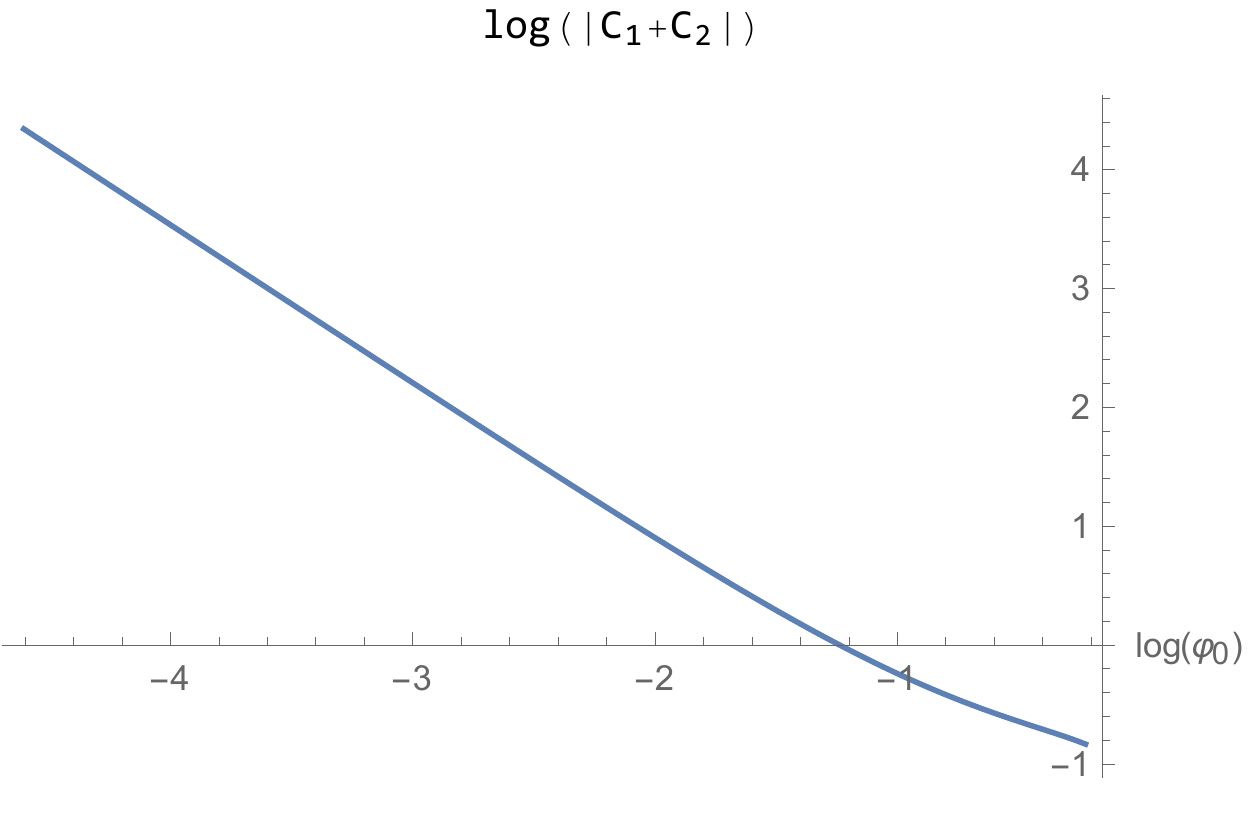}
\end{center}
\caption{$C_1(\f_0)+C_2(\f_0)$ in the case where the sphere 1 shrinks and for $T_{2,0} = 0$. In this situation the space 2 is flat ($\mathcal{R}_2=0$). }
\label{fig:Cpf}
\end{figure}

The scalings of the UV parameters in terms of $\f_0$ when $\f_0 \ll 1$
can then be deduced  from the scaling properties of the EoMs
\eqref{eq:EOM6bis}-\eqref{eq:EOM9bis} under the translation $u \to
u-u_0$. In particular, under such a translation, the leading terms in
the near boundary expansion \eqref{eq:W1gensol}-\eqref{eq:Sgensol}
should have the same scaling.
It implies in particular that the terms at order 1 in curvature in the expansion of $W_1$ and $W_2$ should be invariant under such a translation, which gives the expected scaling in $\f_0$ for $\mathcal{R}_1$ and $\mathcal{R}_2$ when $\f_0\ll 1$ (in which case \eqref{eqf0} can be used to relate $\f_0$ with $u_0$). Knowing this, the expansion of $W_1-W_2$ gives the appropriate scaling for $C_1-C_2$, and that of $S$ \eqref{eq:Sgensol} gives the $\f_0$-dependence for the scalar vev $C_1+C_2$ (which is not the same as $C_1-C_2$):
\be
\label{Cf0} C_1+C_2 \propto \f_0^{\frac{\Delta_- - \Delta_+}{\Delta_-}} \sp \f_0 \ll 1
\ee
Figure \ref{fig:Cpf} represents $C_1+C_2$ as a function of $\f_0$ when $T_{2,0} = 0$.

Note that whereas $\mathcal{R}_1$, $\mathcal{R}_2$ and $C_1-C_2$ tend to 0 when $\f_0 \to 1$ (flat limit), the scalar vev tends to a finite value. This is again coherent with what was found in the $S^4$ case \cite{C}.

\subsection{Efimov spiral and dominant vacuum}
As in the conformal case, in the $Z_2$ symmetric limit ${\mathcal
  R}_2/{\mathcal R}_1  \to 1$ we encounter again a discrete Efimov scaling
and an infinite number of solutions.

\begin{figure}[h!]
\begin{center}
\includegraphics[scale=0.8]{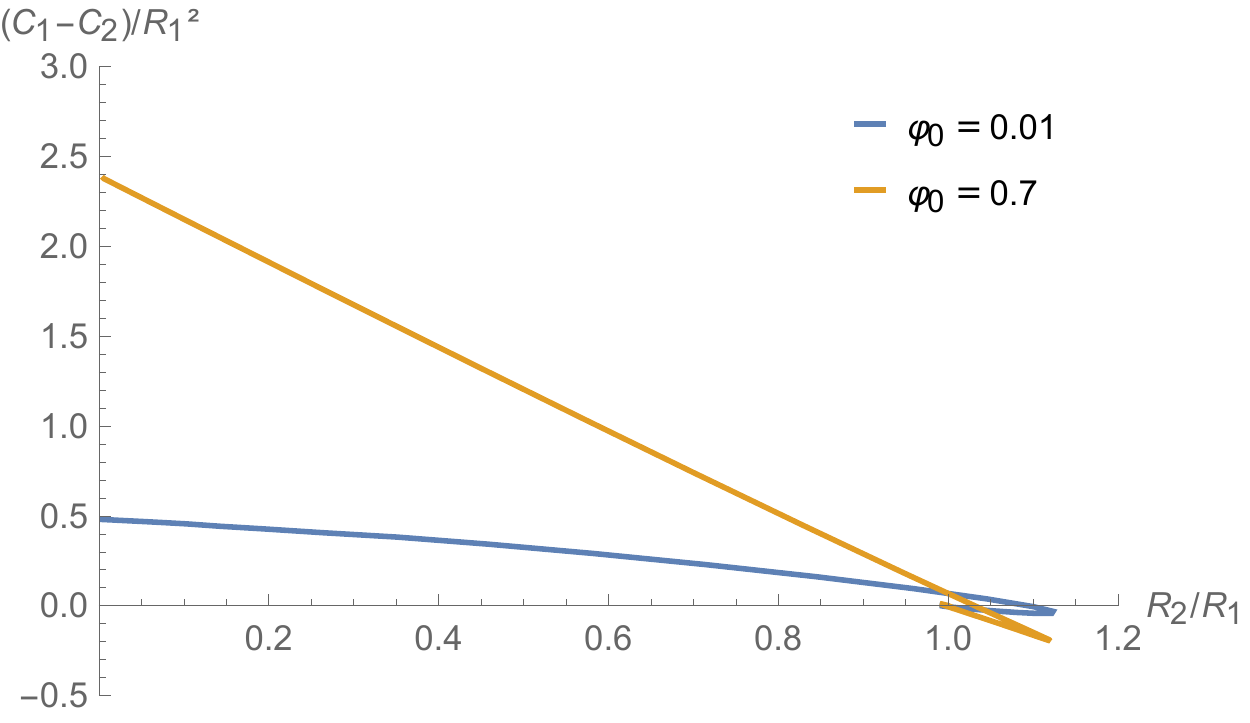}
\end{center}
\caption{The Efimov spiral in the plane
  $(\mathcal{R}_2/\mathcal{R}_1,(C_1-C_2)/\mathcal{R}_1^2)$, in the
  case where the sphere 1 shrinks and for $\f_0 = 0.01$ and $\f_0 =
  0.7$. Although it is difficult to see a ``spiral" in the plot above, it can be inferred as it is similar to the one plotted on the top of figure  \protect\ref{fig:RCf01_m}.
   For values of $\f_0$ intermediate between $0.01$ and $0.8$,
  there is a continuous family of spirals that fills the space between
  the two spirals that are plotted.}
\label{fig:RCf_m}
\end{figure}

Figure \ref{fig:RCf_m} shows the Efimov spiral in the plane
$(\mathcal{R}_2/\mathcal{R}_1,(C_1-C_2)/\mathcal{R}_1^2)$. The
behavior is essentially the same as what was observed without a scalar
field, with the notable property that the amplitudes $(K_C,K_R)$ and
the phases $(\f_C,\f_R)$ defined in equations  \eqref{R2sR1vs} and \eqref{Cvs} are now functions of $\f_0$: there is a continuous family of spirals parametrized by $\f_0$.

\begin{figure}[h!]
\begin{center}
\includegraphics[width=11cm]{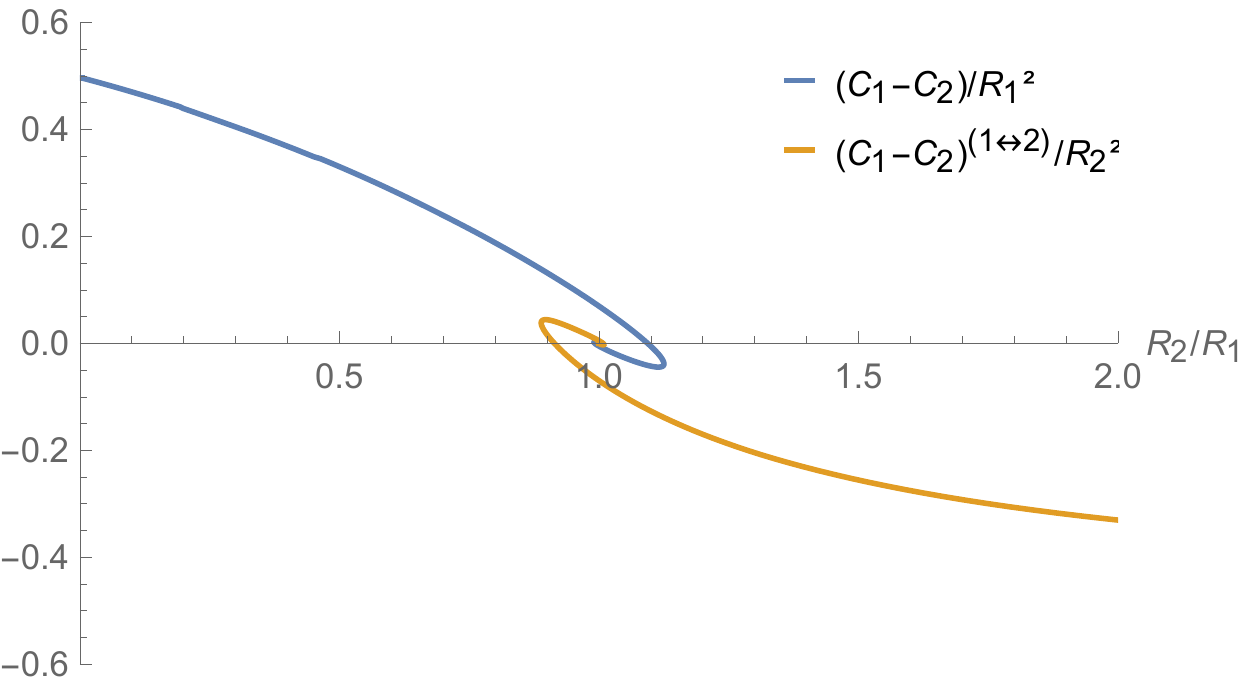}
\vskip 1cm
 \includegraphics[width=11cm]{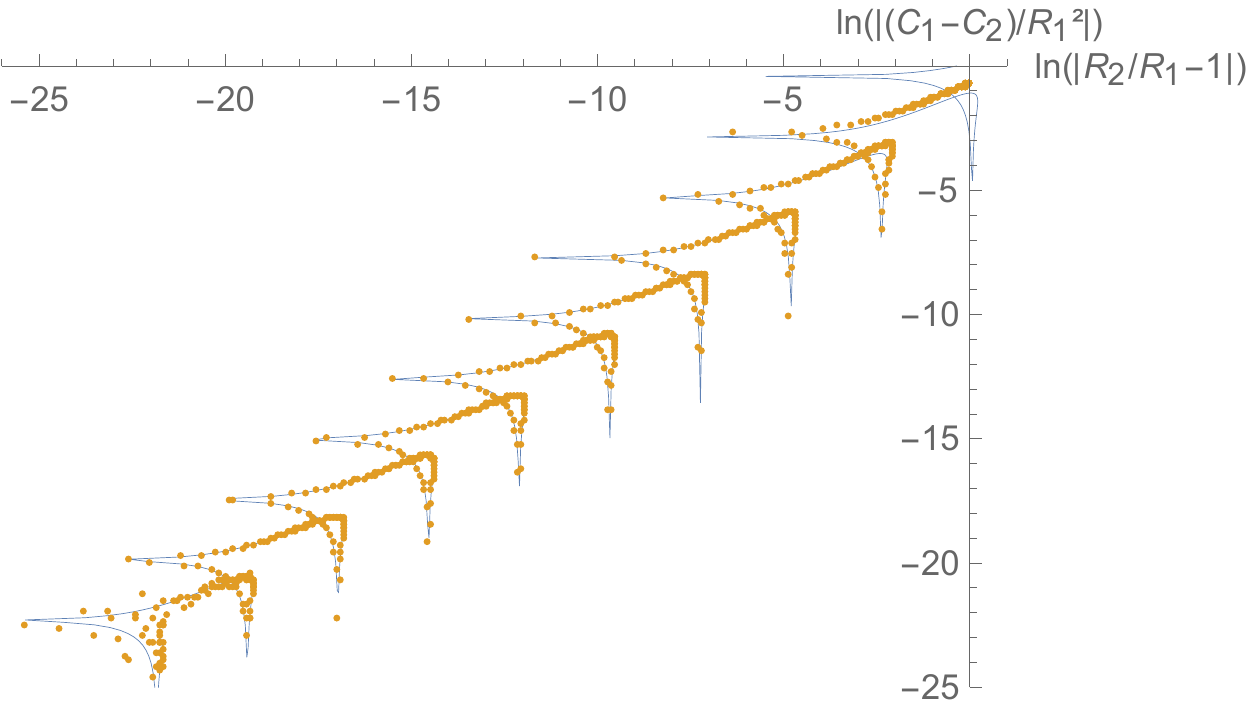}
\end{center}
\caption{{\em Top:}  $(C_1-C_2)/\mathcal{R}_1^2$ in both the case
  where the sphere 1 shrinks (blue) and the case where the sphere 2
  shrinks (orange) and for $\f_0 = 0.1$. {\em Bottom:} The same plot
  in the case where 1 shrinks, where we represent the logarithm of the
  distance to the center of the spiral with coordinates $(1,0)$ for
  each quantity. The orange dots are given by numerical computation
  while the blue curve is the fit found in
  \protect\eqref{R2sR1vs}-\protect\eqref{Cvs}. }
\label{fig:RCf01_m}
\end{figure}

Figure \ref{fig:RCf01_m} shows the spiral for the a fixed endpoint
value, $\f_0 = 0.1$.
With a scalar field, the equation for $\e \equiv A_1-A_2$ \eqref{eqe} (given the same conditions) at leading order is the same:
\be
\label{eqebis}\ddot{\epsilon} + 4\frac{\dot{\e}}{u-u_0} + 6\frac{\e}{(u-u_0)^2} = 0
\ee
The formulae that describe the spiral \eqref{R2sR1vs}-\eqref{Cvs} are
therefore exactly the same as without a scalar field, where the
amplitudes $(K_C,K_R)$ and the phases $(\f_C,\f_R)$ are functions of
$\f_0$.

We use equation \eqref{eq:Fbisbis} to compare the free energy of two vacua with
the same ratio $\mathcal{R}_2/\mathcal{R}_1$ and value of $\f_0$, but
with distinct vevs $(C_1-C_2)/\mathcal{R}_1^2$. The conclusion
is the same as the one reached in section 7 without the scalar field: the stable vacuum corresponds
to the first point that is reached by the spiral in the
$(\mathcal{R}_2/\mathcal{R}_1,(C_1-C_2)/\mathcal{R}_1^2)$-plane. There is therefore a bifurcation at the point $\mathcal{R}_1 = \mathcal{R}_2$, where the sphere that shrinks is exchanged and $C_1-C_2$ changes sign.

\begin{figure}[h!]
\centering
\begin{overpic}
[width=0.65\textwidth]{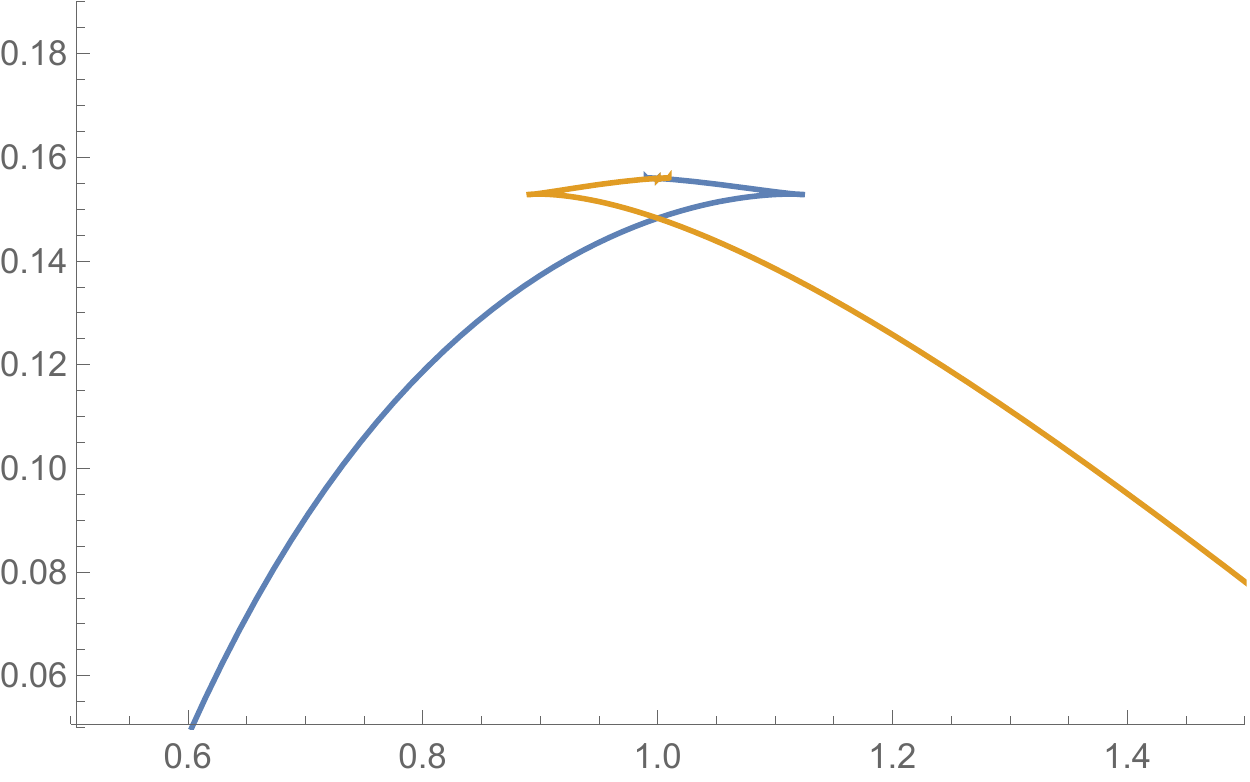}
\put(104,4){${\mathcal{R}_2\over \mathcal{R}_1}$}
\put(4,63){$\bar{\mathcal{F}}$}
\end{overpic}
\caption{Finite part $\bar{\mathcal{F}}$ of the the free energy \protect\eqref{expUVFu} as a function of the ratio $\mathcal{R}_2/\mathcal{R}_1$, on the line of the $(\mathcal{R}_1,\mathcal{R}_2)-$plane where the radius of the sphere which shrinks to zero size in the IR is kept fixed. We represent both the case where the sphere 1 shrinks (blue) and the case where the sphere 2 shrinks (orange). We normalize by the overall volume factor $32\pi^2M_p^3\ell^3$. The point where the two curves first cross at $\mathcal{R}_1 = \mathcal{R}_2$ corresponds to a bifurcation, where the sphere that shrinks is exchanged and the asymmetric vev changes sign.}
\label{fig:FUf}
\end{figure}



\section*{Acknowledgements}\label{ACKNOWL}
\addcontentsline{toc}{section}{Acknowledgements}

We would like to thank A. Golubtsova, Y. Hamada,  L. Witkowski  for
discussions.  We especially thank Junkang Li and Anastasia Golubtsova,  who contributed to  the early stages of this project.

\noindent This work was supported in part  by the Advanced ERC grant SM-grav, No 669288.

\appendix
\renewcommand{\theequation}{\thesection.\arabic{equation}}
\addcontentsline{toc}{section}{Appendix\label{app}}
\section*{Appendix}

\section{Matching to known cases\label{A}}

In this appendix we show that the general equations (\ref{eq:EOM1bis})-(\ref{eq:EOM3bis}) match known special cases.

\begin{enumerate}

\item  When all $d_k=1$ and $A_i=A_j=A$, this is the same as the flat slice case.
We do recover the equations of motion for the flat slicing ansatz:
\begin{align}
\label{eq1} 2(d-1) \ddot{A} + \dot{\f}^2 & =0 \, , \\
\label{eq2} d(d-1) \dot{A}^2 - \frac{1}{2} \dot{\f}^2 + V & =0 \, , \\
\label{eq3} \ddot{\f} +d \dot{A} \dot{\f} - V' & = 0 \, .
\end{align}

\item  When all $d_k=1$, and $A_1$ is distinct from all others $A_{j\not=1}$ which are equal, the metric is
 \be
 ds^2=du^2+e^{2A_1}dt^2+e^{2A_2}d\vec x^2
 \label{12}\ee
 By a change of coordinates, the metric can be put in the black hole form
\be
ds^2={d\tilde u^2\over f}+e^{2A}(fdt^2+d\vec x^2)
\label{13}\ee
We have
\be
A=A_2\sp f=e^{2(A_1-A_2)}\sp d\tilde u=e^{A_1-A_2}du
\label{14}\ee
The equations for (\eqref{13}) are
\be
	2(d-1)\ddot{A}(u)+\dot{\f}^2(u)=0,
\label{15}\ee
\be
	\ddot{f}(u)+d\dot{f}(u)\dot{A}(u)=0\Rightarrow \dot f=C~e^{-dA}
\label{16}\ee
	\be (d-1)\dot{A}(u)\dot{f}(u)+f(u)\left[d(d-1)\dot{A}^2(u)-\frac{\dot{\f}^2}{2}\right]+V(\phi)=0.
\label{17}\ee
\be
\ddot\f(u)+\left(d\dot A(u)+{\dot f(u)\over f(u)}\right)\dot\f(u)-{V'\over f(u)}=0
\label{18}\ee

 From (\eqref{eq:EOM1bis})-(\eqref{eq:EOM3bis}) with $d_k=1$,  $A_1\neq A_j$ with $j=2,\ldots,d$ we obtain the following equations
\begin{align}
\ddot{A}_{2} -  \dot{A}_{2}\left(\dot{A}_{1} - \dot{A}_{2}\right) + \frac{\dot{\varphi}^{2}}{2(d-1)} =0,\\
2((d-2)\ddot{A}_{2} - \dot{A}_{1}\dot{A}_{2} + \ddot{A}_{1} +\dot{A}^{2}_{1}) +\dot{\varphi}^{2} =0,\\
2(d-1)\dot{A}_{1}\dot{A}_{2} + (d-1)(d-2)\dot{A}^{2}_{2} - \frac{1}{2}\dot{\varphi}^{2} + V =0,\\
\ddot{\varphi}  +\dot{A}_{1}\dot{\varphi}  +  (d-1)\dot{A}_{2}\dot{\varphi} - V' =0.
\end{align}
{Implementing the redefinitions in  (\eqref{14}) and converting  to  derivatives with respect to $\tilde{u}$, we obtain eqs. (\eqref{15})-(\eqref{18}).}

\item  When $d_k=d$ we have a single scale factor and this is case analyzed in \cite{C}.
The equations match with those derived there.

\item When $d_1=1$ and $d_2=d-1$ we have the $S^1\times S^{d-1}$ slice. The solution with constant potential should be global AdS$_{d+1}$.\\
{For this case we  have the following equations of motion
 \begin{align}
& (d-1)(d-2)\dot{A}^{2}_{2} + 2 (d-1)\dot{A}_{1}\dot{A}_{2}  - e^{-2A_{2}}R^{\zeta^{2}}  - \frac{1}{2}\dot{\varphi}^{2} + V = 0,\\
& 2(d-1)\left(\ddot{A}_{2} - \dot{A}_{2}\left(\dot{A}_{1}-\dot{A}_{2}\right)\right) + \dot{\varphi}^{2} = 0,\\
& 2(\ddot{A}_{1}  + (d-2)\ddot{A}_{2} + \dot{A}_{1}(\dot{A}_{1}-\dot{A}_{2})+ \frac{1}{d-1} e^{-2A_{2}}R^{\zeta^{2}}) + \dot{\varphi}^{2}= 0,\\
&\ddot{\varphi} + \dot{A}_{1}\dot{\varphi} + (d-1)\dot{A}_{2}\dot{\phi} - V' =0,
 \end{align}
which with $V =-{d(d-1)\over \ell^2}$ and $R^{\zeta^{2}} = {(d-1)(d-2)\over R^2}$ can be reduced to the form
\begin{align}
 &\dot{A}_{2} ((d-2)\dot{A}_{2} + 2 \dot{A}_{1})  - e^{-2A_{2}}{(d-2)\over R^2}-{d\over \ell^2} = 0,\\
 & \ddot{A}_{2} - \dot{A}_{2}\left(\dot{A}_{1}-\dot{A}_{2}\right)  = 0,\\
&\ddot{A}_{1}  + (d-2)\ddot{A}_{2} + \dot{A}_{1}(\dot{A}_{1}-\dot{A}_{2})+ e^{-2A_{2}}{(d-2)\over R^2} = 0.
\end{align}}
The solution to the above equations is indeed AdS space in global coordinates
\be
e^{2A_1}=C_1^2 \cosh^2\left({u\over \ell}\right)\sp
e^{2A_1}={\ell^2\over R^2} \sinh^2\left({u\over \ell}\right)
\ee
 \end{enumerate}

\section{The curvature invariants}
\label{CurvInv}

In this  appendix we shall compute the curvature scalars, $R$, $R_{AB} R^{AB}$ and $R_{ABCD} R^{ABCD}$ and we shall express them in terms of the first order functions $W_1$, $W_2$, $S$, $T_1$ and $T_2$.

\paragraph*{The Ricci scalar}
The Ricci scalar is found to be
\begin{equation}
R = -(4\ddot{A_1} + 4\ddot{A_2} + 6\dot{A_1}^2 + 6\dot{A_2}^2 + 8 \dot{A_1} \dot{A_2}) + R^{\zeta_1} \mathrm{e}^{-2A_1} + R^{\zeta_2}\mathrm{e}^{-2A_2}.
\label{bb1}\end{equation}

Using the equations of motion, this can be written as
\begin{equation}
R = \frac{S^2}{2} + \frac{5}{3} V.
\label{bb2}\end{equation}
As $V$ is regular everywhere for finite $\f$, regularity of the scalar curvature is guaranteed once $\dot\phi$ is regular.

\paragraph*{Ricci squared}
The square of the Ricci tensor is given by
\be
R_{AB} R^{AB} = 4 (\ddot{A_1} + \ddot{A_2} + \dot{A_1}^2 + \dot{A_2}^2)^2 + 16 \dot{A_1}^2 \dot{A_2}^2
\label{bb3}\ee
$$
\mathrel{\phantom{=}} {} + 8 \dot{A_1} \dot{A_2} (\ddot{A_1} + 2 \dot{A_1}^2 -{R^{\zeta_1}\over 2} \mathrm{e}^{-2A_1} + \ddot{A_2} + 2 \dot{A_2}^2 - {R^{\zeta_2}\over 2}\mathrm{e}^{-2A_2})
$$
$$
\mathrel{\phantom{=}} {} + 2(\ddot{A_1} + 2 \dot{A_1}^2 - {R^{\zeta_1}\over 2} \mathrm{e}^{-2A_1})^2 + 2(\ddot{A_2} + 2 \dot{A_2}^2 -{R^{\zeta_2}\over 2} \mathrm{e}^{-2A_2})^2.
$$

Using the equations of motion, this can be written as
\begin{equation}
R_{AB} R^{AB} = \left(\frac{S^2}{2} + \frac{V}{3} \right)^2 + \frac{4V^2}{9}
\label{bb4}\end{equation}
The regularity conditions are as in the case of scalar curvature.

\paragraph*{Riemann squared}
The Kretschmann scalar reads
\begin{equation}
R_{ABCD} R^{ABCD}  = (R^{\zeta_1})^2 \mathrm{e}^{-4A_1} + (R^{\zeta_2})^2\mathrm{e}^{-4A_2} + 12\dot{A_1}^4 + 12\dot{A_2}^4 + 8\ddot{A_1}^2 + 8\ddot{A_2}^2
\label{bb5}\ee
$$
\mathrel{\phantom{=}} {} + 8\dot{A_1}^2 (2\ddot{A_1} -{R^{\zeta_1}\over 2} \mathrm{e}^{-2A_1}) + 8\dot{A_2}^2 (2\ddot{A_2} -{R^{\zeta_2}\over 2} \mathrm{e}^{-2A_2}) + 16 \dot{A_1}^2 \dot{A_2}^2.
$$

In general, we can rewrite the expression as
\be
R_{ABCD} R^{ABCD}  = 8 (\ddot{A_1} + \dot{A_1}^2)^2 + 8 (\ddot{A_2} + \dot{A_2}^2)^2 + 16 \dot{A_1}^2 \dot{A_2}^2
\label{bb6}\ee
$$
\mathrel{\phantom{=}} {} + (R^{\zeta_1})^2 \mathrm{e}^{-4A_1} + 4\dot{A_1}^4 - 4R^{\zeta_1}\dot{A_1}^2 \mathrm{e}^{-2A_1} + (R^{\zeta_2})^2 \mathrm{e}^{-4A_2} + 4 \dot{A_2}^4 - 4R^{\zeta_2} \dot{A_2}^2  \mathrm{e}^{-2A_2}
$$
$$
= 4 (\ddot{A_1} + \dot{A_1}^2 + \ddot{A_2} + \dot{A_2}^2)^2 + 4 (\ddot{A_1} + \dot{A_1}^2 - \ddot{A_2} - \dot{A_2}^2)^2 + 16 \dot{A_1}^2 \dot{A_2}^2
$$
$$ \mathrel{\phantom{=}} {} + 4 (\dot{A_1}^2 -{R^{\zeta_1}\over 2} \mathrm{e}^{-2A_1})^2 + 4 (\dot{A_2}^2 - {R^{\zeta_2}\over 2}\mathrm{e}^{-2A_2})^2.
$$
We also have
\be
4 (\ddot{A_1} + \dot{A_1}^2 + \ddot{A_2} + \dot{A_2}^2)^2 =\left({S^2\over 2} +{1\over 3}V\right)^2
\label{bb7}\ee
\be
4(\ddot{A_1} + \dot{A_1}^2 - \ddot{A_2} - \dot{A_2}^2)^2 = \left( S(W_1'-W_2')-{W_1^2-W_2^2\over 2}\right)^2
\label{bb8}\ee
\be
 4 (\dot{A_1}^2 - {R^{\zeta_1}\over 2}\mathrm{e}^{-2A_1})^2=\left({W_1^2\over 2}-T_1\right)^2=
 \left({W_1^2\over 2}+S^2+W_2^2-W_1W_2-S(W_1'+W_2')\right)^2
 \label{bb9}\ee
 \be
 4 (\dot{A_2}^2 -{R^{\zeta_2}\over 2} \mathrm{e}^{-2A_2})^2=\left({W_2^2\over 2}-T_2\right)^2=
 \left({W_2^2\over 2}+S^2+W_1^2-W_1W_2-S(W_1'+W_2')\right)^2
\label{bb10} \ee
Therefore we can convert
\be
R_{ABCD} R^{ABCD}  = \left(\frac{S^2}{2} + \frac{V}{3} \right)^2 +\left( S(W_1'-W_2')-{W_1^2-W_2^2\over 2}\right)^2 + W_1^2 W_2^2
\label{bb11}\ee
$$
+ \left({W_1^2\over 2}+S^2+W_2^2-W_1W_2-S(W_1'+W_2')\right)^2
 +\left({W_2^2\over 2}+S^2+W_1^2-W_1W_2-S(W_1'+W_2')\right)^2
$$

It is useful to rewrite this equation in terms of $A_1$ and $A_2$:

\bea
&& \label{eq:Kgen}R_{ABCD} R^{ABCD}  = \frac{11}{2}\left( \frac{S^2}{2}
  + \frac{V}{3} \right)^2 - 2S^2V - S^4 \nonumber \\
&& +\,\, 6(\ddot{A_1} + \dot{A_1}^2 - \ddot{A_2} - \dot{A_2}^2)^2 + 48 (\dot{A_1}\dot{A_2})^2 + 8V\dot{A_1}\dot{A_2} - 4S^2\dot{A_1}\dot{A_2}
\eea

When $A_1 = A_2$, this expression reduces to
\be
R_{ABCD} R^{ABCD}  = 8 \mathrm{e}^{-4A} + 40 \dot{A}^4 -16 \dot{A}^2 \mathrm{e}^{-2A} + 16 \ddot{A}^2 + 32 \dot{A}^2 \ddot{A}
\label{bb12}\ee
$$
 = \left(\frac{S^2}{2} + \frac{V}{3} \right)^2 + \frac{1}{24} (S^2 - 2V)^2 + \frac{1}{3} T^2,
$$
where $T = T_1 + T_2 = 4 \exp(-2A)$.
In this case it is singular when $T\to \infty$.

\section{The regularity conditions on the interior geometry}

\label{geomint_a}

We study here the regularity of the solutions near end-points of
the flow where $S\to 0$. As a guiding criterion for regularity we
use the finiteness of the Kretschmann  scalar, whose expression was
derived in the previous appendix, equation (\ref{eq:Kgen}). As we shall
see, this will turn out to be a sufficient (not just necessary)
condition to identify regular geometries.

\subsection{Analysis of the IR behavior of solutions}
\label{sec:reg}

\subsubsection{Leading behavior}

Regular  flows stop at  a point $u_0$ where $\dot\f(u_0)=0$. We want to understand the behavior of the scale factors
near such a point.

We start by assuming a generic power-law leading behavior near $u_0$ of the
form:
$A_1(u)$ and $A_2(u)$ is
\be
\label{ansA1} A_1 = \lambda_1 (u_0-u)^a+\ldots\sp A_2 = \lambda_2
(u_0-u)^b+\ldots
\,   \quad u \to u_0
\ee
where $\lambda_1$,$a$ and $\lambda_2$,$b$ are constants such that $a
\leq 0$, $b \leq 0$ and $\l_1,\l_2 \neq 0$. We further assume the
following ansatz for $\dot \varphi(u)$ near $u_0$:
\be
\label{eq:ansf}\dot\varphi = C_0 (u_0-u)^c +\cdots \sp u \to u_0 \sp
c>0 \sp C_0 \neq 0.
\ee
Substituting the asymptotics (\ref{ansA1}) and  (\ref{eq:ansf}) into
equations   \eqref{23} and \eqref{24} written in terms of the $u$
variable we find that,  at leading  order in $(u_0-u)$, the following
constraints must be obeyed:
\be
\label{23IR} 0= R^{\zeta_1} \ex^{-2\lambda_1 (u_0-u)^a} -
\left\{\begin{array}{ll} 4ba\l_1 \l_2(u_0-u)^{a+b-2} & a<b \\ & \\
    -b^2\l_2^2 (u_0-u)^{2b-2} & a>b \end{array}\right.
\ee
\be
\label{24IR}  0= R^{\zeta_2} \ex^{-2\lambda_2 (u_0-u)^b} -
\left\{\begin{array}{ll} 4ab\l_1 \l_2(u_0-u)^{a+b-2} & a>b \\ & \\
    -a^2\l_1^2 (u_0-u)^{2a-2} & a<b \end{array}\right.
\ee
 For non-zero, negative $a$ and $b$, the exponentials in
 (\ref{23IR}) and (\ref{24IR})  always dominate the power-law
 terms, therefore for non-zero $ R^{\zeta_i}$ the power-law behavior
 assumed in \eqref{ansA1} cannot solve Einstein's equation
 near $u_0$. If (say) $a=0$, then the first equation may be
 consistent (for $b=2$), but the second one fails. Therefore in order
 for (\ref{23}) and (\ref{24}) to be satisfied,  we need both $a$ and
 $b$ to
 vanish\footnote{The same reasoning is easily generalized to an ansatz
   of the form   $$A_{1,2} \underset{u\to u_0}{\sim}
   (u_0-u)^{a_{1,2}}\log((u_0-u)/\ell)^{b_{1,2}}$$ with $a_{1,2} \geq
   0$ and $b_{1,2} > 0 \, , b_{1,2} \neq 1$, and one concludes that $a_1
 = a_2 = 0$.}.

Suppose now $A_1$ and/or  $A_2$ diverge logarithmically at the
endpoint, so that the corresponding scale factors have  a power
law behavior:
\be
\label{ansA} A_1 = \lambda_1 \log\left(\frac{u_0-u}{\ell}\right)+A_{1,0}+\cdots\sp A_2 = \lambda_2 \log\left(\frac{u_0-u}{\ell}\right)+A_{2,0}+\cdots
\ee
where $\lambda_1$,$A_{1,0}$ and $\lambda_2$,$A_{2,0}$ are some
constants, and we suppose that at least one among $\lambda_1$ and
$\lambda_2$ is non-zero. Substituting this ansatz, as well as (\ref{eq:ansf}),   in the EoMs \eqref{eq:EOM6bis} -
\eqref{eq:EOM9bis} one finds,  to leading order in $(u-u_0)$:
\bea
 &&
R^{\zeta_{1}}e^{-2A_{1,0}}\left(\frac{u_0-u}{\ell}\right)^{-2\lambda_1}
+
R^{\zeta_{2}}e^{-2A_{2,0}}\left(\frac{u_0-u}{\ell}\right)^{-2\lambda_2}
\nonumber \\
&& \qquad = 2(u_0-u)^{-2}((\l_1+\l_2)^2 + 2\l_1\l_2) , \label{EOMIR1} \\
&& \nonumber \\
&&
  R^{\zeta_{1}}e^{-2A_{1,0}}\left(\frac{u_0-u}{\ell}\right)^{-2\lambda_1}
  -
  R^{\zeta_{2}}e^{-2A_{2,0}}\left(\frac{u_0-u}{\ell}\right)^{-2\lambda_2}
  \nonumber \\
&&   \qquad =
2(\l_1-\l_2)\,(u_0-u)^{-2}(2(\l_1+\l_2)-1), \label{EOMIR3} \\
&& \nonumber \\
&&  \l_1^2 + \l_2^2 = \l_1+\l_2 , \label{EOMIR2} \\
&& \nonumber \\
&& C_0 \,(u_0-u)^{c-1}(c+2(\l_1+\l_2)) = -V'(\f(u_0)),\label{EOMIR4}
\eea
where $c$ was defined in (\ref{eq:ansf}).

\begin{itemize}
\item Suppose first that at least one among $\lambda_{1,2}$ is
strictly positive.
Then, from  \eqref{EOMIR1}, this implies that either $\lambda_1 = 1$
or $\l_2 = 1$ or both. From the constraint \eqref{EOMIR2}, we deduce then that
there are three possible solutions $(\l_1,\l_2) =
\{(1,0),(0,1),(1,1)\}$. The case   $\l_1=\l_2=1$  however leads to a
singularity at $u_0$: in fact, in this case $\dot{A}_1\dot{A}_2
\underset{u\to u_0}{\sim} (u_0-u)^{-2}$, and the
 Riemann-square invariant  \eqref{eq:Kgen} is dominated by the second
 and third terms, which are both positive and divergent as
 $(u_0-u)^4$. This leaves as only possibilities $\l_1 = 1, \l_2 = 0$
 or $\l_1 = 0, \l_2 = 1$.

\item
Suppose now that both $\l_1$ and $\l_2$ are zero or negative. In this
case, the left  hand side  of \eqref{EOMIR1} vanishes as $u\to u_0$,
 which implies that the coefficient of the (divergent) right hand side
 must vanish too,
\be
\label{c1} 2\l_1\l_2 + (\l_1 + \l_2)^2  = 0
\ee
But this is impossible under the assumption that both $\l_{1,2}$ are
zero or negative, unless they both vanish, $\l_1=\l_2=0$ is therefore
the only solution in this case.
\item In all cases above, equation (\ref{EOMIR4}) implies $c=1$,
  i.e. $\dot{f}$ has to vanish linearly as $u\to u_0$.

\end{itemize}
Thus, with the ansatz of the form (\ref{ansA}), the only solutions
which may possibly be regular correspond to one of the choices below:
\be\label{l12}
\l_1 = 1, \l_2 = 0,  \, \qquad  \l_1 = 0, \l_2 = 1, \, \qquad  \l_1 = \l_2 = 0.
\ee
As we shall see, the first two choices correspond to regular  IR
endpoints (section \ref{sec:IRbc_a}). The last one corresponding to a
{\em bounce}, and will be discussed in  section
\ref{app-bounces}.

\subsubsection{General divergent subleading ansatz}
In  the previous subsection we have assumed that the first subleading terms
(after the logarithmically divergent ones) in equation (\ref{ansA})
are finite constants, and we found that the only solutions are given
in (\ref{l12}).  Here we relax the ansatz (\ref{ansA}) and allow for a
generic subleading (divergent) term. We conclude that the ansatz
(\ref{ansA}) with one of the choices   (\ref{ansA})  is the only
consistent possibility.\\

The general ansatz for the diverging part of $A_{1,2}$ is written:
\be
\label{ansAgen}A_1 = \lambda_1 \log\left(\frac{u_0-u}{\ell}\right)+f_1(u)+ A_{1,0}+\cdots \sp A_2 = \lambda_2 \log\left(\frac{u_0-u}{l}\right)+f_2(u)+A_{2,0}+\cdots \; .
\ee
where we suppose that $f_{1,2}(u) \underset{u\to u_0}{=} o(\log(u_0-u))$ and $1 \underset{u\to u_0}{=} o(f_{1,2}(u))$. We show that it leads to a contradiction.\\
For this ansatz, the EoMs \eqref{eq:EOM6bis} - \eqref{eq:EOM9bis} at leading order in $u-u_0$ are written:
\be
\label{EOMIR1gen} R^{\zeta_{1}}e^{-2A_{1,0}}\left(\frac{u_0-u}{\ell}\right)^{-2\lambda_1}\ex^{-2f_1(u)} + R^{\zeta_{2}}e^{-2A_{2,0}}\left(\frac{u_0-u}{\ell}\right)^{-2\lambda_2}\ex^{-2f_2(u)} +\cdots
\ee
$$
 = 2(u_0-u)^{-2}((\l_1+\l_2)^2 + 2\l_1\l_2)+\cdots ,
$$
\be
\label{EOMIR2gen} \l_1^2 + \l_2^2 = \l_1+\l_2 ,
\ee
\be
\nonumber
  R^{\zeta_{1}}e^{-2A_{1,0}}\left(\frac{u_0-u}{\ell}\right)^{-2\lambda_1}\ex^{-2f_1(u)} - R^{\zeta_{2}}e^{-2A_{2,0}}\left(\frac{u_0-u}{\ell}\right)^{-2\lambda_2}\ex^{-2f_2(u)} +\cdots =
\ee
\be
\label{EOMIR3gen} =2(\l_1-\l_2)\,(u_0-u)^{-2}(2(\l_1+\l_2)-1)+\cdots,
\ee
\be
\label{EOMIR4gen}C_0 \,(u_0-u)^{c-1}(c+2(\l_1+\l_2))+\cdots = -V'(\f(u_0))+\cdots,
\ee
where we suppose for now that the right-hand sides of \eqref{EOMIR1gen}, \eqref{EOMIR3gen} and \eqref{EOMIR4gen} do not vanish, as well as the left-hand side of \eqref{EOMIR3gen}. In this case, \eqref{EOMIR1gen} implies that either $$((u_0-u)/\ell)^{-2\l_1}\exp(-2f_1(u))\underset{u\to u_0}{\sim}(u_0-u)^{-2}$$
 or
 $$((u_0-u)/\ell)^{-2\l_2}\exp(-2f_2(u))\underset{u\to u_0}{\sim}(u_0-u)^{-2}\;,$$ 
which are in contradiction with the hypotheses on $f_{1}$ and $f_2$.
The same reasoning applies to the case where only the right-hand side
of \eqref{EOMIR3gen} does not vanish.

Finally, if both the right-hand side of \eqref{EOMIR1gen} and that of \eqref{EOMIR3gen} vanish, $\l_1$ and $\l_2$ obey the same equations as for the ansatz \eqref{ansA}, with solution $\l_1=\l_2=0$. \eqref{eq:EOM6bis} at leading order then reads:

\be
\label{EOMIR1genbis}R^{\zeta_{1}}e^{-2A_{1,0}}\ex^{-2f_1(u)} + R^{\zeta_{2}}e^{-2A_{2,0}}\ex^{-2f_2(u)} = 2(\dot{f}_1^2+\dot{f}_2^2) + 8\dot{f}_1\dot{f}_2 .
\ee
Depending on whether $f_1$ or $f_2$ dominates in the limit where $u\to u_0$, it implies that $f_1$ or $f_2$ should have a logarithmic behavior in this limit, which is in contradiction with the hypotheses on $f_1$ and $f_2$. Note that it is still true if $f_1/f_2$ remains of order 1.
The conclusion of the above analysis is that, up to order $\mathcal{O}(1)$, the correct regular ansatz for the $A$ variables near a point $u_0$ such that $\dot{\f}(u_0)=0$ is \eqref{ansA} with $(\l_1,\l_2) \in \{(0,0),(1,0),(0,1)\}$.
Finally, \eqref{EOMIR4} in the case where $V'$ does not vanish at $\f(u_0)$ and with $(\l_1,\l_2) \in \{(0,0),(1,0),(0,1)\}$ implies that $c=1$ in (\ref{eq:ansf}).
If $V'$ does vanish at $\f(u_0)$, but there is some minimal $k \geq 2$ such that $V^{(k)}(\f_0)\neq 0$, then \eqref{eq:EOM9bis} at leading order in $u-u_0$ implies:
\be
\label{EOMIR4bis} C_0 \,(u_0-u)^{c-1}(c+2(\l_1+\l_2)) = -V^{(k)}(\f(u_0))\left(-\frac{C_0}{c+1}(u_0-u)^{c+1}\right)^{k-1},
\ee
which leads to a contradiction for $k=2$, and to negative $c$ for $k\geq 3$. So $u_0$ cannot be an end-point of the flow in this case.
The remaining case where the potential $V$ is flat at $\f(u_0)$
implies that the only solution of \eqref{eq:EOM9bis} is the one with
constant scalar field $\f = \f(u_0)$, which corresponds to the CFT
case.

We conclude from this analysis that the only
  consistent behavior at leading order near an endpoint  is
  (\ref{ansA}) with one of the choices (\ref{l12}).

\subsubsection{Subleading terms}
Having determined that the only possible regular solution close to a
point $u_0$ where $\dot\f(u_0) = 0$, are are of the form (\ref{ansA})
with one of the combinations (\ref{l12}) of  coefficients,
we want to determine
the form of the subleading behavior
in such a way that the solution is regular at $u_0$.

We start from the following ansatz, which corresponds to the case
$\l_1=1, \l_2=0$. The discussion of the other cases follows similar
lines and gives the same result.
\be
\label{ansA1fin} A_1 = \log\left(\frac{u_0-u}{\ell}\right)+A_{1,0}+\mu_1(u_0-u)^{a_1}+\b_1(u_0-u)^{b_1}+\cdots ,
\ee
\be
\label{ansA2fin}A_2 = A_{2,0}+\mu_2(u_0-u)^{a_2}+\b_2(u_0-u)^{b_2}+\cdots,
\ee
with $\mu_1,\mu_2,\b_1,\b_2 \neq 0$, $b_1>a_1 > 0$ and $b_2>a_2 > 0$.
The part of the Kretschmann scalar  which
is potentially singular as  $u\to u_0$  is made of  the terms in  the
 second line in equation
(\ref{eq:Kgen}), and it has the following expansion:
\be
\label{Kfin}\mathcal{K} = 8V(\f(u_0))\left[1+\mathcal{O}((u_0-u)^2)\right]\left[\mu_2a_2(u_0-u)^{a_2-2}+
\mathcal{O}((u_0-u)^{b_2-2})+\mathcal{O}((u_0-u)^{a_1+a_2-2})+\cdots\right]
\ee
\be
\nonumber + 48(\mu_2^2a_2^2(u_0-u)^{2a_2-4}+\cdots)
\ee
\be
\nonumber + 6(-2\mu_1\mu_2a_1a_2(a_1+1)(a_2-1)(u_0-u)^{a_1+a_2-4}+\cdots),
\ee
Regularity of the solution demands that acting with further
Laplacians should also give a finite result at $u_0$. This implies
that only integer powers of $(u_0-u)$ should appear in the above
expansion. In particular, if $a_1=a_2=a$, then $a$ is an integer and
$a\geq 2$.

Next, we expand  equations \eqref{23} and \eqref{24} close to $u_0$:
\be
\label{23fin}2(u_0-u)^{-2}(-2\mu_1(u_0-u)^{a_1}+\mathcal{O}((u_0-u)^{b_1})+\mathcal{O}((u_0-u)^{2a_1})) =
\ee
\be
\nonumber -C_0^2(u_0-u)^2 -4\mu_2a_2(a_2-2)(u_0-u)^{a_2-2}-2\mu_1a_1(a_1-1)(u_0-u)^{a_1-2}+\mathcal{O}((u_0-u)^{b_2-2})+\mathcal{O}((u_0-u)^{b_1-2})
\ee
\be
\nonumber +\mathcal{O}((u_0-u)^{2a_2-2})+\cdots,
\ee
 \be
\label{24fin}R^{\zeta_2}\ex^{-2A_{2,0}}(1-2\mu_2(u_0-u)^{a_2}+\mathcal{O}((u_0-u)^{b_2})+\mathcal{O}((u_0-u)^{2a_2})) =
\ee
\be
\nonumber -C_0^2(u_0-u)^2 -2\mu_2a_2(a_2-3)(u_0-u)^{a_2-2}-4\mu_1a_1(a_1+1)(u_0-u)^{a_1-2}+\mathcal{O}((u_0-u)^{b_2-2})+\mathcal{O}((u_0-u)^{b_1-2})
\ee
\be
\nonumber +\mathcal{O}((u_0-u)^{2a_1-2})+\cdots.
\ee
From \eqref{24fin}, there are three possibilities for $a_1$ and
$a_2$:
\begin{itemize}
\item $a_2=2$,  which implies that $a_1 = 2$ from \eqref{23fin}.
\item  $a_2=3$ which implies that $a_1 = 2$ from \eqref{24fin}.
\item $a_2=a_1=a$ which implies that $a$ is an integer from
  \eqref{Kfin}.
\end{itemize}
The conclusion is that $a_1$ and $a_2$ should be integers. We assume that this result can be recursively extended to all exponents in the near $u_0$ expansion of $A_1$ and $A_2$ \eqref{ansA1fin}-\eqref{ansA2fin}, so that the finite parts of $A_1(u)$ and $A_2(u)$ are regular functions near $u_0$ that can be expanded in Taylor series.

\paragraph*{Summary}

The conclusion of the above analysis is that a regular solution for $A_1,A_2$ and $\dot{\f}$
near a point $u_0$ where $\dot{\f}(u_0)=0$ must take the  form:

\be
\label{ansAccl} A_1 = \lambda_1 \log\left(\frac{u_0-u}{\ell}\right)+A_{1,0}+\cdots\sp A_2 = \lambda_2 \log\left(\frac{u_0-u}{\ell}\right)+A_{2,0}+\cdots,
\ee
\be
\label{ansfccl}\dot\varphi = C_0 (u_0-u) +\cdots ,
\ee
where the dots correspond to a Taylor expansion near $u_0$, and $(\l_1,\l_2)\in \{(0,0),(1,0),(0,1)\}$.
We refer to the three different possible choices for the pair $(\l_1,\l_2)$ respectively as a bounce, an IR end-point where the sphere 1 shrinks to zero size and an IR end-point where the sphere 2 shrinks to zero size.

\subsection{The regular IR boundary conditions }
\label{sec:IRbc_a}

We now consider the case of an IR end-point where the sphere 1 shrinks to zero size (the case where the sphere 2 shrinks is symmetric). This corresponds to a point $u_0$ such that near $u=u_0$, $A_1,A_2$ and $\dot\f$ follow the ansatz \eqref{ansAccl}-\eqref{ansfccl} with $\l_1=1$ and $\l_2=0$.  The case    $\l_1 = 0 =\l_2 $ (both spheres stay finite)
 will be discussed in section \ref{app-bounces}.

From the near-IR behavior in equations
\eqref{ansAccl}-\eqref{ansfccl}, it follows that the corresponding expansions for the superpotentials (as functions of $u$) near $u = u_0$ are written:
\begin{align}
W_1 = -\frac{2}{u - u_0} + W_{1,0} + W_{1,1}(u-u_0) + W_{1,2}(u-u_0)^2
+ \mathcal{O}((u - u_0)^3), \label{w1ir} \\
T_1= \frac{2}{(u-u_0)^2} + \frac{T_{1,-1}}{u-u_0} + T_{1,0} + T_{1,1}(u-u_0) + \mathcal{O}((u - u_0)^2),  \label{t1ir}\\
W_2 = W_{2,0} + W_{2,1}(u-u_0) + W_{2,2}(u-u_0)^2 + \mathcal{O}((u - u_0)^3),  \label{w2ir}\\
T_2 = T_{2,0} + T_{2,1}(u-u_0) + \mathcal{O}((u - u_0)^2), \label{t2ir} \\
S = S_{1}(u-u_0) + S_{2}(u-u_0)^2 + \mathcal{O}((u - u_0)^3). \label{sir}
\end{align}
Substituting into the equations of motion \eqref{eq:EOM11bis}-\eqref{eq:EOM14bis} we find the coefficients to be:
\begin{align}
W_{1,0} = 0 \sp W_{1,1} = \frac{1}{27}V(\varphi_0) + \frac{1}{9}T_{2,0} \sp W_{1,2} = 0, \\
T_{1,-1} = 0 \sp T_{1,0} = \frac{1}{27}V(\varphi_0) + \frac{1}{9}T_{2,0} \sp T_{1,1} = 0, \\
W_{2,0} = 0 \sp W_{2,1} = \frac{2}{9}V(\varphi_0) - \frac{1}{3}T_{2,0} \sp W_{2,2} = 0, \\
T_{2,0} = arbitrary \sp T_{2,1} = 0,\\
S_{1} = \frac{1}{3} V^{'}(\varphi_0) \sp S_{2} = 0,
\end{align}
where $\f_0 \equiv \f(u_0)$. Note that in the case of $S^2 \times
S^2$, $T_2(u)$ is always positive, so $T_{2,0}$ can take  any positive
value.

From (\ref{t1ir}) and the definitions (\ref{eq:defT1c}), we can obtain the
constants $A_{1,0}$ and $A_{2,0}$ in equation (\ref{ansAccl}) (recall we are assuming
$\lambda_1=1, \lambda_2 = 0$ ):
\be \label{A0ir}
A_{1,0}  = {1\over 2} \log \left({R^{(\zeta^1)} \ell^2\over 2}\right),
\quad A_{2,0} = {1\over 2} \log \left({R^{(\zeta^2)} \over T_{2,0}}\right).
\ee
The constant $T_{2,0}$ therefore determines the  finite radius $\alpha_{IR}$ of the
sphere-2 at the IR endpoint:
\be\label{radiusir}
\alpha^{IR}_2  = \alpha_2 e^{A_{2,0}} =  \sqrt{{2 \over T_{2,0}}},
\ee
where  $\alpha_2 $ is the radius of the sphere in the fixed fiducial
metric $\zeta^2$ and we have used the relation $ R^{(\zeta^2)} =
2/\alpha_2^2$.

We can now write the superpotentials close to the endpoint in terms of
$\f$: using  $S = \dot{\varphi}$, one finds the behavior of $\varphi(u)$ near $u = u_0$:
\be
\varphi(u) = \varphi_0 + \frac{1}{6}V^{'}(\varphi_0)(u-u_0)^2 + \mathcal{O}((u - u_0)^4),
\ee
\be
u-u_0 = -\sqrt{\frac{6(\varphi - \varphi_0)}{V^{'}(\varphi_0)}} + \mathcal{O}((\varphi - \varphi_0)^{3/2},
\ee
where we assumed $u < u_0$. In terms of $\f$ the expansion therefore reads:

\be
W_1 = \sqrt{\frac{2V^{'}(\varphi_0)}{3(\varphi - \varphi_0)}} - \left(\frac{1}{27}V(\varphi_0) + \frac{1}{9}T_{2,0}\right)\sqrt{\frac{6(\varphi - \varphi_0)}{V^{'}(\varphi_0)}} + \mathcal{O}((\varphi - \varphi_0)^{3/2}), \ee
\be
T_1= \frac{V^{'}(\varphi_0)}{3(\varphi - \varphi_0)} + \frac{1}{27}V(\varphi_0) + \frac{1}{9}T_{2,0} + \mathcal{O}(\varphi - \varphi_0),
\ee
\be
W_2 = -\left(\frac{2}{9}V(\varphi_0) - \frac{1}{3}T_{2,0}\right)\sqrt{\frac{6(\varphi - \varphi_0)}{V^{'}(\varphi_0)}} + \mathcal{O}((\varphi - \varphi_0)^{3/2}),
\ee
\be
T_2 = T_{2,0} + \mathcal{O}(\varphi - \varphi_0),
\ee
\be
S = -V^{'}(\varphi_0)\sqrt{\frac{2(\varphi - \varphi_0)}{3V^{'}(\varphi_0)}} + \mathcal{O}((\varphi - \varphi_0)^{3/2})
\ee
We therefore conclude that $W_1$ and $T_1$ diverge at the IR end-point of the flow, while $S$ and $W_2$ vanish, and $T_2$ remains finite.

The value of the Kretschmann scalar \eqref{eq:Kgen} in the interior (at $\varphi = \varphi_0$) can also be computed from the previous expansions:
\begin{equation}
\label{Kint2} \mathcal{K}(\varphi_0) = \frac{V(\varphi_0)^2}{3}\left(1 + \frac{1}{24}(T_{2,0}\ell^2)^2 + \frac{1}{6}T_{2,0}\ell^2 \right),
\end{equation}
where $\ell$ is the AdS length near the boundary. It is finite for any
value of the constant $T_{2,0}$. Therefore, we conclude that the
ansatz \eqref{ansAccl}-\eqref{ansfccl} with $\l_1 = 1$, $\l_2 = 0$,
and the subleading terms in the expansions determined order by order
by Einstein's equation, gives a regular second-order curvature
invariant.

In fact, one can show without any extra assumptions that  the metric at the endpoint is
completely regular: Near the endpoint $u_0$ we have, using
(\ref{ansAccl}) (with $\l_1 = 1, \l_2=0$):
\be \label{metricir1}
ds^2 \simeq   du^2 + {(u-u_0)^2 \over \ell^2} e^{2A_{1,0}} \alpha_1^2
d\Omega^2 + e^{2A_{2,0}}\alpha_2^2 d\Omega^2  \qquad u \simeq u_0,
\ee
where $\alpha_1$ and $\alpha_2$ are the fiducial radii of spheres 1
and 2, and $ d\Omega^2$ is the metric of the unit 2-sphere. Using the
results (\ref{A0ir}) and the relation $R^{(zeta^i)} = 2/\alpha_i^2$
between the Ricci scalar and the radius, equation (\ref{metricir1}) becomes
\be \label{metricir2}
ds^2 \simeq  du^2 + (u-u_0)^2 d\Omega^2 + {2\over T_{2,0}} d\Omega^2 \qquad u \simeq u_0.
\ee
Changing variables to $\rho = u_0-u$ one can recognize the metric of
$R^3 \times S^2$ at the origin of $R^3$ in spherical coordinates. The
space-time is therefore regular at $u=u_0$.

\subsection{Regular AdS slicings} \label{regads}

Finally, consider the special  case  where $\varphi$ is a constant and
$$V(\varphi) \equiv V_0 = -{12\over \ell^2}\;.$$ From equation  (\ref{Kint2})  one finds:
\begin{equation}
\label{KintSlicing2} \mathcal{K}(u_0) = \mathcal{K}_{AdS^5} \left(1 + \frac{1}{20}(T_{2,0}\ell^2 + 2)^2 \right)
\end{equation}
where
\be
\mathcal{K}_{AdS^5} = {40\over \ell^4} = {5\over 18}V_0^2
 \ee
 is the Kretschmann scalar for the AdS$_5$ space-time. This means that
 the space-time with the metric \eqref{eq:metricS2} is an
 asymptotically AdS$^5$ manifold, but it  deviates from AdS$_5$ in the
 interior. Incidentally, this shows that AdS$_5$ does {\em not} admit a regular
 $S^2\times S^2$ slicing (which has positive $T_{2,0}$). This is
 unlike the case of  other positively curved
 manifolds, like $R\times S^3$ and $S^4$, which provide regular
 slicings of $AdS_5$.

 However, from equation (\ref{KintSlicing2}) we observe that AdS$_5$  may admit instead  a
 special EAdS$_2 \times S^2$ slicing with $T_{2,0} = -2/\ell^2$,
 i.e. such that in the  IR the $S^2$ shrinks to zero size and the $AdS_2$ has
 finite radius $\ell$ (by equation  (\ref{radiusir})). This
 can be explicitly obtained from first principles, from the embedding space
 definition of Euclidean $AdS_5$,
\be \label{adsemb}
-X_{-1}^2 + X_0^2 + X_1^2 + X_2^2 + X_3^2 + X_4^2 = -\ell^2,
\ee
by choosing the following set of local coordinates:
\be \label{adscoord}
\begin{array}{l}X_{-1} = \ell \cosh (u/\ell) \cosh \tau \\
X_0 =\ell \cosh  (u/\ell) \sinh \tau \cos \psi \\
X_1 =\ell \cosh  (u/\ell) \sinh \tau \sin \psi
 \end{array}\;,  \qquad
\begin{array}{l}X_{2} = \ell \sinh  (u/\ell) \cos \theta \\
X_3 =\ell \sinh  (u/\ell) \sin \theta \cos \phi \\
X_4 =\ell \sinh  (u/\ell) \sin \theta \sin \phi
 \end{array}
\ee
the resulting metric is
\be \label{adsmetric}
ds^2 = d u^2 + \ell^2 \cosh^2 {u \over \ell} \left[ d\tau^2 + \sinh^2
  \tau d\psi^2\right] + \ell^2 \sinh^2  {u \over \ell} \left[ d\theta^2 + \sin^2
  \theta d\phi^2\right]
\ee
We recognise the metric in the ansatz (\ref{eq:metric}) with  $EAdS_2 \times
S^2$ sections. At the IR endpoint $u=0$, the $S^2$ shrinks to zero and
the $EAdS_2$ has a finite radius $\ell$, in agreement with the value
we have found above, $T_{2,0} = -2/\ell^2$. In the UV ($u\to -\infty$) both factors
have the same radius $\ell$ (up to the common divergent $e^{-2u/\ell}$ prefactor).

\subsection{Bounces}
\label{app-bounces}

We now consider the case of a bounce. This corresponds to a point $u_0$ such that near $u=u_0$, $A_1,A_2$ and $\dot\f$ follow the ansatz \eqref{ansAccl}-\eqref{ansfccl} with $\l_1=\l_2=0$. The corresponding expansions for the superpotentials (as functions of $u$) near $u = u_0$ are written::

\be
\label{W1bu}W_1 = W_{1,0} + W_{1,1}(u-u_0) + \mathcal{O}((u - u_0)^2),
\ee
\be
\label{T1bu}T_1= T_{1,0} + T_{1,1}(u-u_0) + \mathcal{O}((u - u_0)^2),
\ee
\be
\label{W2bu}W_2 = W_{2,0} + W_{2,1}(u-u_0) + \mathcal{O}((u - u_0)^2),
\ee
\be
\label{T2bu}T_2 = T_{2,0} + T_{2,1}(u-u_0) + \mathcal{O}((u - u_0)^2),
\ee
\be
\label{Sbu}S = S_{1}(u-u_0) + S_{2}(u-u_0)^2 + \mathcal{O}((u - u_0)^3),
\ee
The only qualifying feature of a bounce is the vanishing of $S$.\\
Substituting into the equations of motion \eqref{eq:EOM11bis}-\eqref{eq:EOM14bis} we find the coefficients to be:

\be
W_{1,0} = arbitrary \sp W_{1,1} = T_{2,0} + \frac{1}{2}(W_{1,0}^2 - W_{2,0}^2) - W_{1,0}W_{2,0} - \frac{1}{3}V(\varphi_0)
\ee
\be
T_{1,0} = -T_{2,0} + \frac{1}{2}(W_{1,0}^2 + W_{2,0}^2) + 2W_{1,0}W_{2,0} + V(\varphi_0) \sp T_{1,1} = W_{1,0}T_{1,0}
\ee
\be
W_{2,0} = arbitrary \sp W_{2,1} = -T_{2,0} + W_{2,0}^2 + W_{1,0}W_{2,0} + \frac{2}{3} V(\varphi_0)
\ee
\be
T_{2,0} = arbitrary \sp T_{2,1} = W_{2,0}T_{2,0}
\ee
\be
S_{1} = V^{'}(\varphi_0) \sp S_{2} = \frac{1}{2}V^{'}(\varphi_0)(W_{1,0}+W_{2,0})
\ee

where $\f_0 \equiv \f(u_0)$. Using that $S = \dot{\varphi}$, one finds the behavior of $\varphi(u)$ near $u = u_0$:

\begin{align}
\varphi(u) = \varphi_0 + \frac{1}{2}V^{'}(\varphi_0)(u-u_0)^2 + \frac{1}{6}V^{'}(\varphi_0)(W_{1,0}+W_{2,0})(u-u_0)^3 + \mathcal{O}((u - u_0)^4), \\
u-u_0 = \pm \sqrt{\frac{2(\varphi - \varphi_0)}{V^{'}(\varphi_0)}} - \frac{\varphi - \varphi_0}{3V^{'}(\varphi_0)}(W_{1,0}+W_{2,0}) + \mathcal{O}((\varphi - \varphi_0)^{3/2})
\end{align}

In terms of $\varphi$ the expansion therefore reads:

\be
\label{W1bf}W_1 = W_{1,0} \pm \left(T_{2,0} + \frac{1}{2}(W_{1,0}^2 - W_{2,0}^2) - W_{1,0}W_{2,0} - \frac{1}{3}V(\varphi_0)\right)\sqrt{\frac{2(\varphi - \varphi_0)}{V^{'}(\varphi_0)}} + \mathcal{O}(\varphi - \varphi_0),
\ee
\be
\label{T1bf}T_1= T_{1,0}\left(1 \pm W_{1,0}\sqrt{\frac{2(\varphi - \varphi_0)}{V^{'}(\varphi_0)}} + \mathcal{O}(\varphi - \varphi_0)\right)),
\ee
\be
\label{W2bf}W_2 = W_{2,0} \pm \left(-T_{2,0} + W_{2,0}^2 + W_{1,0}W_{2,0} + \frac{2}{3}V(\varphi_0)\right)\sqrt{\frac{2(\varphi - \varphi_0)}{V^{'}(\varphi_0)}} + \mathcal{O}(\varphi - \varphi_0),
\ee
\be
\label{T2bf}T_2 = T_{2,0}\left(1 \pm W_{2,0}\sqrt{\frac{2(\varphi - \varphi_0)}{V^{'}(\varphi_0)}} + \mathcal{O}(\varphi - \varphi_0)\right),
\ee
\be
\label{Sbf}S = \pm \sqrt{2V^{'}(\varphi_0)(\varphi - \varphi_0)} + \frac{2}{3}(W_{1,0}+W_{2,0})(\varphi - \varphi_0) + \mathcal{O}((\varphi - \varphi_0)^{3/2}).
\ee
The solution \eqref{Sbf} describes two branches, depending on whether $\dot\f$ is positive or negative near $\f_0$, which can be glued together, exactly as in the maximally symmetric case \cite{C}. At $\f_0$, neither is the geometry singular, nor does the flow stop. The singularity of the superpotentials as functions of $\f$ at $\f_0$ is only the sign that $\f$ is not a good coordinate for the flow across $\f_0$: the monotonicity of $\f$ is reversed, but the flow is uninterrupted and regular, as can be seen from the expansions of the superpotentials in $u$ near $u_0$ \eqref{W1bu}-\eqref{Sbu}.

\section{The structure of solutions near the boundary.\label{bc}}

In this appendix we derive the near-boundary expansions of the
flow solutions to equations
(\ref{eq:EOM11bis})-(\ref{eq:EOM14bis}). For the second order
equations in terms of the variable $u$, this corresponds to the limit
where $u \to -\infty$.

The UV expansion can be organized as a double expansion in powers of the
curvatures $R^{(\zeta_1)}, R^{(\zeta_1)}$, and in powers of $\f$:
\begin{itemize}
\item The curvatures enter equations
  (\ref{eq:EOM11bis})-(\ref{eq:EOM14bis}) only through $T_1$ and $T_2$,
  which, from the definition (\ref{eq:defT1c}),  in the UV scale as $e^{-2A_{1,2}} \to 0$. Therefore we can
  solve the equations order by order in $T_1$ and $T_2$. We shall
  loosely use $O(R^n)$ to denote terms in the curvature expansion
  which enter as $T^n \sim e^{-2 n A}$ in the UV expansion.
\item At each order in the curvature, we can expand the corresponding
  functions in powers of $\f$ around the fixed point $\f=0$.
\end{itemize}

The curvature expansion takes the following form, up and including to second order:
\be\label{Sexp}
 S(\f)=S_{0}(\f)+
  S_{1}(\f)+
 S_{2}(\f)  + {\cal O}(R^3),
\ee
\be\label{Wiexp}
W_i(\f)=W_i^{0}(\f)+
W_i^{1}(\f)+
W^{2}_i(\f)  + {\cal O}(R^3),
\ee
\be \label{T1exp-i}
T_i(\f)=
T^{1}_i(\f) +  T^{2}_i(\f)  + {\cal
  O}(R^3)
\ee

Notice that $R^{(\zeta_1)}, R^{(\zeta_1)}$ do not appear at all in
equations  (\ref{eq:EOM11bis})-(\ref{eq:EOM14bis}): they can be
considered as two integration constants of the solutions, prametrizing the
near-boundary behavior of $T_1$ and $T_2$.
Also, we should not forget however that the curvatures are not the
only integration constants, as there should be a total of four. The
dependence on the remaining two parameters should appear as
perturbations to the small curvature expansions.

The goal of the following three subsections is to determine the
functions of $\f$ appearing in equations (\ref{Sexp}-\ref{T1exp-i}) ,
to leading order in an expansion around $\f=0$.  This is done by
writing  the equations of motion (\ref{eq:EOM11bis})-(\ref{eq:EOM14bis}) order by order in the curvature.

\subsection{Order zero in the curvature }

We shall  write (\ref{eq:EOM11bis})-(\ref{eq:EOM14bis}) at leading order  in the curvature expansion as
\begin{align}
\label{E10} (W_1^{0})^2 + (W_2^{0})^2 + 4 W_1^{0} W_2^{0} - S_0^2 + 2 V & = 0, \\
\label{E20} S_0^2 - \frac{3}{2} S_0 ( (W^0_1)^{'}  +  (W^0_2)^{'} ) + \frac{1}{2} (W_1^{0} - W_2^{0})^2 & = 0, \\
\label{E30} (-S_0 (W^0_1)^{'} + (W_1^{0})^2) - (-S_0 (W^0_2)^{'}  + (W_2^{0})^2) & = 0, \\
\label{E40} S_0S'_{0} - S_0(W_1^{0} + W_2^{0}) - V' & = 0.
\end{align}
It is convenient to define the two variables:
\be
\label{defX} X (\f)
 = {1\over 2}(W_1^{0}(\f)+W_2^{0}(\f)),  \quad   f(\f) = W_1^{0}(\f)
 - W_2^{0}(\f).
\ee
We can take the following  as independent equations for the variables
$X,f,S_0$:
\bea
&& -S_0 X' + 2 X^2 + {2\over 3} V = 0  \label{zero1}  \\
&&  S_0  S_0' - 2  S_0 X -V' = 0  \label{zero2} \\
&&   - S_0 f' + 2 X f =  0 \label{zero3}
\eea

We first consider the system of equations (\ref{zero1}-\ref{zero2}),
which does not contain $f(\f)$.  Close to the UV fixed point $\f=0$, where
the potential is approximate by
\begin{equation}
\label{Vexp} V(\varphi) = -\frac{12}{\ell^2} - \frac{m^2}{2}\varphi^2
+ \mathcal{O}(\varphi^3),
\end{equation}
 we  look for  a regular power-series solution for $S_0(\f)$ and $X(\f)$,
\be \label{ansatz0}
S_0 =  s_1 \f + \mathcal{O}(\f^2), \qquad X = x_0 + x_1 \f + x_2 \f^2 + \mathcal{O}(\f^3)
\ee
where we set to zero  the  $O(\f^0)$ term in $S_0$ to ensure we
look at solutions that stop at the UV  fixed point\footnote{In so
  doing, we already chose one of the two integration constants in
  equations (\ref{zero1}-\ref{zero2}). As we shall see, the second one  appears at
  subleading order in a {\em non-analytic} term.}, for which
$S(0)=0$. Substituting (\ref{Vexp}) the ansatz (\ref{ansatz0}) in equations
(\ref{zero1}-\ref{zero2}) and equating terms order by order we obtain
two branches of solutions,


\begin{equation}
\label{defW0} X (\f) = \frac{2}{\ell} + \frac{\Delta_{\pm}}{6\ell}\varphi^2
+ \mathcal{O}(\varphi^3),  \qquad  S_0(\f) =  \frac{\Delta_{\pm}}{\ell} \varphi + \mathcal{O}(\varphi^2)
\end{equation}
where $\Delta_{\pm}$ is one of two choices:
\begin{equation}
\label{Delta}
\Delta_{\pm} = 2 \pm \sqrt{4 + m^2 \ell^2}
\end{equation}
Notice that, to this order, $S = (3/2)(W_1' + W_2')$.

\paragraph*{The non-analytic terms.}
The  power-series solution (\ref{defW0}) can be extended  to higher order
and  does not contain any free parameter.  The remaining integration
constant of equations (\ref{zero1}-\ref{zero2}) enters in  a
subleading  non-analytic term, as it happens in holography  with flat slicing
(see e.g. \cite{exotic}) or $S^4$ slicing \cite{C}.  To identify it,
we look for a small perturbation of the  power series solutions
(\ref{defW0}),
\be \label{pert}
X(\f)  = X^{(\pm)}(\f) + \delta X(\f), \quad S_0(\f) = S_0^{(\pm)}(\f) +
\delta S(\f) .
\ee
We now linearize the system (\ref{zero1}-\ref{zero2}) in  $\delta X,
\delta S$, and at the same time perform an expansion in $\f$ around
$\f=0$ of the coefficient functions. The resulting system of linear
differential equations reads:
\bea
&& - \Delta \f\, \delta X' - {\Delta \over 3} \f \,\delta S + 8\, \delta X =
0\label{pert2a} \\
&& \Delta \f \,\delta S' + (\Delta - 4) \delta S - 2\Delta \f \,\delta X =
0 \label{pert2b}
\eea
where $\Delta \equiv \Delta_{\pm}$ depending on which branch we are
choosing for the unperturbed solution.  The solution to equations (\ref{pert2a}-\ref{pert2b})
reads, to leading order in $\f$:
 \begin{equation} \label{pert3}
\delta X  =  {C\over \ell} |\f|^{4/\Delta}\left(1 + \mathcal{O}(\f) \right) , \qquad \delta S = {12 C \over
  \Delta \ell}   |\f|^{4/\Delta - 1} \left(1 + \mathcal{O}(\f) \right)
\end{equation}
where $C$ is an arbitrary integration constant\footnote{The second integration constant of
the linearized system is unphysical and it can be  fixed by the
requirement that $S = 3X'$ to all orders in $\f$ at at zeroth order in
the curvature.}

Lastly, we turn to the combination $f(\f)$, defined in
(\ref{defX}). This function obeys the linear equation (\ref{zero3}),
whose solution is
\be\label{fsol}
f(\f) = \tilde{C}\exp \int^\f d\f' {2X\over S_0 }
\ee
 where $\tilde{C}$ is one more integration constant. It is enough to
 consider the leading order solutions (\ref{defW0}) to obtain
\be \label{fsol2}
f(\f) = \tilde{C} |\f|^{4/\Delta}(1 + \mathcal{O}(\varphi))
\ee
where this time both $\Delta_{\pm}$ are allowed.

We have therefore two integration constants $C,\tilde{C}$ 
entering respectively the sum and difference or $W_1$ and $W_2$
at order $\f^{4/\Delta_{\pm}}$. 
Thus, defining
\be \label{C1C2}
C_1 = C + {\tilde{C} \over 2}, \qquad C_2 = C - {\tilde{C} \over 2}
\ee
we can write the general near-boundary expansion at zeroth order in
the curvatures:
\begin{align}
W_1^{0} = \frac{2}{\ell} +
\frac{\Delta_{\pm}}{6\ell}\varphi^2(1 + \mathcal{O}(\varphi)) +
\frac{C_1}{\ell}\varphi^{4/\Delta_{\pm}}(1 + \mathcal{O}(\varphi)
),  \label{W1zero}\\
W_2^{0} = \frac{2}{\ell} + \frac{\Delta_{\pm}}{6\ell}\varphi^2(1 +
\mathcal{O}(\varphi)) + \frac{C_2}{\ell}\varphi^{4/\Delta_{\pm}}(1 +
\mathcal{O}(\varphi)), \label{W2zero}\\
S_0 = \frac{\Delta_{\pm}}{\ell}\varphi (1 + \mathcal{O}(\varphi))
+ \frac{6(C_1+C_2)}{\ell\Delta_{\pm}}\varphi^{4/\Delta_{\pm}-1}(1 +
\mathcal{O}(\varphi)) \label{Szero}
\end{align}

\subsection{Order one in the curvature}

We now consider the equations of motion (\ref{eq:EOM11bis}-
\ref{eq:EOM14bis})  at first order  in
$(R^{(\zeta_1)},R^{(\zeta_2)})$ or, which is the same, in  $T_1,T_2$:
\begin{align}
\label{E110} 6 X (W_1^{1}+W_2^{1}) - 2S_0S_{1} -2(T_1^1 + T_2^1) & = 0, \\
\label{E210} 2S_0S_{1} - 3 S_{1} X' - \frac{3}{2}S_0 ((W'_1)^{1,0} + (W'_2)^{1,0}) + \frac{1}{2}T_1^0 & = 0, \\
\label{E310} (-S_0 (W^1_1)' + 2X W_1^{1} - T_1^1) -
(-S_0 (W^1_2)^{'} + 2X W_2^{1}-  T_2^1) & = 0, \\
\label{E410} S_0S'_{1} + S_{1}S'_0 - 2X S_{1}  - S_0(W_1^{1} + W_2^{1})  & = 0.
\end{align}
where we remind the reader that  $X  \equiv (W^{0}_1 +
W^{0}_2)/2$.  
We have also neglected terms involving the difference $W^{0}_1 -
W^{0}_2$ since they are of higher order in $\f$ with respect to the
leading terms in the curvature\footnote{This will be justified a
  posteriori.}.

Again, we want to find a perturbative solution around $\f=0$.  We fist start by determining $T_i(\f)$ from the zeroth order
solutions for $W_{1,2}$ and $S$. Using equation
(\ref{21}), together with the zeroth order expressions
(\ref{W1zero}-\ref{Szero}), we obtain the differential equations
\be
{(T_1^{1})' \over T_1^{1}}  = {2\over \Delta_{\pm} \f} \left(1 +
  \mathcal{O}(\f)\right),  \qquad {(T_2^{1})' \over T_2^{1}}  = {2\over \Delta_{\pm} \f} \left(1 + \mathcal{O}(\f)\right)
\ee
 giving
\be \label{Tlowest}
T_1^{1}(\f) = {\mathcal{R}_1 \over \ell^2}|\f|^{2/\Delta_{\pm}}\left(1 + \mathcal{O}(\f)\right), \qquad   T_{2}^{1}(\f) = {\mathcal{R}_2 \over \ell^2}|\f|^{2/\Delta_{\pm}}\left(1 + \mathcal{O}(\f)\right)
\ee
where $\mathcal{R}_{1,2}$ are (dimensionless) integration constants
which will be related  to the actual curvatures  $(R^{(\zeta_1)},R^{(\zeta_2)})$
in subsection \ref{sec:sum}. As $T_1$ and $T_2$ are proportional at
leading order to $\mathcal{R}_{1}, \mathcal{R}_{2}$, we can use these
constants to count the order in the curvature expansion.

We can now determine $W_{1,2}$ and $S$ at this order.
We introduce again the sum  and difference of the superpotentials as independent
variables,
\be \label{Yg}
Y =  {1\over 2} (W_1^{1} + W_2^{1}) , \qquad g  =  W_1^{1} - W_2^{1}
\ee
Then, adding equation (\ref{E110}) to twice equation  (\ref{E210}) and
using  the zeroth-order relation $S_0 = 3 X'$, we find two
decoupled first or equations for $Y(\f)$ and $g(\f)$:
\begin{equation}
\label{2xE210+E110} S_0 Y ' - 2X Y = -\frac{1}{6}(T_1^1 + T_2^1) , \qquad
S_0 g' - 2 X g = -  (T_1^1 - T_2^1)
\end{equation}
Using the lowest order results  (\ref{defW0}) as well as
(\ref{Tlowest}), we  obtain the solution to leading order in $\f$,
\bea
&& Y(\f) = \frac{\mathcal{R}_1+\mathcal{R}_2}{12\ell}|\varphi|^{2/\Delta_{\pm}} \left( 1 +
  \mathcal{O}(\varphi)\right)+
\frac{C'}{\ell}|\varphi|^{4/\Delta_{\pm}} \left[ 1 +
  \mathcal{O}(\varphi) \right], \label{solY^{1,0}}  \\
&& \nonumber \\
&& g(\f) = {\mathcal{R}_1 - \mathcal{R}_2 \over 2\ell }|\f|^{2/\Delta_{\pm}} \left( 1 +
  \mathcal{O}(\varphi)\right)+
\frac{\tilde{C}'}{\ell}|\varphi|^{4/\Delta_{\pm}} \left[ 1 +
  \mathcal{O}(\varphi) \right].\label{solg}
\eea
The new integration constants $C', \tilde{C}'$ multiply the same
non-analytic term we found in the previous subsection at order zero in
the curvature, and they can be reabsorbed in the definition of  $C,
\tilde{C}$, see equations (\ref{pert3}) and (\ref{fsol2}).

From the definitions  (\ref{Yg}) we obtain, to lowest order in $\f$:
\bea
&& \left( W_1 \right)_{\mathcal{O}(R)}  =
\frac{1}{\ell}\left(\frac{\mathcal{R}_1}{3} - \frac{\mathcal{R}_2}{6}\right)
|\varphi|^{2/\Delta_{\pm}} \left[ 1 +
  \mathcal{O}(\varphi)
\right] + C_1' |\varphi|^{4/\Delta_{\pm}} \left[ 1 +
  \mathcal{O}(\varphi)\right],  \label{W11sol} \\
&& \left( W_2 \right)_{\mathcal{O}(R)}  =
\frac{1}{\ell}\left(\frac{\mathcal{R}_2}{3} - \frac{\mathcal{R}_1}{6}\right)
|\varphi|^{2/\Delta_{\pm}} \left[ 1 +
  \mathcal{O}(\varphi)
\right] +  C_2' |\varphi|^{4/\Delta_{\pm}} \left[ 1 +
  \mathcal{O}(\varphi) \right].\label{W21sol}
\eea
The constants  $C_{1,2}' =C' \pm  {1/2}\, \tilde{C}'$ can be reabsorbed in
 $C_{1,2}$ appearing  in (\ref{W1zero}-\ref{W2zero}).
Finally, we can solve for  $S_1$ algebraically from (\ref{E110}) 
All terms proportional to $|\f|^{2/\Delta_{\pm}}$
cancel, and we are left with:
\begin{equation} \label{S1sol}
\left(S \right)_{\mathcal{O}(R)} = \frac{6(C_1' + C_2')}{\Delta_{\pm}\ell}|\varphi|^{4/\Delta_{\pm}-1} \left[ 1 + \mathcal{O}(\varphi).
\right] 
\end{equation}
Notice that to $O(R)$ we still have the relation $S = (3/2)(W_1 + W_2)'$.
The contribution  (\ref{S1sol}) (as well as the second terms in
(\ref{W11sol}-\ref{W21sol}) can be completely absorbed in the
non-analytic terms of the same order in  (\ref{W1zero}-\ref{Szero}).

\subsection{Order two in the curvature}\label{OR2}

We start by obtaining equations for $T_1$ and $T_2$
from equation (\ref{21}), which   at order $R^2$  reads:
\be \label{Teq2nd}
(T_i^2)' - {T_1^{2} W_i^{0}\over
  S^{0} } = { T^{1}_i W_i^{1}  \over
  S^{0} }, \quad i=1,2,
\ee
where we have used the fact that the contribution to $S$ at linear
order in the curvature vanishes.

Substituting in  equations (\ref{Teq2nd}) the near-boundary behavior
in equations (\ref{defW0}) ,  (\ref{Tlowest}) and (\ref{W11sol}-\ref{W21sol}), we
find, to lowest order in $\f$,
\be\label{T2ndeq}
(T_1^{2})' - {2 \over \Delta_{\pm} \f} T_1^{2} = {2\mathcal{R}_1^2 -
  \mathcal{R}_1 \mathcal{R}_2\over 6 \Delta_{\pm} \ell^2} \f^{4/\Delta_{\pm} -
  1}, \qquad (T_2^{2})' - {2 \over \Delta_{\pm} \f} T_2^{2} = {2\mathcal{R}_2^2 -
  \mathcal{R}_1 \mathcal{R}_2\over 6 \Delta_{\pm} \ell^2} \f^{4/\Delta_{\pm} -
  1},
\ee
 the solutions at order $R^2$ are then:
\be \label{T2ndsol}
(T_1)_{O(R^2)} = {2\mathcal{R}_1^2 -
  \mathcal{R}_1\mathcal{R}_2\over12 \ell^2} |\f|^{4/\Delta_{\pm}}
\left(1 + \mathcal{O}(\f)\right),
\quad (T_2)_{O(R^2)} = {2\mathcal{R}_2^2 -
  \mathcal{R}_1\mathcal{R}_2\over12 \ell^2} |\f|^{4/\Delta_{\pm}}\left(1 + \mathcal{O}(\f)\right).
\ee
plus a solution of the homogeneous equation $\sim \f^{2/\Delta_{\pm}}$
which can be reabsorbed into the leading order terms (\ref{Tlowest}).

Next, we introduce the variables
\be \label{Zh2}
Z (\f)={1\over2} \left(W_1  + W_2 \right)_{O(R^2)}, \qquad
h(\f) = \left( W_1 - W_2 \right)_{O(R^2)},
\ee
We  consider again the sum of equation \eqref{eq:EOM11bis} plus
twice \eqref{eq:EOM12bis}. At order $R^2$ we obtain:
\be
\label{eqZ}
 S^0 Z' - 2 X Z = - {1\over 6} \left(T_1^{2} + T_2^{2}
      \right) + {1\over 3} \left(W_1^{1}\right)^2 +  {1\over
        3}\left(W_2^{1}\right)^2 +  {1\over 3} W_1^{1} W_2^{1} .
\ee
The functions appearing on the
right hand side can be found in equations (\ref{W11sol}-\ref{W21sol}) and
(\ref{T2ndsol}), and one can easily  check that the right hand side
of equation (\ref{eqZ}) vanishes identically. Using the leading order
power-series expansion of $X$ and  $S^{0,0}$ from equation
(\ref{defW0}), equation (\ref{eqZ}) reduces to the usual homogeneous linear
equation,
\be
\Delta_{\pm} \f Z'  - 4  Z = 0.
\ee
This results in:
\be
\label{W1pW2O2}(W_1+W_2)_{\mathcal{O}(R^2)} = 
(C_1'' + C_2'' )\f^{4/\Delta_{\pm}}[1+\mathcal{O}(\f)]
\ee
whose solution can be reabsorbed once again into a redefinition of
the integration constant $C$ already introduced in equation
(\ref{pert3}).



We now consider  the difference $h = W_1-W_2$. Writing  equation
\eqref{eq:EOM13bis} at second  order  in the curvature, we obtain the
following linear  differential equation:
\be \label{eqh}
S_0 h' - 2 X h = - (T_1 - T_2)_{O(R^2)}
+ (W_1 + W_2)_{O(R)} \times  (W_1 - W_2)_{O(R)}
\ee
Using the leading order expansion of $S_0$ and $X$ on the left
hand side, as well as the results (\ref{T2ndsol}) and (\ref{W11sol}-\ref{W21sol}) on the
right hand side,  equation (\ref{eqh}) becomes, at leading order:
\be
\Delta_{\pm} \f h ' - 4 h = - {1\over 12 \ell^2}\left(\mathcal{R}_1^2 -
  \mathcal{R}_2^2 \right)
\ee
whose solution contains the usual homogeneous term $\sim
\f^{4/\Delta}$ plus a new logarithmic correction:
\be
\label{solf} (W_1(\f) - W_2(\f))_{\mathcal{O}(\mathcal{R}^2)} = \left[\left(C_1'' - C_2''\right)
|\f|^{4/\Delta_{\pm}} -\frac{(\mathcal{R}_1^2 - \mathcal{R}_2^2)}{12\Delta_{\pm} \ell}\f^{4/\Delta_{\pm}}\log|\f|\right](1 + \mathcal{O}(\f)) \, .
\ee

Lastly, we can obtain $S(\f)$ at order $R^2$  from equation
(\ref{eq:EOM11bis}), using the results obtained so far for all the
other quantities. The result is:
\be
(S)_{\mathcal{O}(R^2)} = \left[ \frac{6(C_1'' +
  C_2'')}{\Delta_{\pm}\ell} - {1\over 24 \Delta_{\pm} \ell} \left( 5 \mathcal{R}_1^2 + 5
  \mathcal{R}_2^2 - 8 \mathcal{R}_1 \mathcal{R}_2
\right)\right] |\varphi|^{4/\Delta_{\pm}-1}  \left( 1 +
  \mathcal{O}(\varphi)\right) \label{S2sol}
\ee
Notice that there are no logarithmic terms but, to this order, the
relation $S = (3/2) (W_1 + W_2)'$ is violated by the second term.

It turns out to be  more convenient to redefine the integration constants
$C_1$ and $C_2$ in  such a way that the second line in (\ref{S2sol})
appears in $W_1 + W_2$ but not in $S(\f)$, because this simplifies the
expression for the scale factors and has a clearer physical meaning.
\subsection{Near-boundary RG flow solution: full
  result} \label{sec:sum}

Here we collect the results of the previous two sections, and we
obtain the near-boundary behavior of the scale factors $A_{1,2}(u)$
and scalar field profile $\f(u)$.

It is convenient to redefine the integration
constants appearing in front of  $\f^{4/\Delta_{\pm}}$ term in such a
way that there are no explicit $\mathcal{R}^2$ terms appearing in $S$:
more explicitly,  with respect to the definitions in the previous
sections, we redefine:
\be
C_{1,2}+ C_{1,2}' + C_{1,2}'' - {1\over 48 \ell} \left( {5\over 6}
  \mathcal{R}_1^2 + {5\over 6}
  \mathcal{R}_2^2 - {4\over 3} \mathcal{R}_1 \mathcal{R}_2
\right) \longrightarrow  C_{1,2}
\ee
This redefinition does not affect the difference $C_1-C_2$.

Combining the results   (\ref{W1zero}-\ref{Szero}),  (\ref{Tlowest}),
(\ref{S1sol}),   (\ref{W11sol}-\ref{W21sol}), (\ref{T2ndsol}),
(\ref{W1pW2O2}), (\ref{solf}) and  (\ref{S2sol}),
the  expressions for $W_i$, $S$
and $T_i$ in the vicinity of an extremum of V,  and up to order $\mathcal{O}(R^2)$ are given by
\begin{eqnarray}
 W_1^{\pm}(\f) & =& \frac{1}{\ell} \left[2 + \frac{\Delta_\pm}{6} \f^2
   + \mathcal{O}(\f^3) \right] \,+ \,\frac{2\mathcal{R}_1 -
   \mathcal{R}_2}{6 \ell} \, |\f|^{\frac{2}{\Delta_\pm}} \ [1+
 \mathcal{O}(\f)] \nonumber \\
\label{eq:W1genap} & \hphantom{=}& \,
-\, \frac{\mathcal{R}_1^2-\mathcal{R}_2^2}{24\Delta_{\pm}\ell}|\f|^{\frac{4}{\D_{\pm}}}\log\left|\f\right|[1+\mathcal{O}(\f)
] \, \\
\nonumber &\hphantom{=}& + \,\left[\frac{C_1}{\ell} \,+
  \,\frac{\mathcal{R}_1^2+4\mathcal{R}_2^2-4\mathcal{R}_1\mathcal{R}_2}{144\ell}\right]
|\f|^{\frac{4}{\Delta_{\pm}}}[1+\mathcal{O}(\f)] \, , \\
& & \nonumber \\
 W_2^{\pm}(\f) & = &\frac{1}{\ell} \left[2 +
  \frac{\Delta_\pm}{6} \f^2 + \mathcal{O}(\f^3) \right] \,+
\,\frac{2\mathcal{R}_2-\mathcal{R}_1 }{6 \ell} \,
|\f|^{\frac{2}{\Delta_\pm}} \ [1+ \mathcal{O}(\f)]
\nonumber  \\
\label{eq:W2genap} & \hphantom{=}& \,
+\, \frac{\mathcal{R}_1^2-\mathcal{R}_2^2}{24\Delta_{\pm}\ell}|\f|^{\frac{4}{\D_{\pm}}}\log\left|\f\right|
[1+\mathcal{O}(\f)
] \, \\
\nonumber &\hphantom{=}&  + \, \left[ \frac{C_2}{\ell} +
  \frac{4\mathcal{R}_1^2+\mathcal{R}_2^2-4\mathcal{R}_1\mathcal{R}_2}{144\ell}
  \right]|\f|^{\frac{4}{\Delta_{\pm}}}[1+\mathcal{O}(\f)] \, , \\
& & \nonumber \\
\label{eq:Sgenap} S_{\pm}(\f) & =& \frac{\Delta_{\pm}}{\ell} \f \ [1+ \mathcal{O}(\f)] + \frac{6(C_1+C_2)}{\Delta_{\pm} \ell} \, |\f|^{\frac{4}{\Delta_{\pm}}-1} \ [1+ \mathcal{O}(\f)
]  \\
& & \nonumber \\
\label{eq:T1genap} T_1^{\pm}(\f) & =&  {\mathcal{R}_1 \over \ell^2} \, |\f|^{\frac{2}{\Delta_{\pm}}} [1+ \mathcal{O}(\f) 
] \, + \,   \frac{2\mathcal{R}_1^2-\mathcal{R}_1\mathcal{R}_2}{12 \ell^2}\,|\f|^{\frac{4}{\Delta_{\pm}}}[1+ \mathcal{O}(\f) 
]  \\
& & \nonumber \\
\label{eq:T2genap} T_2^{\pm}(\f) & = & {\mathcal{R}_2 \over
\ell^2} \, |\f|^{\frac{2}{\Delta_{\pm}}} [1+ \mathcal{O}(\f)
] \, + \,
\frac{2\mathcal{R}_2^2-\mathcal{R}_1\mathcal{R}_2}{12 \ell^2}\,|\f|^{\frac{4}{\Delta_{\pm}}}[1+ \mathcal{O}(\f) 
] \,
\end{eqnarray}

Up to now we have not made any distinction between  the $+$ and $-$
branch, but from the expansions (\ref{eq:W1genap}-\ref{eq:T2genap}) we can infer some
some important differences between the two. The main observation is
that, in obtaining  the expressions above, we have systematically
assumed that the {\em leading} terms as $\f \to 0$ in $W_1$, $W_2$ and
$S$ are  given by
\be \label{WSlead}
W_1, W_2 \simeq {2\over \ell} + \ldots, \qquad S\simeq {\Delta_{\pm}
  \over \ell}\f +\ldots
\ee
This requirement  imposes some constraints on the rest of the terms in the
expansion, which  depends on the branch one chooses.
\begin{itemize}
\item {\bf $(-)$-branch, $\Delta_->0$}\\
This is the case the extremum of the potential is a
maximum. The non-analytic subleading terms in
$S$ and $W$ are at  of order
\be
W_{non-analytic} \sim \mathcal{R}|\f|^{1+ (2-\Delta_-)/\Delta_-} ,
\qquad S_{non-analytic} \sim (C_1+ C_2) |\f|^{1 + 2(2-\Delta_-)/\Delta_-}
\ee
and since  $0<\Delta_-< 2$, these are subleading with respect to the
terms in (\ref{WSlead}). Therefore the analysis of the previous
sections  goes through.
\item {\bf $(-)$-branch, $\Delta_-<0$}\\
In this case the extremum of the potential is a
maximum. Because $\Delta_-<0$, the non-analytic subleading terms in
$S$ and $W$, as well as the leading term in $T_{1,2}$  {\em diverge}
and the solution does not exist as an expansion around $\f=0$, unless we set ${\mathcal R}_1 =
{\mathcal R}_2 = C_1 =C_2 = 0$. If  this is  case,  we find the flat IR
solution where $\f = 0$ corresponds to an IR fixed point. This
 shows that the flat IR fixed point cannot be reached in the
presence of curvature, like in the more symmetric $S^4$ case \cite{C}.
\item {\bf $(+)$-branch}\\
Since $\Delta_+> 2$,  it does not make a difference whether the
extremum of $V$ is a maximum or a minimum. The
non-analytic subleading terms in $S$ and $W$ are of the order
\be
W_{non-analytic} \sim \mathcal{R}|\f|^{1-(\Delta_+-2)/\Delta_+} ,
\qquad S_{non-analytic} \sim (C_1+ C_2) |\f|^{1 - 2(\Delta_+-2 )/\Delta_+}
\ee
The non-analytic term in $W$ is subleading, but  the one in  $S$
potentially dominates over the leading term in (\ref{WSlead}). Therefore, in the $+$ branch, the integration constants
controlling the $\f^{4/\Delta_+}$ terms have to obey the constraint
\be
(+)-\text{branch:} \qquad  C_1+ C_2 = 0.
\ee
On the other hand, the combination $C_1-C_2$ is unconstrained.
\end{itemize}

Next, we obtain the  expansions of $\varphi(u)$ and $A_{1,2}(u)$ near
the boundary. This can be achieved by integrating order by order in
$\f$ the first order flow
equations:
\be \label{floweq}
\dot{f}(u) = S(\f), \quad \dot{A}_1 = -2 W_1(\f(u)), \quad \dot{A}_2 =
-2 W_2(\f(u)).
\ee

The result of the integration is, in the $(-)$-branch:
\be
\label{eq:phimapS2} \f(u)  = \f_- \ell^{\Delta_-}\ex^{\Delta_-u / \ell} \left[ 1+ \mathcal{O} \left(\ex^{2u/\ell} \right)  \right]+
 \frac{6(C_1+C_2) \, |\f_-|^{\Delta_+ /
    \Delta_-}}{\Delta_-(4-2 \Delta_-)} \, \ell^{\Delta_+}
\ex^{\Delta_+ u / \ell} \left[ 1+ \mathcal{O} \left(\ex^{2u/\ell}
  \right) \right]  \, ,
  \ee
  \be
\label{eq:A1map} A_1(u) = \bar{A_1} -\frac{u}{\ell} - \frac{\f_-^2 \, \ell^{2 \Delta_-}}{24} \ex^{2\Delta_- u / \ell} \, [1+\mathcal{O}(\ex^{\Delta_-u/\ell})]-
\ee
$$
-\frac{|\f_-|^{2/\Delta_-} \, \ell^2}{24} (2\mathcal{R}_1 - \mathcal{R}_2) \ex^{2u/\ell} \, [1+\mathcal{O}(\ex^{\Delta_-u/\ell})+\mathcal{O}(\ex^{(\Delta_+-\Delta_-)u/\ell})]+$$
$$
 +\frac{1}{192}(\mathcal{R}_1^2-\mathcal{R}_2^2)|\f_-|^{4/\Delta_-} \, \ell^4 \frac{u}{\ell} \, \ex^{4u/\ell} [1+\mathcal{O}(\ex^{\Delta_-u/\ell})]-
$$
$$
 - \frac{1}{8}|\f_-|^{4/\Delta_-} \, \ell^4\ex^{4u/\ell}\left( (C_1+C_2)\frac{\Delta_+}{4-2\Delta_-}+\frac{C_1-C_2}{2} \right. +
$$
$$+
\left. \frac{1}{48}\left[\frac{5}{6}(\mathcal{R}_1^2+\mathcal{R}_2^2)-\frac{4}{3}\mathcal{R}_1\mathcal{R}_2\right]
  -
  \frac{\mathcal{R}_1^2-\mathcal{R}_2^2}{24\D_-}\log\left(\f_-\ell^{\D_-}\right)\right)
+\ldots \, ,
$$
\be
\label{eq:A2map} A_2(u)  = \bar{A_2} -\frac{u}{\ell} - \frac{\f_-^2 \, \ell^{2 \Delta_-}}{24} \ex^{2\Delta_- u / \ell} \, [1+\mathcal{O}(\ex^{\Delta_-u/\ell})]-
\ee
$$
-\frac{|\f_-|^{2/\Delta_-} \, \ell^2}{24} (-\mathcal{R}_1 + 2\mathcal{R}_2) \ex^{2u/\ell} \, [1+\mathcal{O}(\ex^{\Delta_-u/\ell})+\mathcal{O}(\ex^{(\Delta_+-\Delta_-)u/\ell})] -
$$
$$
 - \frac{1}{192}(\mathcal{R}_1^2-\mathcal{R}_2^2)|\f_-|^{4/\Delta_-} \, \ell^4 \frac{u}{\ell} \, \ex^{4u/\ell} [1+\mathcal{O}(\ex^{\Delta_-u/\ell})]-
$$
$$
- \frac{1}{8}|\f_-|^{4/\Delta_-} \, \ell^4\ex^{4u/\ell}\left( (C_1+C_2)\frac{\Delta_+}{4-2\Delta_-}-\frac{C_1-C_2}{2} \right. +
$$
$$
+ \left. \frac{1}{48}\left[\frac{5}{6}(\mathcal{R}_1^2+\mathcal{R}_2^2)-\frac{4}{3}\mathcal{R}_1\mathcal{R}_2\right] + \frac{\mathcal{R}_1^2-\mathcal{R}_2^2}{24\D_-}\log\left(\f_-\ell^{\D_-}\right)\right)  +\ldots \, ,
$$
where $\f_-$, $\bar{A}_1$ and $\bar{A}_2$ are new  integration
constant which parametrize initial conditions for the flow.
According to the discussion above,   we are considering
$\Delta_{\pm } >0$  therefore to be close to $\f=0$ we need $u\to
-\infty$. Therefore, the $\f$-expansion in
(\ref{eq:W1genap}-\ref{eq:T2genap}) becomes an expansion in
$\exp(u/\ell) \ll 1$ in (\ref{eq:phimapS2}-\ref{eq:A2map}). According
to the standard holographic dictionary, the
parameter $\f_-$ represents in the field theory the source of
operator dual to the field $\f$, while the combination $C_1+C_2$
parametrizes the vev of the same operator.

In the $(+)$-branch we have similar expressions, except for the
fact that we have to set $C_1+C_2=0$:
\be
\label{eq:phipapS2} \f(u)  = \f_+ \ell^{\Delta_+}\ex^{\Delta_+ u /
  \ell} \left[ 1+ \mathcal{O} \left( \ex^{2u/\ell} \right) \right]  \,
,
\ee
\be
\label{eq:A1pap} A_1(u)  = \bar{A_1} -\frac{u}{\ell} - \frac{\f_+^2 \, \ell^{2 \Delta_+}}{24} \ex^{2\Delta_+ u / \ell} \, [1+\mathcal{O}(\ex^{\Delta_+u/\ell})]-
\ee
$$  -\frac{|\f_+|^{2/\Delta_+} \, \ell^2}{24}
(2\mathcal{R}_1 - \mathcal{R}_2) \ex^{2u/\ell} \,
[1+\mathcal{O}(\ex^{\Delta_+u/\ell})]
+ \frac{1}{192}(\mathcal{R}_1^2-\mathcal{R}_2^2)|\f_+|^{4/\Delta_+} \, \ell^4 \frac{u}{\ell} \, \ex^{4u/\ell} [1+\mathcal{O}(\ex^{\Delta_+u/\ell})]-
$$
$$
 - \frac{1}{8}|\f_+|^{4/\Delta_+} \, \ell^4\ex^{4u/\ell}\left( \frac{C_1-C_2}{2} \right. +
\left. \frac{1}{48}\left[\frac{5}{6}(\mathcal{R}_1^2+\mathcal{R}_2^2)-\frac{4}{3}\mathcal{R}_1\mathcal{R}_2\right]
  -
  \frac{\mathcal{R}_1^2-\mathcal{R}_2^2}{24\D_+}\log\left(|\f_+|\ell^{\D_+}\right)\right)
+\ldots \, ,
$$
\be
\label{eq:A2pap} A_2(u) = \bar{A_2} -\frac{u}{\ell} - \frac{\f_+^2 \, \ell^{2 \Delta_+}}{24} \ex^{2\Delta_+ u / \ell} \, [1+\mathcal{O}(\ex^{\Delta_+ u/\ell})]-
\ee
$$
 -\frac{|\f_+|^{2/\Delta_+} \, \ell^2}{24} (-\mathcal{R}_1 + 2\mathcal{R}_2) \ex^{2u/\ell} \, [1+\mathcal{O}(\ex^{\Delta_+u/\ell})]
 - \frac{1}{192}(\mathcal{R}_1^2-\mathcal{R}_2^2)|\f_+|^{4/\Delta_+} \, \ell^4 \frac{u}{\ell} \, \ex^{4u/\ell} [1+\mathcal{O}(\ex^{\Delta_+u/\ell})]-
$$
$$
- \frac{1}{8}|\f_-|^{4/\Delta_-} \, \ell^4\ex^{4u/\ell}\left( -\frac{C_1-C_2}{2} \right. +\left. \frac{1}{48}\left[\frac{5}{6}(\mathcal{R}_1^2+\mathcal{R}_2^2)-\frac{4}{3}\mathcal{R}_1\mathcal{R}_2\right] + \frac{\mathcal{R}_1^2-\mathcal{R}_2^2}{24\D_+}\log\left(|\f_+|\ell^{\D_+}\right)\right)  +\ldots \, ,
$$
In this case, $\f_+$ parametrizes  the vev of the operator dual to
$\f$.

From the expressions above, we can obtain the field theory
interpretation of the integration constants $\mathcal{R}_i$:
substituting (\ref{eq:phimapS2}) or  (\ref{eq:phimapS2}) into the
leading order expression for $T_i$ in (\ref{Tlowest}) and comparing
with the definition  (\ref{eq:defT1c}), we find, in either $\pm$ branch:
\be\label{curlyR1}
\mathcal{R}_i = {R^{(\zeta^i)} \over (\f_\pm)^{2/\Delta_{\pm}}} e^{-2\bar{A}_i}.
\ee
Finally, remember that $R^{(\zeta^i)}$ are just placeholders,
 and the physical parameters are the UV scalar curvatures
  $R_{i}^{UV}$ of the metric seen by the UV QFT. We can relate  these
  quantities to (\ref{curlyR1}) by recalling that the metric on which
  the QFT is defined can be read-off from the leading term in the
  near-boundary expansion as $u\to -\infty$,
\be
ds^2 = du^2 + e^{-2u/\ell} (ds^2_{QFT} +\ldots)  =  du^2 + e^{-2u/\ell} \left[
  e^{2\bar{A}_1}\zeta^1_{\alpha\beta} dx^\alpha dx^\beta +
  e^{2\bar{A}_2}\zeta^2_{\alpha\beta} dx^\alpha dx^\beta  + \ldots \right]
\ee
Therefore the QFT metric on each 2-sphere is given by
$e^{2\bar{A}_i}\zeta^i_{\alpha\\beta}$ and its curvature is $R_{UV}^i
= e^{-2\bar{A}_i}R^{(\zeta^i)} $. Therefore equation (\ref{curlyR1})
becomes:
\be
\mathcal{R}_i = {R^{(UV)}_i \over (\f_\pm)^{2/\Delta_{\pm}}}.
\ee
This equation relates the dimensionless curvature parameters
$\mathcal{R}_i$ introduced as integration constants, to the physical
parameters of the UV QFT, namely the curvatures of the two spheres and
the source parameter $\f_-$ (in the $(-)$-branch) or vev parameter
$\f_+$ (in the $(+)$-branch).

For simplicity, one can  make the choice $\bar{A_i} = 0$, to identify the UV
metrics with the fiducial metrics $\zeta^i$.

\section{The on-shell action}
\label{app:FE}

In this appendix, we provide details on the evaluation of the
on-shell action for the flows we consider in this paper.

We start with the action (\ref{eq:action}),
\be
\label{action}S[g_{\mu \nu}, \f] = M_p^3 \int \mathrm{d}u\mathrm{d}^4x
\sqrt{|g|}\left(\mathcal{R}^{(g)}-\frac{1}{2}\partial_a\f \partial^a\f
  - V(\f)\right) + S_{GHY},
\ee
with $\mathcal{R}^{(g)}$ the Ricci scalar for the full metric and $S_{GHY}$ the Gibbons-Hawking-York boundary term. The various terms in the action \eqref{action} are written in terms of the holographic quantities:
\begin{align}
\mathcal{R}^{(g)} & = \frac{1}{2}\dot{\f}^2 + \frac{5}{3} V(\f) \, , \\
\partial_a\f \partial^a\f & = \dot{\f}^2 \, , \\
\sqrt{|g|} & = \ex^{2(A_1+A_2)}\sqrt{|\zeta^{1}||\zeta^{2}|} \, .
\end{align}
where $\zeta^{1,2}$ are the fiducial ($u$-independent) metrics of the
two spheres. The first of the  above identities  is the trace of Einstein's
equation.   Substituting into (\ref{action}) we obtain the  expression:
\be
\label{Sonshell}S_{on-shell} = \frac{2}{3} M_p^3 \, V_{2\times 2}
\int_{UV}^{IR}\mathrm{d}u \, \ex^{2(A_1+A_2)} V(\f(u)) + S_{GHY},
\ee
where $V_{2\times 2}$ is the volume of the two 2-spheres,
\be
 \label{defVol}V_{2\times 2} \equiv \int\mathrm{d}x^4
 \sqrt{|\zeta^{1}||\zeta^{2}|} = Vol(S^1)\times Vol(S^2) =
 \frac{64\pi^2}{R^{\zeta^1}R^{\zeta^2}}.
 \ee

The potential $V(\f)$ can also be expressed as a function of $A_1$ and $A_2$ using \eqref{eq:EOM6bis} and \eqref{eq:EOM7bis}:
\be
\label{VvA}V = -\frac{3}{2}(\ddot{A}_1+\ddot{A}_2) +
\frac{3}{4}(R^{\zeta^1}\ex^{-2A_1} + R^{\zeta^2}\ex^{-2A_2}) -
3(\dot{A}_1+\dot{A}_2)^2,
\ee
so that the on-shell effective action in (\ref{Sonshell}) reads
\begin{align}
\label{Sonshellbis}S_{on-shell} & =  M_p^3 \, V_{2\times 2} \left[\ex^{2(A_1+A_2)}(\dot{A}_1+\dot{A}_2)\right]_{IR}^{UV} + S_{GHY} \\
\nonumber & \hphantom{=} \, + \frac{M_p^3\, V_{2\times 2}}{2}\left[R^{\zeta^1}\int_{UV}^{IR}\mathrm{d}u \, \ex^{2 A_2} + R^{\zeta^2}\int_{UV}^{IR}\mathrm{d}u \, \ex^{2A_1}\right]
\end{align}
Note that the IR contribution to the first term vanishes for the
appropriate regular boundary conditions (see section \ref{sec:IRbc}),
since either $e^{A_1}$ or $e^{A_2}$ vanishes in the IR.

 The GHY term needs also be expressed in terms of $A_1$ and $A_2$. It is given by:
\be
\label{SGHY}S_{GHY} = 2M_p^3\left[\int\mathrm{d}x^4\sqrt{|\gamma|} K \right]^{UV}
\ee
where $\gamma_{\mu \nu}(u)$ is the metric on a S$^2 \times$ S$^2$
slice at a fixed UV value of $u$ and $K$ is the extrinsic curvature of
the slice. They are given by:
\be
K  = -2(\dot{A}_1+\dot{A}_2) \sp
\sqrt{|\gamma|}  = \ex^{2(A_1+A_2)}\sqrt{|\zeta^{1}||\zeta^{2}|} \, .
\ee
This gives
\be\label{SGHYbis} S_{GHY} = -4M_p^3\, V_{2\times 2} \left[\ex^{2(A_1+A_2)}(\dot{A}_1+\dot{A}_2)\right]^{UV}
\ee
Substituting in equation (\ref{Sonshellbis}) we obtain for the on-shell  action:

\begin{align}
\label{Sonshellbisbis-app} S_{on-shell} & = -3 M_p^3 \, V_{2\times 2} \left[\ex^{2(A_1+A_2)}(\dot{A}_1+\dot{A}_2)\right]^{UV}  \\
\nonumber & \hphantom{=} \, + \frac{M_p^3\, V_{2\times 2}}{2}\left[R^{\zeta^1}\int_{UV}^{IR}\mathrm{d}u \, \ex^{2 A_2} + R^{\zeta^2}\int_{UV}^{IR}\mathrm{d}u \, \ex^{2A_1}\right]
\end{align}
Using   the
relation between the volume and curvature of the 2-spheres
(\ref{defVol}),  as well as the definitions  (\ref{eq:defW1c}-\ref{eq:defT1c}),
 equation (\ref{Sonshellbisbis-app})  above can be written in
terms of the superpotentials:
\be
\label{eq:F-app} S_{on-shell}= 32\pi^2M_p^3 \left(3\left[\frac{W_1(\f)+W_2(\f)}{T_1(\f)T_2(\f)}\right]^{UV}+\int_{UV}^{IR}\mathrm{d\f\over S(\f)}\;\left[\frac{1}{T_1(\f)}+\frac{1}{T_2(\f)}\right]\right)
\ee

Remember that the unrenormalized free energy $\mathcal{F}$ is equal to
$-S_{on-shell}$. Inserting the expression of $V_{2\times 2}$
\eqref{defVol} and expressing everything in terms of the first order
superpotentials using the definitions (\ref{eq:defW1c}-\ref{eq:defT1c})  $\mathcal{F}$ can be expressed in the following way:

\be
\label{eq:F_a}\mathcal{F} = -32\pi^2M_p^3 \left(3\left[\frac{W_1(\f)+W_2(\f)}{T_1(\f)T_2(\f)}\right]^{UV}+\int_{UV}^{IR}\mathrm{d}\f\;\left[\frac{1}{S(\f)T_1(\f)}+\frac{1}{S(\f)T_2(\f)}\right]\right)
\ee

\paragraph*{Comparison with the S$^4$ case}
In the S$^4$ case, the expression for the free energy is
given by \cite{C}:

\be
\label{eq:F4} \mathcal{F}_4 = 6 M_p^3 \, V_{4} \left[\ex^{4A}
  \dot{A}\right]^{UV} - \frac{M_p^3\, V_{4}}{2}
R^{\zeta}\int_{UV}^{IR}\mathrm{d}u \, \ex^{2 A} \, ,
\ee
where $R^{\zeta}$ is the curvature of the 4-sphere in the UV (with the appropriate choice of boundary condition for $A$)and $V_4$ its volume. And when $A_1 = A_2$, $R^{\zeta_1} = R^{\zeta_2}$:
\be
\label{eq:Feq} \mathcal{F} = 6 M_p^3 \, V_{2\times 2} \left[\ex^{4A_1}
  \dot{A}_1\right]^{UV} - \frac{M_p^3\, V_{2\times 2}}{2}
2R^{\zeta_1}\int_{UV}^{IR}\mathrm{d}u \, \ex^{2 A_1} \, ,
\ee
where $R^{\zeta_1}$ is the curvature of the 2-spheres in the UV. By considering the 4-sphere with curvature:
\be
\label{R4vR2} R_4 = 2 R^{\zeta_1} \, .
\ee
The previous expression \eqref{eq:Feq} can be written as:
\be
\label{eq:Feq2} \mathcal{F} = 6 M_p^3 \, V_{2\times 2} \left[\ex^{4A_1}
  \dot{A}_1\right]^{UV} - \frac{M_p^3\, V_{2\times 2}}{2} R_4
\int_{UV}^{IR}\mathrm{d}u \, \ex^{2 A_1} \, ,
\ee
which is of the same form as \eqref{eq:F4}. So, up to an overall factor of $2/3$ equal to the ratio of the volumes, the free energy of the solutions where both spheres have equal radius and $A_1 = A_2$ reduces to the free energy of a boundary theory defined on a 4-sphere.

\subsection{The U superpotentials}

We introduce the superpotentials $U_1$ and $U_2$ defined by the differential equation they satisfy:

\be
\label{eqUi_a} SU_i' - W_iU_i = -1
\ee
This equation implies:
\be \label{eqUi2_a}
\frac{\mathrm{d}\f}{S(\f)T_i(\f)} =
-\mathrm{d}\left(\frac{U_i(\f)}{T_i(\f)}\right),
\ee
which makes it possible to express $\mathcal{F}$ as:
\be
\label{eq:Fbis_a}\mathcal{F} = -32\pi^2M_p^3 \left(3\left[\frac{W_1(\f)+W_2(\f)}{T_1(\f)T_2(\f)}\right]^{UV}+\left[\frac{U_1(\f)}{T_1(\f)}+\frac{U_2(\f)}{T_2(\f)}\right]_{IR}^{UV}\right)
\ee
Each equation (\ref{eqUi_a}) defines $U_i$  up to a single
integration constant. Although it may seem that the expression
(\ref{eq:Fbis_a}) depends on which the choice of the particular
solutions for $U_i$, this is clearly not the case: indeed, the
original expression (\ref{eq:F_a}) is unambiguous, and so must be
(\ref{eq:Fbis_a}): in fact, by construction, the integration constant
enters in $U_i$ in such a way that it exactly  cancels when  we take
the difference between the UV and IR contributions, since integrating
back equation (\ref{eqUi2_a}), the two integrated terms term in
(\ref{eq:F_a}) become:
\be \label{eq-diff}
\int_{IR}^{UV}\frac{\mathrm{d}\f}{S(\f)T_i(\f)}  =
\left[\frac{\bar{U}_i(\f)}{T_i(\f)} + {\mathcal B}_i \right]_{UV} - \left[\frac{\bar{U}_i(\f)}{T_i(\f)} + {\mathcal B}_i \right]_{IR}
\ee
where $\bar{U}_i$ are two references solutions and ${\mathcal B}_i$
are the two integration constants of each of equations
(\ref{eqUi_a}). It is clear from the above equation that   ${\mathcal
  B}_i$ cancel out when taking the difference between the UV and IR
contributions.

\paragraph*{Near-boundary expansion}
We can see explicitly how the integration constants ${\mathcal B}_i$
appear in  the near-boundary ($u\to -\infty$) expansion, by  inserting
the UV-expansions of the superpotentials:
\eqref{eq:W1gensol}-\eqref{eq:Sgensol} into \eqref{eqUi_a}: as $u\to
-\infty$ we obtain:

\be
\label{U1UV_a}U_1(\f) \underset{\f \to 0^+} = \ell \left[\frac{1}{2} +
  \left(\mathcal{B}_1 +\frac{2\mathcal{R}_1-\mathcal{R}_2}{12\D_-}
    \log|\f| \right)   |\f|^{2/\Delta_{\pm}}[1+ \ldots ]\right]
\ee
 \be
\label{U2UV_a}U_2(\f) \underset{\f \to 0^+} = \ell \left[\frac{1}{2} +
  \left(\mathcal{B}_2 +\frac{2\mathcal{R}_2-\mathcal{R}_1}{12\D_-}
    \log|\f| \right)  |\f|^{2/\Delta_{\pm}}  [1+ \ldots ]\right]
\ee
where we have omitted higher order terms which vanish as $\f \to 0$.

Up to this point the constants $\mathcal{B}_i$ can be chosen
arbitrarily, however there is a specific choice which is very
convenient, as  it allows us to write the free energy purely as an UV
contribution, as discussed below.

\paragraph*{Boundary condition in the interior} To compute the free-energy \eqref{eq:Fbis}, a boundary condition is required for $U_1$ and $U_2$ in the IR. To derive it we inject \eqref{eq:W1gensol}-\eqref{eq:Sgensol} into \eqref{eqUi} and find:
\begin{align}
\label{U1IR_a}U_1(\f) & \underset{\f \to \f_0^-} = \frac{b_1}{\f-\f_0} + U_0\sqrt{|\f-\f_0|} + \mathcal{O}(|\f-\f_0|) \, , \\
\label{U2IR_a}U_2(\f) & \underset{\f \to \f_0^-} = b_2 +
U_0\sqrt{|\f-\f_0|} + \mathcal{O}(|\f-\f_0|) \, ,
\end{align}
where $b_1$ and $b_2$ are two integration constants and:
\be
\label{defU02}U_0 \equiv \sqrt{\frac{6}{|V'(\f_0)|}}
\ee
Going back to the expression in terms of $A_1$ and $A_2$
\eqref{Sonshellbisbis-app}, we know that $\mathcal{F}$ only depends on
$\mathcal{R}_2/\mathcal{R}_1$, $C$ and the UV cutoff. There is
therefore no dependence on any other independent integration constants
such as $b_1$ and $b_2$. This fixes the way $\mathcal{B}_1$ and
$\mathcal{B}_2$ should respectively depend on $b_1$ and $b_2$. In
general the terms  in \eqref{eq-diff}  have the following expansion:
\begin{align}
\label{U1T1UVIR_a}\left[\frac{U_1(\f)}{T_1(\f)}\right]_{IR}^{UV} & = \frac{\ell^3}{2\mathcal{R}_1}|\f|^{-2/\Delta_{\pm}} + \ell^3 \frac{\mathcal{B}_1}{\mathcal{R}_1}-\frac{3b_1}{V'(\f_0)} + \cdots \, , \\
\label{U2T2UVIR_a}\left[\frac{U_2(\f)}{T_2(\f)}\right]_{IR}^{UV} & =
\frac{\ell^3}{2\mathcal{R}_2}|\f|^{-2/\Delta_{\pm}} + \ell^3
\frac{\mathcal{B}_2}{\mathcal{R}_2}-\frac{b_2}{T_{2,0}} + \cdots \, ,
\end{align}
where we brought out in each case the leading term that depends on
$b_i$. Note in particular that there may be other terms at order
$\mathcal{O}(1)$ that do not depend on $b_i$, as well as subleading
terms that do depend on $b_i$. The conclusion is that:

\begin{align}
\label{B1_a}\mathcal{B}_1(\mathcal{R}_2, \mathcal{R}_1, b_1) &= \mathcal{B}_1(\mathcal{R}_2, \mathcal{R}_1, 0) + \frac{3b_1\mathcal{R}_1}{\ell^3 V'(\f_0)} \, , \\
\label{B2_a}\mathcal{B}_2(\mathcal{R}_2, \mathcal{R}_1, b_2) &= \mathcal{B}_1(\mathcal{R}_2, \mathcal{R}_1, 0) + \frac{b_2\mathcal{R}_2}{\ell^3 T_{2,0}} \, .
\end{align}
In  the expression above, the  ambiguity  in the definition of  $U_1$
and $U_2$  is made explicit in terms of  $b_1$ and $b_2$. As we have
discussed, the choice
of these constants is   irrelevant as long as the evaluation of the
free energy $\mathcal{F}$ is concerned. In the following, we make the
convenient choice $b_1 = b_2 = 0$. This choice enables to express the free energy exclusively in terms of UV quantities:
\be
\label{eq:Fbisbis_a}\mathcal{F}(\Lambda,\mathcal{R}_1,\mathcal{R}_2,C_1,C_2) = -32\pi^2M_p^3 \left[3\frac{W_1(\f)+W_2(\f)}{T_1(\f)T_2(\f)}+\frac{U_1(\f)}{T_1(\f)}+\frac{U_2(\f)}{T_2(\f)}\right]^{UV}
\ee



\subsection{The UV-regulated free energy}
We now evaluate expression (\ref{eq:Fbisbis_a})  at the UV
boundary.  This quantity is divergent, because $e^{A_{i}} \to \infty$
in the UV. Therefore, we regulate the boundary at $u= \ell \log
\epsilon$ with $\epsilon\ll 1$, and introduce a dimensionless
``energy scale'' cut-off,
\be
\label{deflambda-app}\Lambda \equiv \left. \frac{\ex^{\frac{A_1(u)+A_2(u)}{2}}}{\ell (R_1^{UV}R_2^{UV})^{1/4}}\right|_{\frac{u}{\ell}=\log(\e)}
\ee
We now write \eqref{eq:Fbisbis_a} as an expansion in inverse powers of
$\Lambda$, using the UV expansions
\eqref{eq:W1gensol}-\eqref{eq:T2gensol} and
\eqref{eq:phimsolS2}-\eqref{eq:A2msol}

We consider each term of equation  (\ref{eq:Fbisbis_a})  separately,  and start with the first one:
 \be
  \label{expUVF1} 3\frac{W_1(\f)+W_2(\f)}{T_1(\f)T_2(\f)} =
  3\left(\ell \Lambda\right)^{4}\left(\frac{4}{\ell}+\frac{\Delta_-}{3\ell}\f^2(1+\mathcal{O}(\f))+\frac{\mathcal{R}_1+\mathcal{R}_2}{6\ell}\f^{2/\Delta_-}(1+\mathcal{O}(\f))\right.
 \ee
 $$ \left. +\frac{1}{\ell}\left(C_1+C_2 + \frac{5(\mathcal{R}_1-\mathcal{R}_2)^2+2\mathcal{R}_1\mathcal{R}_2}{144}\right)\f^{4/\Delta_-}(1+\mathcal{O}(\f)+\mathcal{O}(\mathcal{R})+\mathcal{O}(C))\right)
 $$
In the expression above, and in all the expression that follow, it is
understood that $\f$
is evaluated on the regulated boundary, even when the argument
$\epsilon$ is omitted.  We remind the reader here  that the $\mathcal{O}(\mathcal{R})$ and
$\mathcal{O}(C)$ always come with the appropriate power of $\f$, that
is respectively $\f^{2/\Delta_-}$ and $\f^{4/\Delta_-}$.

The leading terms in equations \eqref{eq:phimsolS2}-\eqref{eq:A2msol} allow to express
$\f$ in terms of $\Lambda$,
\be
\label{fvL} \f = \frac{\Lambda^{-\Delta_-}}{(\mathcal{R}_1\mathcal{R}_2)^{\D_-/4}}\left(1 + \mathcal{O}(\Lambda^{-2})\right) 
 \, ,
\ee
where we have used the definition of the UV ``dimensionless''
curvatures ${\cal R}_i  = R^{UV}_i \f_-^{-2/\Delta_-} $.
With equation (\ref{fvL}),   the expression  (\ref{expUVF1}) becomes:
\be
 \label{expUVF1bis} 3\frac{W_1(\f)+W_2(\f)}{T_1(\f)T_2(\f)} = 3\ell^3 \left[4\Lambda^4\left(1+\frac{\Delta_-}{12(\mathcal{R}_1\mathcal{R}_2)^{\D_-/2}}\Lambda^{-2\Delta_-}(1+\mathcal{O}(\Lambda^{-\Delta_-})+ \mathcal{O}(\L^{-(\D_+ - \D_-)})\right)\right. \ee
$$
+\frac{\mathcal{R}_1+\mathcal{R}_2}{6(\mathcal{R}_1\mathcal{R}_2)^{1/2}}\Lambda^2
\left(1+\mathcal{O}(\Lambda^{-\Delta_-})+ \mathcal{O}(\L^{-(\D_+ - \D_-)})\right)+
$$
$$
  \left. \left(\frac{C_1+C_2}{\mathcal{R}_1\mathcal{R}_2}+
      \frac{5(\mathcal{R}_1-\mathcal{R}_2)^2+2\mathcal{R}_1\mathcal{R}_2}{144\mathcal{R}_1\mathcal{R}_2}\right)\left(1+\mathcal{O}(\Lambda^{-\Delta_-})+\mathcal{O}(\L^{-(\D_+
      - \D_-)})
  \right)
  \right]
$$

We  now consider the last  two terms in \eqref{eq:Fbisbis_a}. Using
the expansions (\ref{U1UV_a}-\ref{U2UV_a}), as well as equation
(\ref{fvL}), they have the following expansions:
\begin{align}
\label{expUVF2} \frac{U_1(\f)}{T_1(\f)} &=
\frac{\ell^3}{2}\left(\frac{\mathcal{R}_2}{\mathcal{R}_1}\right)^{1/2}\left(\Lambda^2
  +
  \frac{1}{(\mathcal{R}_1\mathcal{R}_2)^{1/2}}\left(2\mathcal{B}_1+\frac{\mathcal{R}_2-\mathcal{R}_1}{8}
    - \frac{2\mathcal{R}_1-\mathcal{R}_2}{6}\log\left(\L
      \left({\mathcal R}_1{\mathcal R}_2\right)^{1/4}\right)\right) \right. \times \\
\nonumber &\hphantom{=} \left(1+\mathcal{O}(\Lambda^{-\Delta_-})+\mathcal{O}(\L^{-(\D_+ - \D_-)})
\right) \Big)  \, , \\
\label{expUVF3} \frac{U_2(\f)}{T_2(\f)} &= \frac{\ell^3}{2}\left(\frac{\mathcal{R}_1}{\mathcal{R}_2}\right)^{1/2}\left(\Lambda^2  + \frac{1}{(\mathcal{R}_1\mathcal{R}_2)^{1/2}}\left(2\mathcal{B}_2+\frac{\mathcal{R}_1-\mathcal{R}_2}{8} - \frac{2\mathcal{R}_2-\mathcal{R}_1}{6}\log\left(\L \left({\mathcal R}_1{\mathcal R}_2\right)^{1/4}\right)\right) \right. \times \\
\nonumber &\hphantom{=}
\left.(1+\mathcal{O}(\Lambda^{-\Delta_-})+\mathcal{O}(\L^{-(\D_+ -
    \D_-)}) \right)
\Big)  \, .
\end{align}

Adding together all terms (\ref{expUVF1}), (\ref{expUVF2}) and (\ref{expUVF3}),  we obtain the final expression for the expansion of the regularized free energy as $\Lambda \gg 1$:
\begin{align}
\label{expUVFbisbis}\mathcal{F}& = -32\pi^2M_p^3\ell^3
\left[12\Lambda^4\left(1+\frac{\Delta_-}{12(\mathcal{R}_1\mathcal{R}_2)^{\D_-/2}}\L^{-2\Delta_-}(1+
    \ldots )
  \right)\right. \\
 \nonumber
 &\hphantom{=}+\left(\left(\frac{\mathcal{R}_1}{\mathcal{R}_2}\right)^{1/2}+\left(\frac{\mathcal{R}_2}{\mathcal{R}_1}\right)^{1/2}\right)\Lambda^2(1+ \ldots
 ) \\
 \nonumber &\hphantom{=} + \frac{1}{\mathcal{R}_1\mathcal{R}_2} \left(\frac{\mathcal{R}_1^2+\mathcal{R}_2^2-4\mathcal{R}_1\mathcal{R}_2}{12} \log\left(\L \left({\mathcal R}_1{\mathcal R}_2\right)^{1/4}\right)+\frac{4(\mathcal{R}_1-\mathcal{R}_2)^2+\mathcal{R}_1\mathcal{R}_2}{24} \right.  \\
  \nonumber & \hphantom{=}
  \left.+3\left(C_1+C_2\right)  +
    \mathcal{B}_1\mathcal{R}_2+\mathcal{B}_2\mathcal{R}_1 \Big)\left(1+ \ldots 
        ) \right)
    \right],
\end{align}
where the ellipsis represent subleading terms which are either
$\mathcal{O}(\Lambda^{-\Delta_-})$ or $\mathcal{O}(\L^{-(\D_+ -
  \D_-)})$.

Notice that  the UV-divergent part, (terms in
(\ref{expUVFbisbis}) proportional to $\Lambda^4,
\Lambda^2$ and $\log\Lambda$) is universal and  depends only
on the  $\mathcal{R}_1$ and $\mathcal{R}_2$ but
not on vev parameters $C_i, \mathcal{B}_i$. The latter quantities give
a finite contribution, given by the last line of equation
(\ref{expUVFbisbis}).
This implies that if we take two   distinct solutions with the same UV
boundary conditions (i.e. two solutions with the same UV curvatures
${\cal R}_i$ but
distinct sets of vevs $(C_1,C_2,\mathcal{B}_1, \mathcal{B}_2)$ and $(\tilde{C}_1,\tilde{C}_2,\tilde{\mathcal{B}}_1, \tilde{\mathcal{B}}_2)$, their  free energy
{\em difference} has a finite limit $\Lambda \to \infty$, to which  only the
difference in the last line in equation (\ref{expUVFbisbis}) contributes, and given by:
\begin{equation}
\label{DF-app} \Delta\mathcal{F} =
-\frac{32\pi^2M_p^3\ell^3}{\mathcal{R}_1\mathcal{R}_2}\,\left[
  3\Big((C_1+C_2)-(\tilde{C}_1+\tilde{C}_2)\Big) + (\mathcal{B}_1-\tilde{\mathcal{B}}_1)\mathcal{R}_2+(\mathcal{B}_2-\tilde{\mathcal{B}}_2)\mathcal{R}_1
   \right] ,
\end{equation}

\subsection{Conformal case} If there is no scalar field, or if the
latter is stuck at an extremum of the potential  the same expression,  \eqref{eq:Fbisbis_a} can be used after replacing the superpotentials by their expression in terms of $A_1$ and $A_2$:
\be
\label{eq:Fnof_a}
\mathcal{F}
= -32\pi^2M_p^3
\left[-6\frac{\dot{A}_1(u)+\dot{A}_2(u)}{R^{\zeta_1}R^{\zeta_2}\ex^{-2(A_1(u)+A_2(u))}}+\frac{U_1(u)}{R^{\zeta_1}\ex^{-2A_1(u)}}+\frac{U_2(u)}{R^{\zeta_2}\ex^{-2A_2(u)}}\right]^{\ln(\e)},
\ee
where now $U_1(u)$ and $U_2(u)$ are solutions of the following ODEs:
\be
\label{eqUiu_a} \dot{U}_i +2\dot{A}_i U_i = -1.
\ee

The expression of $\mathcal{F}$  in the limit where $\Lambda\to +\infty$ (the equivalent of \eqref{expUVFbisbis}) reads:
\be
\label{expUVFu}\mathcal{F} = -32\pi^2M_p^3\ell^3 \left[12\Lambda^4 +\left(\left(\frac{\mathcal{R}_1}{\mathcal{R}_2}\right)^{1/2}+\left(\frac{\mathcal{R}_2}{\mathcal{R}_1}\right)^{1/2}\right)\Lambda^2 +\right.
\ee
$$
 + \frac{1+(\mathcal{R}_2/\mathcal{R}_1)^2-4\mathcal{R}_2/\mathcal{R}_1}{12 \mathcal{R}_2/\mathcal{R}_1} \log\left(\L \left(\mathcal{R}_2\mathcal{R}_1\right)^{1/4}\right) \left(1+\mathcal{O}(\Lambda^{-2})\right)
$$
$$ \left.+ \left(\frac{\mathcal{B}_1}{\mathcal{R}_1}+\frac{\mathcal{B}_2}{\mathcal{R}_2}+
\frac{4(1-(\mathcal{R}_2/\mathcal{R}_1))^2+\mathcal{R}_2/\mathcal{R}_1}{24 \mathcal{R}_2/\mathcal{R}_1}\right)
\left(1+\mathcal{O}(\Lambda^{-2})\right)\right].
$$
As in the RG-flow case, the divergent terms of order $\Lambda^4$,
$\Lambda^2$ and $\log\Lambda $ are universal, as they depend only on
$\mathcal{R}_2$ and $\mathcal{R}_1$ (and in fact, only on their ratio,
as expected because of conformal invariance). The last line in
expression (\ref{expUVFu})  contains
the finite terms, and the dependence on the vevs
$\mathcal{B}_i$. Notice that there is no explicit contribution from the vev
$C$ (defined in (\ref{A1-A2}), and  corresponding in the
non-conformal case to $C_1-C_2$). This is similar to the running scalar
field case, where the free energy (\ref{DF-app}) depends only on the sum of
the $C_i$'s but not on their difference it was the case in the

From equation  (\ref{expUVFu}) The free energy difference between two solutions with same ratio
$\mathcal{R}_2/\mathcal{R}_1$ is given then given by:
\be
\label{DFu_a} \Delta\mathcal{F} = -\frac{32\pi^2M_p^3\ell^3}{\mathcal{R}_2/\mathcal{R}_1}\,\left[\frac{\mathcal{R}_2}{\mathcal{R}_1}\left(\frac{\mathcal{B}_1}{\mathcal{R}_1}-\widetilde{\left(\frac{\mathcal{B}_1}{\mathcal{R}_1}\right)}\right)+ \frac{\mathcal{B}_2}{\mathcal{R}_1}-\widetilde{\left(\frac{\mathcal{B}_2}{\mathcal{R}_1}\right)}\right]
\ee
\section{The vev of the stress-energy tensor}

\label{sec:Tij}

We explain in this appendix the steps that lead to the expression for the vev of the stress-energy tensor of the boundary QFT \eqref{Tab}-\eqref{Taa}.

\subsection{CFT case}

We start by deriving the expression for the stress-energy tensor in the case where the boundary theory is a CFT \eqref{Tabcft}-\eqref{Taacft}, that is when we set the scalar field $\f$ to a constant. As explained in Skenderis et al., 2000, the vev of the stress-energy tensor can be directly related to a quantity that appears in the Fefferman-Graham expansion of the metric near the boundary. For an asymptotically AdS space-time the metric near the boundary can be brought into the form:
\be
\label{ds2FG} \mathrm{d}s^2 = \ell^2 \left[\frac{\mathrm{d}u^2}{u^2} + \frac{1}{u^2}g_{ij}(u^2,x)\mathrm{d}x^i\mathrm{d}x^j\right] \, ,
\ee
where $g_{ij}(u^2,x)$ has the following expansion when $u\to 0$:
\be
\label{gijFG} g_{ij}(u^2,x)=g^{(0)}(x)+u^2 g^{(2)}(x)+u^4[ g^{(4)}(x)+\log u^2~
h^{(4)}(x)]+\cdots \, ,
\ee
where $g^{(0)}(x)$ corresponds to the boundary condition for the metric.
The equations of motion determine recursively the functions
$g^{(2n)}$ and $h^{(4)}$ in terms of $g^{(0)}(x)$ except for $g^{(4)}$.
This is in accordance with the fact that the
second order equations of motion have two independent bulk
solutions. The two independent functions are $g^{(4)}$ and $g^{(0)}$.
$g^{(4)}$  turns out to be related to the expectation
value of the stress-energy  tensor in the dual field theory, \cite{HS}.
\index{stress (energy) tensor!vev}
After solving the equations of motion recursively in powers of $\rr$, $g^{(2n)}$ will
be a functional of $g^{(0)}(x)$ involving $2n$ derivatives.
The logarithmic term proportional to $h^{(4)}(x)$
is determined by $g^{(0)}$ and turns out to be the metric
variation of the conformal anomaly of the dual field theory.

The general expression for the various terms in \eqref{gijFG} in the case of a 4-dimensional boundary is found to be:
\be
\label{gij2} g^{(2)}_{ij}={1\over 2}R_{ij}-{1\over 12}R~g^{(0)}_{ij} \, ,
\ee
\be
\label{gij4} g^{(4)}_{ij}={1\over 8}g^{(0)}_{ij}\left[({\rm Tr} g^{(2)})^2
-{\rm Tr}[(g^{(2)})^2]\right]
+{1\over 2}(g^{(2)})^2_{ij}-{1\over 4}g^{(2)}_{ij}({\rm Tr}
g^{(2)})+T_{ij} \, ,
\ee
\be
\label{hij4} h^{(4)}_{ij}= {1\over 16\sqrt{g^{(0)}}}{\delta \over \delta
g^{(0),ij}}~\int d^4x \sqrt{g^{(0)}} \left[R_{ij}R^{ij}-{1\over 3}R^2\right] \, ,
\ee
where $T_{ij}(x)$ is an ``integration constant" satisfying
\be \nabla^iT_{ij}=0\sp {T_i}^i= -{1\over 4}\left[(\mathrm{Tr} g^{(2)})^2-\mathrm{Tr}[(g^{(2)})^2]\right] \, .
\label{Tijap}\ee
where the covariant derivative is taken with respect to $g^{(0)}$. It turns out that it is proportional to the vev of the stress-energy tensor of the boundary CFT:
\be
\label{vevT} \left<T_{ij}\right> = 4(M_p \ell)^3 T_{ij} \, ,
\ee
with $\ell$ the AdS length. The vev \eqref{Tabcft}-\eqref{Taacft} is then found by comparing the above expressions in the case of a boundary CFT defined on the product manifold S$^2\times$S$^2$, with the near-boundary expansion of the metric obtained by solving the equations of motion \eqref{eq:EOM1cft}-\eqref{eq:EOM3cft} for the ansatz \eqref{eq:metricS2}. The latter is given by expanding $\ex^{2A_1}$ and $\ex^{2A_2}$, where the expansions for $A_1$ and $A_2$ are given by \eqref{eq:A1msolu-i} and \eqref{eq:A2msolu-i}, respectively, where we set $\bar{A}_{1}$ and $\bar{A}_{2}$ to 0. We obtain the following expression for $g_{ij}^{(4)}$:

\be
\label{gij4bis} g_{ij}^{(4)} =  \left( \begin{array}{cc}
 \frac{\left(\frac{11}{4}(R_1^{UV})^2 - \frac{1}{4}(R_2^{UV})^2-2 R_1^{UV} R_2^{UV}-72 \frac{C}{\ell^2}\right)\zeta^1}{3\times 96} & 0  \\
0 &  \frac{\left(\frac{11}{4}(R_2^{UV})^2 - \frac{1}{4}(R_1^{UV})^2-2 R_1^{UV} R_2^{UV}+72 \frac{C}{\ell^2}\right)\zeta^2}{3\times 96}   \end{array} \right) \, .
\ee
Subtracting
\be
\label{g2term}{1\over 8}g^{(0)}_{ij}\left[({\rm Tr} g^{(2)})^2 -{\rm Tr}[(g^{(2)})^2]\right]+{1\over 2}(g^{(2)})^2_{ij}-{1\over 4}g^{(2)}_{ij}({\rm Tr}g^{(2)}) =
\nonumber \frac{(R_1^{UV}+R_2^{UV})^2}{6\times 96} \left( \begin{array}{cc}
 \zeta^1 & 0  \\
0 &  \zeta^2   \end{array} \right) \, .
\ee
gives the expression for $T_{ij}$:
\be
\label{Tijapbis} T_{ij} = \left( \begin{array}{cc}
 \frac{\left(\frac{3}{4}(R_1^{UV})^2 - \frac{1}{4}(R_2^{UV})^2- R_1^{UV} R_2^{UV}-24 \frac{C}{\ell^2}\right)\zeta^1}{96} & 0  \\
0 &  \frac{\left(\frac{3}{4}(R_2^{UV})^2 - \frac{1}{4}(R_1^{UV})^2- R_1^{UV} R_2^{UV}+24 \frac{C}{\ell^2}\right)\zeta^2}{96}   \end{array} \right) \, .
\ee
from which \eqref{Taacft}-\eqref{Tcft-traceless} is derived.
The constant $C$ is the vev parameter  appearing as  in  in (\ref{A1-A2}).

\subsection{With a scalar perturbation}

We now reintroduce a scalar operator in the boundary theory, dual to the field $\f$ in the holographic description. Because the scalar vev vanishes in the (+)-branch, we consider the (-)-branch here.

The same formula \eqref{gij4} can be used to derive the vev of the stress-energy tensor. From the fact that the $g^{(2)}$-dependent term depends only on $\f$ through its spatial derivatives on constant $u$ slices, \cite{papad}, which are null in the case we consider, and using the expansion of $A_1$ and $A_2$ \eqref{eq:A1msol}-\eqref{eq:A2msol} to compute $g^{(4)}$, it is manifest that the scalar vev should contribute to the vev of the stress-energy tensor in the following way:
\be
T_{ij}=T^{\mathcal R}_{ij}+T^{C}_{ij}
\ee
with
\be
\label{Tijapf} \frac{T^{\mathcal R}_{ij}}{|\f_-|^{\frac{4}{\Delta_-}}} = {1\over 384}\left( \begin{array}{cc}
{\left({3}\mathcal{R}_1^2 - \mathcal{R}_2^2- 4\mathcal{R}_1 \mathcal{R}_2\right)\zeta^1_{ij}} & 0  \\
0 &  {\left({3}\mathcal{R}_2^2 - \mathcal{R}_1^2- 4\mathcal{R}_1 \mathcal{R}_2\right)\zeta^2_{ij}} \end{array} \right),
\ee

\be
\label{Tijapf2} \frac{T^C_{ij}}{|\f_-|^{\frac{4}{\Delta_-}}} = -{1\over 4}\left( \begin{array}{cc}
{\left(\frac{\D_+}{4-2\D_-}(C_1+C_2) + \frac{C_1-C_2}{2}\right)\zeta^1_{ij}} & 0  \\
0 &  {\left(\frac{\D_+}{4-2\D_-}(C_1+C_2) -\frac{C_1-C_2}{2}\right)\zeta^1_{ij}} \end{array} \right),
\ee

from which \eqref{Tab}-\eqref{Taa} is derived.

\subsection{The vev of the stress-energy tensor on S$^4$}

We derive in this subsection the expression for the vev of the stress-tensor for a boundary theory defined on the maximally symmetric space S$^4$ (or dS$^4$ in Lorentzian signature), in the case where there is no scalar operator (that is when the dual bulk space-time is AdS$^5$).

We follow the same steps as above and use the expression derived in \ref{FGgen} for $g_{ij}^{(4)}$ in the case of S$^4$:
\be
 g^{(4)}_{ij} = -\frac{1}{8}\frac{R^2}{144} g^{S^4}_{ij} + T_{ij} \, .
\ee
$T_{ij}$ is found using the other expression of $g^{(4)}_{ij}$ in terms of $A$ given by \eqref{eq:metric}:
\be
g_{ij}(u,x) = g^{S^4}_{ij}(x)\left(1 - \frac{R\ell^2}{24} \ex^{2u/\ell} + \frac{R^2\ell^4}{48^2}\ex^{4u/\ell}\right) \, .
\ee
Upon a change of variable $\ex^{u/\ell} = \tilde{u}/\ell$:
\be
g_{ij}(\tilde{u}^2,x) = g^{S^4}_{ij}(x)\left(1 - \frac{R}{24} \tilde{u}^2 + \frac{R^2}{48^2}\tilde{u}^4\right) \, .
\ee
So we identify:
\be
g_{ij}^{(4)}(x) = \frac{1}{16} \frac{R^2}{144} g^{S^4}_{ij}(x) \, ,
\ee
and obtain for the stress tensor vev
\be
T_{ij} = \frac{R^2}{48 \times 16} g^{S^4}_{ij}(x) \, .
\ee

\section{General product of spheres}
In this appendix we generalize our formalism to an arbitrary product
of spheres.

\subsection{Fefferman-Graham expansion}

\label{FGgen}

We derive in this appendix expressions for $g_{ij}^{(2)}$, $h_{ij}^{(4)}$ and the $g^{(2)}$-dependent term of $g_{ij}^{(4)}$ in the case where $g^{(0)}_{ij}$ describes a product of $n$ Einstein manifolds with dimension $d_i$ and with curvatures $k_i$, $i=1,\cdots,n$
\be
g^{(0)}_{ij}=\left(\begin{matrix} g^{1}_{ij}& & &\\
& g^{2}_{ij}&&&\\
&&\ddots &\\
&&& g^n_{ij}\end{matrix}\right)\sp R_{ij}=\left(\begin{matrix} R^{1}_{ij}& & &\\
& R^{2}_{ij}&&&\\
&&\ddots &\\
&&& R^n_{ij}\end{matrix}\right)=\left(\begin{matrix} k_1~g^{1}_{ij}& & &\\
& k_2~g^{2}_{ij}&&&\\
&&\ddots &\\
&&& k_n~g^n_{ij}\end{matrix}\right) \, ,
\ee
so that the square of the Ricci tensor and the Ricci scalar are given by:
\be
R=\sum_{i=1}^n d_ik_i\sp R_{kl}R^{kl}=\sum_{i=1}^n d_ik_i^2
\ee
\be
R^2_{ij}=\left(\begin{matrix} k_1^2~g^{1}_{ij}& & &\\
&k_2^2~g^{2}_{ij}&&&\\
&&\ddots &\\
&&& k_n^2~g^n_{ij}\end{matrix}\right) \,
\ee
and

\begin{eqnarray}
g^{(2)}_{ij}&=& {1\over 2}R_{ij}-{1\over 12}R~g^{(0)}_{ij} \\
\nonumber &=& \frac{1}{2}\left(\begin{matrix} \left(k_1-{R\over 6}\right)g^{1}_{ij}& & &\\
& \left(k_2-{R\over 6}\right)g^{2}_{ij}&&&\\
&&\ddots &
\end{matrix}\right)
\end{eqnarray}

\begin{eqnarray}
g^{(4)}_{ij}&=& {1\over 8}g^{(0)}_{ij}\left[({\rm Tr} g^{(2)})^2
-{\rm Tr}[(g^{(2)})^2]\right]
+{1\over 2}(g^{(2)})^2_{ij}-{1\over 4}g^{(2)}_{ij}({\rm Tr}
g^{(2)})+T_{ij} \\
\nonumber &=& \frac{1}{8}\left(\begin{matrix} \left(k_1^2-\frac{(8-d)R}{6} k_1+\frac{(d^2 - 13d + 52)R^2}{144}-{R_{kl}R^{kl}\over 4}\right)g^{1}_{ij}& & \\
&&\ddots &
\end{matrix}\right) + T_{ij}
\end{eqnarray}

\begin{eqnarray}
h^{(4)}_{ij}&=& {1\over 8}\left[\left(R^2_{ij}-{g^{(0)}_{ij}\over 4}R_{kl}R^{kl}\right)-{1\over 3}R\left(R_{ij}-{g^{(0)}_{ij}\over 4}R\right)\right] \\
\nonumber &=& \frac{1}{8}\left(\begin{matrix} \left(k_1^2-{R\over 3}k_1+\left({R^2\over 12}-{R_{kl}R^{kl}\over 4}\right)\right)g^{1}_{ij}& & &\\
& \left(k_2^2-{R\over 3}k_2+\left({R^2\over 12}-{R_{kl}R^{kl}\over 4}\right)\right)g^{2}_{ij}&&&\\
&&\ddots \, .
\end{matrix}\right)
\end{eqnarray}

Note that $h_{ij}^{(4)}$ depends on general on second derivatives of $R$ and $R_{ij}$ on the boundary, which vanish in this case.

\subsection{The Efimov spiral for a general product of spheres}

\label{sec:genprod}

We consider in this appendix the general case of a slicing by $S_1^{d_1}\times S_2^{d_2} \times \cdots \times S_n^{d_n}$, n spheres with respective dimension $d_1, d_2, \cdots, d_n$ and $\sum_{k=1}^n d_k = d$, in the case where there is no scalar field.

We suppose only the sphere 1 shrinks in the interior, and that the boundary curvature sources satisfy $(1/d_1)\mathcal{R}_1 = (1/d_2)\mathcal{R}_2$. This corresponds to setting the source for $A_1-A_2$ to 0 (Note that if one of the spheres is flat, the condition is that the curvature of the other sphere be 0. It is the case in particular if one of the spheres is of dimension 1). All the other spheres are supposed to be such that $A_1-A_k$ has a non-vanishing source for $k>2$. \\
We assume that the argument of \ref{sec:reg} can be generalized so that near the IR end-point $A_1 \sim \log((u_0-u)/\ell)$. Also, we consider a situation similar to the one we considered in the paragraph about the Efimov spiral in \ref{spiralcft}:

\begin{itemize}
\item $A_1 - A_2$ is infinitesimal. We denote:
\be
\label{defegen}\e = A_1 - A_2
\ee

\item We are away from the UV, so that $\dot{A}_1 \sim 1/(u-u_0)$, but not too close to the IR end-point where we know that $A_1 - A_2 \to -\infty$. The precise condition is that $\alpha_{IR} \ll u_0-u \ll \alpha_{UV}$, where $\alpha_{IR}$ and $\alpha_{UV}$ respectively refer to the radius of the sphere 2 in the IR and in the UV:
\begin{eqnarray}
\label{aUVgen} \a_{UV} &=& \sqrt{\frac{2\ell^2}{\mathcal{R}_2}} = \sqrt{\frac{2}{\mathcal{R}_2(u_0=0)}}\, \ell\,\ex^{u_0/\ell} \, , \\
\label{aIRgen} \a_{IR} &=& \lim_{u\to u_0^-} \sqrt{\frac{2}{\ex^{-2u/\ell} T_2(u)}} =   \sqrt{\frac{2}{T_{2,0}\ell^2}} \, \ell \, \ex^{u_0/\ell}  \, .
\end{eqnarray}

\item Here we also require that all the derivatives of $A_k$ for $k > 2$ are negligible compared to the corresponding derivatives for $A_1$. This hypothesis is consistent because for $k > 2$ we supposed that the source of $A_1-A_k$ does not vanish, and for every $j \neq 1$, $A_j$ tends to a constant in the IR. Therefore there should be an interval of values of $u$ for which $A_1\sim A_2$ but the derivatives of every other scale factors are negligible.
\end{itemize}

We now solve the equations of motion \eqref{eq:EOM1bis},\eqref{eq:EOM4bis} and \eqref{eq:EOM5bis} within this set of hypotheses. We first write \eqref{eq:EOM4bis} for $i=1$ and $j=2$:
\be
\label{317i1j2} \ddot{A_1} + \dot{A_1} \sum_{k=1}^n d_k \dot{A_k} - \frac{1}{d_1} \mathrm{e}^{-2A_1} R^{\zeta^1} = \ddot{A_2} + \dot{A_2} \sum_{k=1}^n d_k \dot{A_k} - \frac{1}{d_2} \mathrm{e}^{-2A_2} R^{\zeta^2} \, .
\ee
This gives for $\e$:
\be
\label{eqegen1}\ddot{\e} + \dot{\e}\sum_kd_k\dot{A}_k + \frac{2}{d_1}T_1\e = 0 \, ,
\ee
where $T_i = R^{\zeta_i}\ex^{-2A_i}$. To express $T_1$ we first multiply \eqref{eq:EOM4bis} for $i=1$ and general $j$ by $d_j$ and sum over $j$:
\be
\label{sum317} d\ddot{A}_1 + d\dot{A}_1\sum_kd_k\dot{A}_k-\frac{d}{d_1}T_1 = \sum_jd_j\ddot{A}_j + \left(\sum_kd_k\dot{A}_k\right)^2 - \sum_jT_j \, .
\ee
Multiplying by $2/d$:
\be
\label{sum317bis} 2\ddot{A}_1 + 2\dot{A}_1\sum_kd_k\dot{A}_k-\frac{2}{d_1}T_1 = \frac{2}{d}\sum_jd_j\ddot{A}_j + \frac{2}{d}\left(\sum_kd_k\dot{A}_k\right)^2 - \frac{2}{d}\sum_jT_j \, ,
\ee
where $(2/d)\sum_jT_j$ is given by \eqref{eq:EOM4bis}:
\be
\label{316}-\frac{2}{d}\sum_jT_j = \frac{1}{d} \sum_{i,j}d_id_j(\dot{A}_i-\dot{A}_j)^2 + 2\left(1-\frac{1}{d}\right)\sum_kd_k\ddot{A}_k \, .
\ee
Substituting into \eqref{sum317bis} finally gives the general expression of $T_1$:
\be
\label{T12}T_1 = d_1\ddot{A}_1 - d_1\sum_kd_k\ddot{A}_k + d_1\dot{A}_1\sum_kd_k\dot{A}_k - \frac{d_1}{d} \left(\sum_kd_k\dot{A}_k\right)^2 -\frac{d_1}{2d}\sum_{i,j}d_id_j(\dot{A}_i-\dot{A}_j)^2 \, .
\ee

We now use the hypothesis that we can ignore the derivatives of $A_k$ for $k>2$ and that $A_1\sim A_2$ and $\dot{A}_1 \sim 1/(u-u_0)$. At leading order the terms in \eqref{eqegen1} are then given by:
\begin{eqnarray}
\sum_kd_k\dot{A}_k &=& \frac{d_1+d_2}{u-u_0} + \cdots \, , \\
\frac{2}{d_1}T_1 &=& 2\frac{d_1+d_2-1}{(u-u_0)^2} + \cdots + \mathcal{O}(\e) \, ,
\end{eqnarray}
where the dots refer to subleading terms in the expansion in $u-u_0$. Substituting into \eqref{eqegen1} gives the general equation verified by $\e$:
\be
\label{eqegen}\ddot{\e} + \frac{d_1+d_2}{u-u_0}\dot{\e} + 2\frac{d_1+d_2-1}{(u-u_0)^2}\e = 0 \, .
\ee
If $d_1+d_2 \leq 9$, the solution is:
\be
\label{esolgen}(A_1-A_2)(u) \sim \left(\frac{u_0-u}{\alpha}\right)^{-(d_1+d_2-1)/2} \sin\left(\frac{\sqrt{|(d_1+d_2-1)(d_1+d_2-9)|}}{2}\ln\left(\frac{u_0-u}{\alpha}\right) + \phi \right) \, ,
\ee
where $\alpha$ and $\phi$ a real constants. Proceeding similarly to \ref{spiralcft} we find the Efimov spiral to be described by:
\begin{eqnarray}
\label{R2sR1vsgen} \frac{d_1\mathcal{R}_2}{d_2\mathcal{R}_1}-1 & =
{K_{IR} \over K_R} \; \frac{\sin\left(\phi_R + \frac{\sqrt{|(d_1+d_2-1)(d_1+d_2-9)|}}{4}s\right)}{\sin(\phi_R - \phi_C)}\; \ex^{-((d_1+d_2-1)/4)\;s} \, , \\
\label{Cvsgen} \frac{C}{\mathcal{R}_1^2} & = {K_{IR} \over K_C} \; \frac{\sin\left(\phi_C + \frac{\sqrt{|(d_1+d_2-1)(d_1+d_2-9)|}}{4}s\right)}{\sin(\phi_C - \phi_R)}\; \ex^{-((d_1+d_2-1)/4)\;s} \, ,
\end{eqnarray}
with $s =  \ln\left(\left(\alpha_{UV}/\alpha_{IR}\right)^2\right)$ and $C$ the vev for $A_1-A_2$. The amplitudes $K_R,K_C$ and the phases $\phi_R,\phi_C$ are real numbers. Note that for $d_1+d_2 \geq 9$ the sinus should be replaced by a hyperbolic sinus, so that the spiral reduces to a line.

 If  $d_1+d_2 \geq 9$, the independent solutions of equation
 (\ref{eqegen}) are real power-laws:
\be \label{realexp}
(A_1-A_2)(u) \sim  c_\pm (u_0 - u)^{-\delta_\pm}  \quad \delta_{\pm} =  {d_1 + d_2 -1  \over 2} \pm
{1\over 2} \sqrt{(d_1 + d_2 -1)(d_1 + d_2 -9)}
\ee
where $ c_\pm $ are integration constants. Notice that  $\delta_{\pm} >0 $, therefore the
deviation from the  symmetric solution $A_1=A_2$ always grows
in the IR. However, in this case there are no oscillations, therefore
one does not expect a discrete infinite family of solutions, nor an
Efimov spiral.

In the limiting case $d_1 +d_2 = 9$  the independent solutions are:
\be \label{degen}
(A_1-A_2)(u) \sim c_1  (u_0 - u)^{-4}  + c_2  (u_0 - u)^{-4} \log (u_0
- u),
\ee
and here too we find no discrete scaling structure.

The critical value $d_1 + d_2 =9$  is reminiscent of the BF bound for a
scalar field close to an $AdS$
extremum: there too, when the BF bound is violated the solutions are
oscillating, whereas above the BF bound both solutions have real
exponents and  grow monotonically  away from the extremum. The
difference is that here the perturbation $\epsilon$ is around an
unphysical solution since the geometry with $\epsilon=0$ is singular.

Finally, in the case where one of the spheres is flat, the solution
\be
 \mathcal{R}_2 = \mathcal{R}_1 = 0 \sp A_1 = A_2 = -\frac{u}{\ell},
\ee
is regular, as can be seen by evaluating the Riemann square \eqref{eq:Kgen}.

\end{document}